\documentclass[11pt]{article}
\usepackage{jheppub}

\usepackage{amsmath}
\usepackage{amssymb}
\usepackage{verbatim}

\usepackage{dsfont}

\usepackage[normalem]{ulem}

\usepackage[T1,OT1]{fontenc}

\usepackage{graphicx}

\allowdisplaybreaks[4]

\usepackage[hang]{footmisc}
\renewcommand{\d}{\mathrm{d}}
\newcommand{\e}{\mathrm{e}}
\newcommand{\w}{\wedge}

\newcommand{\nl}{\notag \\ &\quad\,}
\newcommand{\nll}{\notag \\ &}

\title{Control Problem of de Sitter Vacua with T-brane Uplift}

\author{Daniel Junghans}

\affiliation{Max-Planck-Institut f\" ur Physik (Werner-Heisenberg-Institut), Boltzmannstr.~8, 85748 Garching,
Germany}

\emailAdd{junghans @ mpp.mpg.de}

\notoc

\abstract{
T-branes have been argued to provide a consistent uplift to
de Sitter vacua in the context of the LARGE-volume scenario of type IIB string theory.
We point out that this construction suffers from a serious control problem: the string-frame volume is bounded from above by a number that depends only on the topology of the Calabi-Yau orientifold and is often $\mathcal{O}(1)$ or smaller.
This observation immediately rules out controlled de Sitter vacua in many models with $h^{1,1}=2$ if the non-perturbative superpotential is generated by D3-brane instantons.
We give evidence that similar conclusions apply for $h^{1,1}>2$ and show this explicitly for two models with $h^{1,1}=4$.
Our volume bound is seemingly less severe in models where the non-perturbative superpotential is generated by gaugino condensation.
However, we study $\mathcal{O}(600)$ such models and
find that, although de Sitter solutions with large volumes are naively possible there, they suffer from uncontrolled warping corrections whose volume suppression is cancelled by other factors, in particular large charges.
We also observe that the models with the smallest warping corrections within our set have very small tadpoles and at the same time require a fine-tuned string coupling and flux superpotential, which may indicate an additional problem with the axio-dilaton/complex-structure-moduli stabilization. Finally, we point out an independent challenge related to the stabilization of the brane moduli. Our various results lead us to conclude that it is unlikely that the scenario of T-brane uplifting can be realized in string theory in a self-consistent manner.

}

\begin{document}

\begin{flushright}
MPP-2026-184
\end{flushright}

\numberwithin{equation}{section}

\maketitle

\newpage

\section{Introduction}

The nature of dark energy
is one of the major puzzles of contemporary physics. In the simplest cosmological models, dark energy is a positive cosmological constant, implying that the vacuum state is a de Sitter (dS) spacetime. On the other hand, recent measurements suggest that dark energy may instead vary over time \cite{DESI:2024mwx, DESI:2025zgx}. An intriguing possibility is that such a behavior is a consequence of some yet to be understood principle of quantum gravity.
In particular, a concrete research question one may ask is whether string theory forbids dS vacua for some reason and thus predicts that dark energy is dynamical.

Naively, the answer to this question is no since a variety of scenarios for dS vacua have been proposed in string theory (see \cite{Andriot:2026lac} for a recent review). One of the most popular scenarios is the LARGE-volume scenario (LVS) \cite{Balasubramanian:2005zx, Conlon:2005ki}. Its central claim is that compactifying type IIB string theory on certain types of Calabi-Yau orientifolds yields vacuum solutions (either AdS or dS) in which the volume is stabilized at an exponentially large value so that an excellent control over $\alpha'$ and warping corrections (which are suppressed by powers of the volume) is achieved. However, while this claim is plausible in the AdS case, it was shown in \cite{Junghans:2022exo} (and further substantiated in \cite{Gao:2022fdi, Junghans:2022kxg, Hebecker:2022zme, Schreyer:2022len, Schreyer:2024pml, ValeixoBento:2023nbv}) that it is not true for dS vacua, at least when they are obtained using an anti-brane uplift. Indeed, one then observes that not all types of corrections can be made small at the same time but attempting to control one of them leads to another one blowing up and vice versa. This suggests that LVS dS vacua with anti-brane uplift cannot be constructed self-consistently.

Similar control problems have also been observed in a variety of other dS scenarios in the last few years. For example, the classical-dS scenario \cite{Hertzberg:2007wc, Silverstein:2007ac} was shown to have control problems in \cite{Junghans:2018gdb, Banlaki:2018ayh, Horer:2024hgy}\footnote{An interesting model circumventing the assumptions of the no-go of \cite{Junghans:2018gdb} was proposed in \cite{Andriot:2024cct} but is expected to suffer from uncontrolled O5-plane backreaction, as pointed out in \cite[footnote 1]{Horer:2024hgy}.}, the KKLT scenario with anti-brane uplift \cite{Kachru:2003aw} was shown to have control problems in \cite{Carta:2019rhx, Gao:2020xqh, Blumenhagen:2022dbo}\footnote{One can construct explicit KKLT dS vacua \cite{McAllister:2024lnt} which avoid the backreaction issues pointed out in \cite{Carta:2019rhx, Gao:2020xqh}, but at the cost of a small 3-cycle with a volume of order unity and thus uncontrolled curvature corrections to the anti-brane uplift. See also \cite{Farakos:2025shl} for a study of a 3d analogue of KKLT with similar control issues.}, the supercritical dS scenario \cite{Silverstein:2001xn, Maloney:2002rr} was shown to have control problems in \cite{Junghans:2023lpo}, and for models with a Casimir-energy uplift such problems were shown in \cite{Parameswaran:2024mrc} (see, however, \cite{Bento:2025wuu, Aparici:2026wdm}).
Recently, control issues were also observed in the KKLT scenario with $F$-term uplift \cite{Hebecker:2025tui}.

In view of these various results, it is natural to suspect that it is a general property of string theory that dS vacua are not possible, at least in the weakly curved and weakly coupled regimes which are computationally accessible.
This suspicion also resonates with several swampland conjectures against dS, see, e.g., \cite{Obied:2018sgi, Ooguri:2018wrx, Garg:2018reu, Andriot:2018wzk, Hebecker:2018vxz, Danielsson:2018ztv, Brennan:2017rbf} (see also \cite{Dvali:2017eba, Dvali:2018jhn} for a different line of reasoning against dS). However, since these conjectures were mainly motivated by very simple examples, it is a priori unclear under which assumptions they are true and, in particular, why something should be expected to go wrong with more sophisticated constructions like KKLT or the LVS.
On the other hand, that something goes wrong is precisely what is suggested by the growing evidence for control problems mentioned above.
It is therefore important to understand whether there is a general lesson behind these observations
or whether counter-examples exist.

In this paper, we will investigate the claim of LVS dS vacua with a \emph{T-brane uplift}. For this type of uplift, no issues have been reported so far, which is why it is often presented as a counter-example against seemingly too pessimistic views on dS in string theory. T-branes are non-commutative brane configurations introduced in \cite{Donagi:2003hh, Cecotti:2010bp}. They were argued in \cite{Cicoli:2015ylx} to provide a consistent dS uplift in the context of the LVS. See also, e.g., \cite{Cicoli:2017shd, Cicoli:2021dhg} for other works using this uplift mechanism.

The main observation of this paper is that \emph{the T-brane uplift has a very similar control problem as the anti-brane uplift}.
In particular, we will show that the string-frame volume $\mathcal{V}_s$ in a T-brane-uplifted dS vacuum satisfies an upper bound
which only depends on the topological data of the Calabi-Yau orientifold. This bound is highly constraining and yields $\mathcal{V}_s \lesssim \mathcal{O}(1)$ in many models in which the non-perturbative superpotential is generated by D3-brane instantons. This means that $\alpha'$ and warping corrections are not suppressed by powers of the volume in such models and therefore not under control.

There are also types of models in which our bound is less severe so that $\mathcal{V}_s$ can be stabilized at values significantly larger than unity. In particular, this can happen when the non-perturbative superpotential is due to gaugino condensation with a large-rank gauge group.
To assess the viability of such models, we perform a detailed study of 28 different Calabi-Yau orientifolds with $h^{1,1}=2$.
We then scan through all consistent T-brane configurations in these orientifolds, leading to $\mathcal{O}(600)$ different T-brane models with naive dS vacua.
We find that, in spite of the large volumes, warping corrections are not under control in any of these models. This is reminiscent of the LVS with anti-brane uplift, where the impossibility to control warping corrections (simultaneously with other requirements) is also the main issue \cite{Junghans:2022exo, Gao:2022fdi, Junghans:2022kxg, Hebecker:2022zme, Schreyer:2022len, Schreyer:2024pml, ValeixoBento:2023nbv}. In the T-brane case, we argue that the warping corrections are controlled by the parameter $\lambda_W = \frac{3^{2/3} 190}{9} \frac{a_s \kappa_{sss}^{1/3}}{\hat\xi^{1/3}} \frac{|Q_3|}{\mathcal{V}_s^{2/3}}$ (see the main text for definitions) up to an unknown numerical coefficient which we argue to generically be $\mathcal{O}(1)$. We then show that $\lambda_W$ cannot be made small within our set of models since the naive suppression due to the inverse volume scaling is always cancelled by some of the other factors, in particular the tadpole $Q_3$. The smallest possible values for $\lambda_W$ we find in our scan are $\mathcal{O}(1)$. Hence, the usual LVS claim of a suppression of unknown corrections by an exponentially small control parameter is false in these models.

We also discuss models with $h^{1,1}>2$ and argue that we expect similar control problems there. While we do not perform an exhaustive scan of such models, we confirm our expectation by showing that $\mathcal{V}_s \ll 1$ in two explicit dS models with $h^{1,1}=4$ constructed previously in \cite{Cicoli:2017shd, Cicoli:2021dhg}.

We stress that our finding of a large control parameter $\lambda_W \gtrsim \mathcal{O}(1)$ does not rigorously rule out the possibility of numerically small warping corrections since there might be models in which the unknown numerical prefactors multiplying the corrections are smaller than generically expected.
However, since computing all possible corrections explicitly is usually not feasible, moduli-stabilization scenarios in string theory are traditionally based on the assumption of a large-volume and small-coupling regime since then all potentially dangerous corrections are suppressed by powers of these parameters and the exact numerical prefactors are unimportant.\footnote{Since the IIB flux landscape is finite for Calabi-Yau orientifolds \cite{Douglas:2003um, Ashok:2003gk, Denef:2004ze, Grimm:2020cda, Bakker:2021uqw}, it is impossible to take limits in which the inverse volume and the string coupling go to zero and the corrections become arbitrarily well controlled. Therefore, the best one can hope for are very small but still finite control parameters.
This is different from Calabi-Yau flux vacua in type IIA, where there is evidence that limits of arbitrarily good control are possible, at least in AdS \cite{DeWolfe:2005uu, Camara:2005dc}.}
On the other hand, a construction which purely relies on the hope that something we cannot compute might be very small is much weaker than a construction in which one can plausibly tune a control parameter small.
As mentioned before, the paradigm of the LVS is that an exponentially small control parameter is provided by the inverse of the volume, but we show in this paper that this rationale fails for the T-brane uplift. This behavior is reminiscent of anti-brane-uplifted dS vacua in the LVS (and in KKLT) where it was also found that not all corrections can be simultaneously suppressed by small control parameters.

Aside from the warping corrections, we also study the tadpole $Q_3$ generated by the O3-planes and curved O7-planes/D7-branes in our set of models. The size of $Q_3$ is important since, through the tadpole condition, it determines how many fluxes can be turned on in order to stabilize the moduli and adjust the parameters $g_s$, $W_0$ (which have to be fine-tuned to get a dS vacuum). An interesting correlation we find in this context is that smaller warping corrections (i.e., smaller $\lambda_W$) correlate with a smaller $|Q_3|$. This is opposite to the anti-brane uplift, where smaller corrections correlate with \emph{larger} $|Q_3|$ \cite{Gao:2022fdi, Junghans:2022kxg}. In the T-brane case, we find that, in models for which $\lambda_W$ is near its lower bound, the (upstairs) tadpole is $|Q_3| \le 30$.\footnote{By ``upstairs'', we mean that the tadpole is measured on the Calabi-Yau covering space. The ``downstairs'' tadpole, i.e., the one on the orientifolded space, is $\le 15$.}
We find it doubtful that such a small tadpole suffices to stabilize all complex-structure moduli and at the same time perform the required fine-tuning in the $(g_s, W_0)$-plane.
Hence, even if one assumes optimistically that the warping corrections are still borderline under control near $\lambda_W\approx 1$, models which have a $\lambda_W$ near this value may still not be consistent simply because the moduli cannot be stabilized at the necessary values. While we do not prove this, our observation of small tadpoles suggests that the true lower bound on $\lambda_W$ is actually significantly larger than 1.

This paper is organized as follows. In Section \ref{sec:lvs}, we provide a detailed review of various aspects of the LVS with T-brane uplift.
In Section \ref{sec:problem}, we state the main problem of the T-brane uplift, namely the upper bound on $\mathcal{V}_s$ which must be respected in all dS vacua.
In Section \ref{sec:lvs-corr}, we discuss a number of potentially relevant corrections to the LVS potential.
In Section \ref{sec:h112}, we perform a scan of T-brane models in Calabi-Yau orientifolds with $h^{1,1}=2$. We show that $\lambda_W \gtrsim \mathcal{O}(1)$ in all of these models and that models in which $\lambda_W$ is near its lower bound have small tadpoles. In Section \ref{sec:largeh11}, we explain why we expect that models with $h^{1,1}>2$ will have similar control problems as those with $h^{1,1}=2$. We further study two explicit examples with $h^{1,1}=4$ which indeed turn out to have very small volumes. In Section \ref{sec:modstab}, we point out an additional challenge for vacua with T-branes, namely that they have unstabilized brane moduli if the assumptions of the scenario in \cite{Cicoli:2015ylx} hold. In Section \ref{sec:concl}, we conclude with a few remarks.
In App.~\ref{app:tbrane}, we verify the correct normalization of the T-brane-uplift term.
In App.~\ref{app:toric}, we provide some details regarding the algebraic construction of the various Calabi-Yau orientifolds used in our scan.

\section{LVS with T-branes}
\label{sec:lvs}

In this section, we review various topics relevant for the LVS with T-brane uplift. We also spell out a few useful details which to our knowledge have not been explicitly discussed before. 
See also, e.g., \cite{Balasubramanian:2005zx, Conlon:2005ki, Cicoli:2015ylx, Cicoli:2017shd, Cicoli:2021dhg, Crino:2020qwk} for discussions of the LVS and \cite{Cecotti:2010bp, Marchesano:2017kke, Marchesano:2019azf, Cicoli:2015ylx, Cicoli:2017shd, Cicoli:2021dhg} for discussions of T-branes.

\subsection{K\"ahler-moduli Stabilization}
\label{sec:lvs-km}

We consider the simplest version of the LVS where type IIB string theory is compactified on an orientifold of a Calabi-Yau three-fold $X$ with $h^{1,1}=h^{1,1}_+=2$ (see Section \ref{sec:largeh11} for the more general case with $h^{1,1}>2$).
We take $X$ to be of Swiss-cheese type, i.e., there is a basis of divisors $D_b$, $D_s$ such that
\begin{equation}
\mathcal{V} = \kappa_b \tau_b^{3/2}-\kappa_s \tau_s^{3/2},
\end{equation}
where $\tau_b = \frac{1}{2}\int_{D_b} \iota^*_b J^2= \frac{1}{2}\int_X J^2\w D_b$, $\tau_s = \frac{1}{2}\int_{D_s} \iota^*_s J^2= \frac{1}{2}\int_X J^2\w D_s$ are the divisor volumes, $\mathcal{V}=\frac{1}{6}\int_{X} J^3$ is the Calabi-Yau volume (all in the Einstein frame) 
and $J=t_bD_b+t_sD_s$ is the K\"ahler form.\footnote{Here and in the following, we denote by $D_i$ the divisors and also the Poincar\'e-dual 2-forms on $X$. Furthermore, $\iota_{i}^*$ denotes the pullback onto the divisor $D_i$.} The constants $\kappa_b = \frac{\sqrt{2}}{3\sqrt{\kappa_{bbb}}}$, $\kappa_s = \frac{\sqrt{2}}{3\sqrt{\kappa_{sss}}}$ are defined in terms of the triple-intersection numbers
\begin{equation}
\kappa_{bbb}=\int_X D_b^3, \qquad \kappa_{sss}=\int_X D_s^3, \label{uogjokpjtkj}
\end{equation}
and all other triple intersections of $D_b$ and $D_s$ vanish.

The (complex) bulk moduli are the axio-dilaton, $h^{2,1}_-$ complex-structure moduli and $h^{1,1}_+=2$ K\"ahler moduli $T_b=\tau_b+i\beta_b$ and $T_s=\tau_s+i\beta_s$, where the $\beta_i$'s are axions which will not be important for our analysis. We ignore D7-brane moduli for the moment but will come back to them in the Sections \ref{sec:lvs-d7}--\ref{sec:lvs-up2}.
The axio-dilaton and the complex-structure moduli are assumed to be stabilized supersymmetrically using fluxes as in \cite{Giddings:2001yu, Kachru:2003aw}. The dynamics of the two K\"ahler moduli approximately decouples at large volumes \cite{Achucarro:2008sy, Gallego:2008qi} and is governed by an effective $\mathcal{N}=1$ supergravity theory with K\"ahler potential
\begin{equation}
K = - 2 \ln \left( \mathcal{V} + \frac{\hat\xi}{2g_s^{3/2}} \right) - \ln\left(\frac{2}{g_s}\right) + K_\text{cs}, \label{k}
\end{equation}
where $\mathcal{V}$ should be read as a function of $T_b$ and $T_s$, and $K_\text{cs}$ is an irrelevant constant that will be set to zero in the following. We included in $K$ the leading $\alpha'$ correction
with
\begin{equation}
\hat \xi = -\frac{ \zeta(3)}{2(2\pi)^3} \hat \chi(X) = -\frac{\zeta(3)}{2(2\pi)^3} \left(\chi(X) + 2\int_X D_\text{O7}^3\right). \label{zjghjgjg}
\end{equation}
Here, the first term is the BBHL term \cite{Becker:2002nn}, where $\zeta(3)\approx 1.20$ and $\chi(X)$ is the Euler characteristic of the Calabi-Yau manifold. The second term is an orientifold correction derived in \cite{Minasian:2015bxa} which arises in the presence of an O7-plane wrapped on a divisor $D_\text{O7}$.

The superpotential is assumed in the LVS to be of the form
\begin{equation}
W = W_0 + A_s \e^{-a_s T_s}, \label{w}
\end{equation}
where $W_0$, $A_s$ and $a_s$ are constants. The $W_0$ term comes from stabilizing the axio-dilaton and the complex-structure moduli with fluxes \cite{Gukov:1999ya, Giddings:2001yu}, while the exponential term $W_\text{np}\equiv A_s \e^{-a_s T_s}$ is generated by a non-perturbative effect on $D_s$ (see Section \ref{sec:lvs-np}).

From \eqref{k} and \eqref{w}, we can compute the $F$-term scalar potential $V = \e^K ( |DW|^2-3|W|^2)$ with the result
\begin{equation}
V = \frac{4a_s^2|A_s|^2g_s\sqrt{\tau_s}\e^{-2a_s\tau_s}}{3\kappa_s\mathcal{V}} 
- \frac{2a_s|A_s| |W_0| g_s \tau_s \e^{-a_s\tau_s}}{\mathcal{V}^2}
+ \frac{3 \hat\xi |W_0|^2}{8\sqrt{g_s}\mathcal{V}^3}
+\frac{C_\text{up}}{\mathcal{V}^{x}}. \label{lvspotential}
\end{equation}
Here we already integrated out the two axions $\beta_b$, $\beta_s$ so that the potential only depends on the two real moduli $\tau_b$ and $\tau_s$. We also added by hand an uplift term $C_\text{up} \mathcal{V}^{-x}$ with (so far) arbitrary coefficient and volume scaling. We will discuss and justify this term in more detail in Section \ref{sec:lvs-up2}.

To determine the conditions for dS minima of $V$, we will follow the approach used in \cite{Crino:2020qwk, Junghans:2022exo} for the case of an anti-brane uplift, which we slightly generalize here to arbitrary uplift terms. We first solve the equations of motion for $\tau_b$ and $\tau_s$, which yields
\begin{align}
\mathcal{V} &=\frac{3(a_s\tau_s -1)\kappa_s|W_0| \sqrt{\tau_s}}{(4a_s\tau_s -1)a_s|A_s|} \e^{a_s\tau_s}, \label{sglisgljs}\\
\tau_s &= \frac{\hat\xi^{2/3}}{(2\kappa_s)^{2/3} g_s} + \frac{1+\frac{x}{3-x}\alpha}{3a_s} + \mathcal{O}(g_s). \label{sglisgljs2}
\end{align}
Here we expanded in $g_s$ to find the solution for $\tau_s$ and introduced the uplift parameter
\begin{equation}
\alpha = \frac{48(3-x)}{2^{2/3}27} \frac{C_\text{up}a_s \mathcal{V}^{3-x}}{\sqrt{g_s} |W_0|^2\kappa_s^{2/3}\hat\xi^{1/3}}. \label{alpha}
\end{equation}

Substituting the solution into \eqref{lvspotential}, we obtain the vacuum energy
\begin{equation}
V_0=\frac{\hat\xi^{1/3}\kappa_s^{2/3}\sqrt{g_s} |W_0|^2}{a_s\mathcal{V}^3} \frac{2^{2/3}3}{16} \left(\alpha-1\right) \label{dsgsgsfsgjsf}
\end{equation}
up to subleading terms in $g_s$.
The (canonically normalized) eigenvalues of the mass matrix $\partial_i\partial_j V$ in the $\tau_b,\tau_s$ moduli space are
\begin{align}
m_1^2 &= \frac{a_s^2|W_0|^2\hat\xi^{4/3}}{2^{1/3}\kappa_s^{4/3} g_s \mathcal{V}^2}, \\
m_2^2 &= \frac{81\hat\xi^{1/3} \kappa_s^{2/3} \sqrt{g_s} |W_0|^2}{2^{1/3} 16 a_s\mathcal{V}^3} \left(1-\frac{x}{3}\alpha\right)
\end{align}
up to subleading terms in $g_s$ and $1/\mathcal{V}$. For meta-stable dS vacua, $\alpha$ thus has to lie in the range
\begin{equation}
1 < \alpha < \frac{3}{x}.
\end{equation}
We therefore need an uplift term with $x<3$, i.e., one that scales less than cubically with the inverse volume in \eqref{lvspotential}. The anti-brane uplift satisfies this with $x=\frac{4}{3}$ \cite{Crino:2020qwk, Junghans:2022exo}, whereas the T-brane uplift has $x=\frac{8}{3}$, as we will review in Section \ref{sec:lvs-up2}.

A common lore is that the LVS construction guarantees exponentially large volumes and therefore an excellent control over $\alpha'$ and warping corrections (since they are suppressed by powers of the volume). Indeed, according to \eqref{sglisgljs}, $\mathcal{V} \sim \e^{a_s\tau_s}$ so that it seems that huge volumes are possible as long as we can ensure moderately large $a_s\tau_s$. However, this is actually only true for AdS vacua. In dS, the volume is constrained in addition by \eqref{alpha} (since $C_\text{up}\neq 0$, $\alpha=\mathcal{O}(1)$) and therefore \emph{not} exponentially large unless we are able to tune the uplift coefficient $C_\text{up}$ exponentially small. We will see below that this does not work for the T-brane uplift. In particular, the string-frame volume $\mathcal{V}_s=g_s^{3/2}\mathcal{V}$ which controls the $\alpha'$ and warping corrections will turn out to be tightly constrained by \eqref{alpha}, to the effect that corrections are not well controlled at all in dS.

\subsection{D7-brane Worldvolume Theory}
\label{sec:lvs-d7}

To generate a T-brane uplift as in \cite{Cicoli:2015ylx, Cicoli:2017shd, Cicoli:2021dhg}, we assume a compactification in which a stack of $N \in 2\mathbb{N}$ D7-branes is wrapped on a divisor $D_\text{T}$ with a non-vanishing part along $D_b$, i.e.,
\begin{equation}
D_\text{T} = p_b D_b+p_sD_s, \qquad p_b\neq 0. \label{dt}
\end{equation}
The divisor should furthermore be effective so that the D7-branes are supersymmetric (before turning on worldvolume fluxes) \cite{Becker:1995kb, Marino:1999af, Jockers:2004yj, Denef:2008wq}. For Calabi-Yau manifolds obtained as hypersurfaces of an ambient toric variety $A$, effective divisors descend from positive integral linear combinations of the irreducible divisors of $A$. This implies $p_b \in \mathbb{N}$, $p_s \in \mathbb{Z}$ in the models we will study in this paper.\footnote{
In the models we will consider, the irreducible divisors are combinations of positive integer multiples of $D_b$ and integer multiples of $D_s$, and therefore the effective divisors are as well (cf.~App.~\ref{app:toric}). In principle, Calabi-Yau manifolds can have additional effective divisors which are not effective on the ambient space and need not be positive linear combinations of the irreducible ones (see, e.g., \cite{Demirtas:2018akl, McAllister:2024lnt} for discussions). It is not known how to compute them in general so that we will not consider them.} Since $D_\text{T}$ is an orientifold-even divisor, the gauge group is either SO$(N)$ (if the divisor is pointwise invariant, i.e., the branes lie on top of an O7-plane) or USp$(N)$ (if the divisor is invariant but not pointwise, i.e., the branes do not lie on top of an O7-plane) \cite{Blumenhagen:2008zz}.
Note that $N$ in our convention counts the ``upstairs'' brane number in the stack, i.e., including image branes.

The local physics of the D7-branes is described by a twisted 8d super-Yang-Mills theory with an internal $(2,0)$-form $\Phi$ (i.e., its legs are along $D_\text{T}$) and a 1-form gauge field $A$ \cite{Beasley:2008dc}. In a 4d vacuum, the internal background fields satisfy the $F$-term and $D$-term equations
\begin{equation}
\mathcal{F}^{2,0} = \mathcal{F}^{0,2} = 0, \qquad \bar\partial \Phi +i[A,\Phi]=0, \qquad \iota^*_\text{T} J \w \mathcal{F} + \frac{1}{2\pi}[\Phi, \Phi^\dagger] = 0, \label{d7eom}
\end{equation}
where $\iota^*_\text{T}: H^{2}(X) \to H^{2}(D_\text{T})$ denotes the pullback of harmonic 2-forms on $X$ onto the brane worldvolume $D_\text{T}$.
Furthermore, $\mathcal{F}$ is the gauge-invariant field strength\footnote{We work in units where $2\pi\sqrt{\alpha'}=1$.}
\begin{equation}
\mathcal{F} = \frac{1}{2\pi}F_2 + \iota^*_\text{T} B_2\, \mathds{1}_N, \label{dsgssgjgsg}
\end{equation}
where $F_2$ is the usual Yang-Mills field strength and $B_2$ is the NSNS 2-form. We use here the conventions of \cite{Junghans:2026jza} where \eqref{d7eom} was derived from the non-Abelian DBI action.

Note that \eqref{d7eom} is valid for small $g_s$, large $\mathcal{V}$ and small brane fields since the equations ignore backreaction effects (warping, varying dilaton, etc.) and $\alpha'$ corrections. They furthermore do not take into account the effect of bulk fluxes. The latter couple to $\Phi$ in the D7-brane action \cite{Camara:2004jj, Gomis:2005wc, Junghans:2026jza} and thus yield a correction to the equations of motion that will be responsible for the T-brane uplift \cite{Cicoli:2015ylx}. We will discuss this and other corrections in more detail later but assume for the moment that \eqref{d7eom} holds in order to study the leading-order dynamics of the brane fields.

Before we do so, let us also make a few comments on the orientifold projection. Without orientifolding, $\Phi$, $A$ and $\mathcal{F}$ are valued in the adjoint of U$(N)$. We can thus expand
\begin{equation}
\Phi = \Phi^a T^a, \qquad A = A^a T^a, \qquad \mathcal{F} = \mathcal{F}^a T^a,
\end{equation}
where the $\Phi^a$'s are (2,0)-forms, the $A^a$'s are real 1-forms, the $\mathcal{F}^a$'s are real 2-forms and the $T^a$'s with $a=0,\ldots, N^2-1$ are the U$(N)$ generators, which in our convention are Hermitian.
The orientifolding imposes constraints on these fields so that some of their components are projected out \cite{Gimon:1996rq, Douglas:1996sw} (see also, e.g., \cite{Jockers:2004yj, Collinucci:2008pf, Blumenhagen:2008zz, Collinucci:2008sq} for discussions of the D7-brane case). One finds
\begin{equation}
\Phi = - M \sigma^*(\Phi^T) M^{-1}, \qquad A = - M \sigma^* (A^T) M^{-1}, \qquad \mathcal{F} = - M \sigma^* (\mathcal{F}^T) M^{-1}, \label{gimon}
\end{equation}
where $\sigma$ is the orientifold involution and $M$ is an $N\times N$ matrix which, up to a basis change, equals either
$M = \left(\begin{smallmatrix} 0 & \mathds{1}_{N/2} \\ \mathds{1}_{N/2} & 0  \end{smallmatrix} \right)$ (for a D7 stack on top of an O7-plane)\footnote{A common choice in the literature is to take instead $M = \mathds{1}_N$, which is related to our choice by $\Phi \to U \Phi U^{-1}$, $A \to U A U^{-1}$, $\mathcal{F} \to U \mathcal{F} U^{-1}$, $M\to UMU^T$ with $U= \frac{i-1}{2}\left(\begin{smallmatrix} \mathds{1}_{N/2} & i\mathds{1}_{N/2} \\ i\mathds{1}_{N/2} & \mathds{1}_{N/2}  \end{smallmatrix} \right) $.} or $M = \left(\begin{smallmatrix} 0 & \mathds{1}_{N/2} \\ - \mathds{1}_{N/2} & 0  \end{smallmatrix} \right) $ (for a D7 stack on an orientifold-even but not pointwise invariant divisor).\footnote{Determining the correct sign in the constraint for $\Phi$ is somewhat subtle. To verify it, consider a Calabi-Yau manifold with local complex coordinates $z^1$, $z^2$, $z$ and an orientifold-invariant D7 stack located at $z=0$. We then have $\Phi \sim \Omega_{12z}|_{z=0}\phi^z \d z^1 \w \d z^2$ in terms of the holomorphic $(3,0)$-form $\Omega_3$ and the adjoint worldvolume scalar $\phi^z$ describing the normal deformations of the branes \cite{Jockers:2004yj, Junghans:2026jza}. Since $\Omega_3 = -\sigma^* \Omega_3$ for O3/O7 orientifolds \cite{Acharya:2002ag, Brunner:2003zm}, we further have $\Omega_{12z} \d z^1 \w \d z^2 = \pm \sigma^* (\Omega_{12z} \d z^1 \w \d z^2)$, where the upper sign is for a stack on top of an O7-plane (i.e., for an involution $\sigma: z\to -z$) and the lower sign is for a transversally intersecting stack. Using this, \eqref{gimon} is equivalent to $\phi^z = \mp M \sigma^*(\phi^z)^T M^{-1}$.
Taking as a simple example two D7-branes on top of an O7-plane and assuming a constant scalar vev along the worldvolume, we obtain $\phi^z = \left(\begin{smallmatrix} +c & 0 \\ 0 & - c \end{smallmatrix} \right)$ for some constant $c$. Hence, the position of a brane moved off the O7-plane is constrained to be minus the position of its image, consistently with what one would expect. For a transversally intersecting stack, we instead get $\phi^z = \left(\begin{smallmatrix} +c & 0 \\ 0 & +c \end{smallmatrix} \right)$ so that the two branes can only move as a pair, which is consistent with observations in \cite{Sen:1996xx, Braun:2008ua, Collinucci:2008pf}.}
Note that the constraint on $A$ only holds up to (large) gauge transformations shifting $F_2$ and $B_2$, which we will come back to in Section \ref{sec:lvs-flux}.

The projection with symmetric $M$ is sometimes called SO projection since, for vanishing $\Phi$ vev and background flux, it yields a massless 4d gauge field $A_\mu$ in the adjoint of SO$(N)$ (as follows from \eqref{gimon} using that $A_\mu = \sigma^* (A_\mu)$ on a pointwise-invariant divisor). When a $\Phi$ vev or flux is turned on, some of the 4d gauge bosons become massive and the gauge group is broken to a subgroup of SO$(N)$.
The projection with anti-symmetric $M$ is called Sp projection since, for vanishing $\Phi$ and flux, it yields a massless 4d gauge field valued in the adjoint of USp$(N)$ (note that the zero mode of $A_\mu$ does not depend on the internal coordinates and thus satisfies $A_\mu = \sigma^* (A_\mu)$).
To work out how the projection constrains $\Phi$ and the internal components of $A$, $\mathcal{F}$, we have to take into account that they split into orientifold-even and odd forms (e.g., $\Phi=\Phi_+ + \Phi_-$ with $\Phi_\pm= \pm\sigma^*(\Phi_\pm)$), which are in different representations. In particular, for the Sp projection, \eqref{gimon} implies that orientifold-even forms are valued in the adjoint of USp$(N)$, while orientifold-odd forms are valued in the two-index anti-symmetric representation. For the SO projection, we only have to consider orientifold-even forms since there are no odd ones on a pointwise-invariant divisor. The former are valued in the adjoint of SO$(N)$. See, e.g., \cite{Ibanez:2007rs} for a review.

As an example, consider the case $N=2$. The gauge group before orientifolding is then U$(2)$ so that we can write
\begin{align}
& \text{U}(2): && \Phi = \begin{pmatrix} \Phi^0 + \Phi^1 & \Phi^2 + i\Phi^3 \\ \Phi^2 - i\Phi^3 & \Phi^0 - \Phi^1 \end{pmatrix}, \qquad
\mathcal{F} = \begin{pmatrix} \mathcal{F}^0 + \mathcal{F}^1 & \mathcal{F}^2 + i\mathcal{F}^3 \\ \mathcal{F}^2 - i\mathcal{F}^3 & \mathcal{F}^0 - \mathcal{F}^1 \end{pmatrix},
\end{align}
and analogously for $A$. After the orientifold projection, the surviving components are
\begin{align}
& \text{SO}(2): && \Phi = \begin{pmatrix} \Phi^1_+ & 0 \\ 0 & -\Phi^1_+ \end{pmatrix}, && \mathcal{F} = \begin{pmatrix} \mathcal{F}^1_+ & 0 \\ 0 & -\mathcal{F}^1_+ \end{pmatrix}, \\
& \text{USp}(2)\simeq \text{SU}(2): && \Phi = \begin{pmatrix} \Phi^0_- + \Phi^1_+ & \Phi^2_+ + i\Phi^3_+ \\ \Phi^2_+ - i\Phi^3_+ & \Phi^0_- - \Phi^1_+ \end{pmatrix}, &&
\mathcal{F} = \begin{pmatrix} \mathcal{F}^0_- + \mathcal{F}^1_+ & \mathcal{F}^2_+ + i\mathcal{F}^3_+ \\ \mathcal{F}^2_+ - i\mathcal{F}^3_+ & \mathcal{F}^0_- - \mathcal{F}^1_+ \end{pmatrix},
\end{align}
where the subscripts $\pm$ indicate orientifold-even/odd 2-forms.

For general $N$, \eqref{gimon} implies that, up to basis changes, $\Phi$ and $\mathcal{F}$ take the form
\begin{align}
& \text{SO}(N): && \Phi = \begin{pmatrix} P_+ & A_{1+} \\
A_{2+} & -P_+^T \\ \end{pmatrix}, && \mathcal{F} = \begin{pmatrix} \mathcal{H}_+ & \mathcal{A}_+ \\
\mathcal{A}_+^\dagger & -\mathcal{H}_+^T \\ \end{pmatrix}, \label{evenodd1} \\
& \text{USp}(N): &&\Phi = \begin{pmatrix} P_-+P_+ & A_{1-} + S_{1+} \\
A_{2-} + S_{2+} & P_-^T-P_+^T \\ \end{pmatrix}, && \mathcal{F} = \begin{pmatrix} \mathcal{H}_- + \mathcal{H}_+ & \mathcal{A}_-+\mathcal{S}_+ \\
\mathcal{A}_-^\dagger+\mathcal{S}_+^\dagger & \mathcal{H}_-^T-\mathcal{H}_+^T \\ \end{pmatrix} \label{evenodd2}
\end{align}
in terms of $\frac{N}{2}\times\frac{N}{2}$ matrices (whose entries are 2-forms), where $\mathcal{H}_\pm$ are Hermitian, $A_{i\pm}$, $\mathcal{A}_\pm$ are anti-symmetric, $S_{i+}$, $\mathcal{S}_+$ are symmetric, $P_\pm$ are unconstrained and the subscripts $\pm$ again stand for orientifold-even/odd.
We thus see that, on general D7 stacks, both even and odd forms arise in the non-Abelian worldvolume fields. The T-brane vacua we will study in this paper have diagonal fluxes with $\mathcal{H}_+\neq 0$ and off-diagonal $\Phi$ vevs with $A_{1\pm},S_{1+}\neq 0$. Hence, the fluxes we will consider are orientifold-even, while $\Phi$ can in general acquire orientifold-even or odd vevs in our setup.

\subsection{T-branes}
\label{sec:lvs-up}

We now want to construct a T-brane solution to the $F$-term and $D$-term equations \eqref{d7eom}. T-branes are solutions in which $\Phi$ acquires a vev in the upper-right \emph{triangle} above the diagonal (hence the ``T'') such that $[\Phi,\Phi^\dagger]\neq 0$ \cite{Cecotti:2010bp}. This can be interpreted as a bound state of the $N$ D7-branes in the stack. To obtain such a configuration, we turn on a worldvolume flux valued in the Cartan subalgebra of the gauge algebra.
For convenience, we choose a basis of generators such that the Cartan generators are diagonal (and traceless) matrices.\footnote{
For SO$(N)$, this would not be the case in the standard basis where the generators are anti-symmetric matrices. However, in the basis we chose in Section \ref{sec:lvs-d7}, we can take diagonal fluxes for both SO$(N)$ and USp$(N)$, cf.~\eqref{evenodd1}, \eqref{evenodd2}.}
We further take the flux on the $\frac{N}{2}$ branes and their $\frac{N}{2}$ orientifold images to be the same up to a minus sign as in \cite{Cicoli:2015ylx, Cicoli:2017shd, Cicoli:2021dhg}. Our ansatz is thus
\begin{equation}
\mathcal{F} = -\hat{\mathcal{F}} \begin{pmatrix} \mathds{1}_{N/2} & 0 \\
0 & -\mathds{1}_{N/2} \\ \end{pmatrix}, \label{tzptzitjli0}
\end{equation}
where $\hat{\mathcal{F}}$ is a closed 2-form on $D_\text{T}$. The non-trivial gauge bundle in this case is a direct sum of $N$ line bundles $\mathcal{L}\oplus \mathcal{L} \oplus \ldots \oplus \mathcal{L}^{-1} \oplus \mathcal{L}^{-1}$ with curvatures $c_1(\mathcal{L}^{\pm 1})\sim \mp \hat{\mathcal{F}}$.
The first equation in \eqref{d7eom} implies that $\hat{\mathcal{F}} \in H^{1,1}(D_\text{T})$ in cohomology \cite{Marino:1999af, Beasley:2008dc} and,
according to \eqref{evenodd1}, \eqref{evenodd2}, $\hat{\mathcal{F}}$ must furthermore be orientifold-even.
We also require $\hat{\mathcal{F}}$ to not be primitive,
\begin{equation}
\iota^*_\text{T} J \w \hat{\mathcal{F}} \neq 0,
\end{equation}
since otherwise the $D$-term equation (the last one in \eqref{d7eom}) yields $[\Phi,\Phi^\dagger]=0$ and no T-brane is possible.

In order to solve \eqref{d7eom} in the presence of the flux, we can locally make the ansatz
\begin{equation}
\mathcal{F} = \frac{1}{2\pi}\d A = \frac{i}{2\pi} \partial \bar\partial f \begin{pmatrix} \mathds{1}_{N/2} & 0 \\
0 & -\mathds{1}_{N/2} \\ \end{pmatrix}, \qquad A =  \frac{i\bar\partial f}{2} \begin{pmatrix} \mathds{1}_{N/2} & 0 \\
0 & -\mathds{1}_{N/2} \\ \end{pmatrix} + \text{c.c.} \label{ansatzfa}
\end{equation}
for some real $f=f(z^i,\bar z^{\bar\jmath})$, where $z^1$, $z^2$ are complex coordinates on $D_\text{T}$.\footnote{Here we assume $\iota^*_\text{T}B_2=0$ for simplicity so that $\mathcal{F} = \frac{1}{2\pi}F_2 = \frac{1}{2\pi}\d A$. For $\iota^*_\text{T}B_2\neq 0$, we instead have $\frac{1}{2\pi}\d A =$ $ \mathcal{F} - \iota^*_\text{T}B_2 \mathds{1}_N = \frac{i}{2\pi} \partial \bar\partial f \left(\begin{smallmatrix} \mathds{1}_{N/2} & 0 \\
0 & -\mathds{1}_{N/2} \\ \end{smallmatrix}\right) - \iota^*_\text{T}B_2 \mathds{1}_N$. Hence, $A$ acquires an extra term $\sim \mathds{1}_N$ compared to \eqref{ansatzfa}. This does not affect the solution for $\Phi$ since it cancels out in \eqref{d7eom} (in particular, in $[A,\Phi]$).}
We further make the ansatz (valid again in a local patch of $D_\text{T}$)
\begin{equation}
\Phi = \e^f\omega \begin{pmatrix} 0 & \mathcal{M} \\
0 & 0 \\ \end{pmatrix}, \label{ansatzphi}
\end{equation}
where $\omega$ is a closed $(2,0)$-form and $\mathcal{M}$ is an $\frac{N}{2}\times\frac{N}{2}$ matrix satisfying $\mathcal{M} \mathcal{M}^\dagger = \mathcal{M}^\dagger \mathcal{M} = \mathds{1}_{N/2}$. Globally, $\e^f \bar \omega$ is a cohomology class in $H^{2}(D_\text{T},\mathcal{L}^{-2})$ (see \cite[Secs.~5 \& 6]{Junghans:2026jza} for more details).
According to \eqref{evenodd1}, \eqref{evenodd2}, it depends on the gauge group and the choice of $\mathcal{M}$ whether $\e^f\omega$ is orientifold-even or odd: In the USp$(N)$ case, $\e^f\omega$ has to be orientifold-odd for anti-symmetric $\mathcal{M}$ and orientifold-even for symmetric $\mathcal{M}$. In the SO$(N)$ case, only orientifold-even $\e^f\omega$ and anti-symmetric $\mathcal{M}$ is allowed.\footnote{Note that this implies that T-branes can only exist for $N\ge 4$ in the SO$(N)$ case, consistent with the fact that SO$(2)$ is an Abelian group and thus cannot yield solutions with $[\Phi,\Phi^\dagger]\neq 0$. In the USp$(N)$ case, T-branes are possible for any even $N$, the simplest example being $N=2$ with gauge group USp$(2)\simeq$ SU$(2)$.}

Substituting \eqref{ansatzfa} and \eqref{ansatzphi} into \eqref{d7eom}, we see that the two $F$-term equations are automatically solved and the $D$-term equation reduces to a differential equation determining $f$:
\begin{equation}
\iota^*_\text{T} J \w i \partial \bar\partial f +\e^{2f}\omega \w \bar \omega = 0. \label{df}
\end{equation}
We refer to \cite{Cecotti:2010bp, Marchesano:2017kke} for more details on how to solve this equation. For the purpose of this paper, it will be sufficient to know that we consider T-brane backgrounds with a diagonal flux of the form \eqref{tzptzitjli0} and an off-diagonal $\Phi$ vev of the form \eqref{ansatzphi}.

\subsection{Constraints on the Worldvolume Flux}
\label{sec:lvs-flux}

Let us now discuss in more detail which fluxes $\hat{\mathcal{F}}$ can be turned on.
In general, $H^{1,1}(D_\text{T})$ contains elements which are pulled back from harmonic forms on the ambient Calabi-Yau (i.e., forms in $H^{1,1}(X)$) and elements which cannot be obtained from such pullbacks.
See, e.g., \cite{Jockers:2005zy, Buican:2006sn, Blumenhagen:2008zz, Grimm:2011dj} for discussions.
Pullback $(1,1)$-forms are necessarily orientifold-even in our setup (recall that we assume $h^{1,1}_-(X)=0$), while $(1,1)$-forms that are not pullbacks can have even and odd pieces in general.
One can furthermore choose a basis of $(1,1)$-forms such that the pullback basis forms are orthogonal to (i.e., wedge to zero with) the non-pullback basis forms. We will call any $(1,1)$-form that is a linear combination of the latter type of basis elements ``variable'', following the terminology of \cite{Grimm:2011dj}. The cohomology group thus splits into two subgroups: $H^{1,1}(D_\text{T},\mathbb{R})= H^{1,1}_\text{pull}(D_\text{T},\mathbb{R}) \oplus H^{1,1}_\text{var}(D_\text{T},\mathbb{R})$. Since a T-brane requires $\iota^*_\text{T} J \w \hat{\mathcal{F}} \neq 0$, it is clear that $\hat{\mathcal{F}}$ must contain a pullback piece (consistently with the fact that, as explained previously, $\hat{\mathcal{F}}$ is orientifold-even).
We can thus write
\begin{equation}
\hat{\mathcal{F}} = f_b \,\iota^*_\text{T} D_b + f_s \,\iota^*_\text{T} D_s + \hat{\mathcal{F}}_\text{var} \label{tzptzitjli}
\end{equation}
up to exact terms\footnote{As explained in \cite{Marchesano:2017kke}, $\hat{\mathcal{F}}$ is not harmonic in a T-brane background.}, where $f_b$, $f_s$ are some (yet to be determined) flux numbers and $\hat{\mathcal{F}}_\text{var}$ denotes possible variable fluxes that wedge to zero with the two pullback terms. Note that, in the case $p_s=0$, there is no $\iota^*_\text{T} D_s$ component since the pullback of the form $D_s$ onto the divisor $D_\text{T}=p_bD_b$ then vanishes.

We now discuss the quantization conditions for $f_b$, $f_s$.
The worldvolume flux is quantized as \cite{Freed:1999vc, Collinucci:2010gz, Blumenhagen:2008zz}
\begin{equation}
\mathcal{F} -\left( \iota^*_\text{T} B_2 + \frac{\iota^*_\text{T} D_\text{T}}{2} \right) \mathds{1}_N \in H^{2}(D_\text{T},\mathbb{Z})_{N\times N}, \label{dsgjsdjglisgj}
\end{equation}
where we used that $c_1(D_\text{T})=-\iota^*_\text{T} D_\text{T}$ on a Calabi-Yau manifold and the notation on the right-hand side means that all elements of the $N\times N$ matrix on the left-hand side are in $H^{2}(D_\text{T},\mathbb{Z})$.\footnote{Note that this formula holds for Abelian flux and only in a basis where the Cartan generators are diagonal, as the matrix $H^{2}(D_\text{T},\mathbb{Z})_{N\times N}$ is not basis-invariant.}
More specifically, they are in $H^{1,1}(D_\text{T},\mathbb{Z})$ since $\mathcal{F}$, $\iota^*_\text{T} B_2$ and $\iota^*_\text{T} D_\text{T}$ are all $(1,1)$-forms.
Using furthermore \eqref{tzptzitjli0} in \eqref{dsgjsdjglisgj} yields
\begin{equation}
\hat{\mathcal{F}} \pm \left(\iota^*_\text{T} B_2 + \frac{\iota^*_\text{T} D_\text{T}}{2} \right) \in H^{1,1}(D_\text{T},\mathbb{Z}). \label{sdfdlifaj}
\end{equation}
Determining the elements of $H^{1,1}(D_\text{T},\mathbb{Z})$ is somewhat subtle.
One may think that they are simply given by $m_b \iota^*_\text{T} D_b + m_s \iota^*_\text{T} D_s + \eta$ for some $m_b,m_s\in \mathbb{Z}$ and some variable form $\eta$.
However, this is in general not correct and instead also \emph{fractional} values of $m_b$, $m_s$ can occur. This can happen for three reasons:
\begin{itemize}
\item If $D_b$ and $D_s$ are not an integral basis of $H^{1,1}(X,\mathbb{Z})$, the group contains elements involving fractions of them. Pulling these forms back to $D_\text{T}$ then yields elements in $H^{1,1}(D_\text{T},\mathbb{Z})$ with $m_b,m_s\in \mathbb{Q}$ and $\eta=0$. See, e.g., \cite{Cicoli:2017shd, Crino:2020qwk} for explicit examples.

\item Furthermore, the pullback forms in $H^{1,1}(D_\text{T},\mathbb{Z})$ need in principle not be inherited from $H^{1,1}(X,\mathbb{Z})$ but may also come from $H^{1,1}(X,\mathbb{Q})$. In the latter case, $H^{1,1}(D_\text{T},\mathbb{Z})$ has elements with $m_b,m_s\in \mathbb{Q}$ and $\eta=0$ even when $D_b$, $D_s$ are an integral basis of $H^{1,1}(X,\mathbb{Z})$. It was argued in \cite{Denef:2007vg} that this cannot happen for divisors for which the Lefschetz hyperplane theorem holds. However, this requires $D_\text{T}$ to be ample (i.e., to lie in the K\"ahler cone), which is in general not true for $D_\text{T}$.\footnote{As stated before, one should however impose that $D_\text{T}$ is effective.}

\item Even when the above two caveats do not apply -- i.e., when all pullback forms in $H^{1,1}(D_\text{T},\mathbb{Z})$ have integral $m_b$, $m_s$ -- it can still happen that the group contains \emph{non-pullback} forms with $m_b, m_s \in \mathbb{Q}$ and $\eta\neq 0$. Such group elements mixing fractional pullback and variable terms were called ``glue vectors'' in \cite{Denef:2007vg}. Their appearance means that $H^{1,1}(D_\text{T},\mathbb{Z})$ -- unlike $H^{1,1}(D_\text{T},\mathbb{R})$ -- does typically not split into two subgroups: $H^{1,1}(D_\text{T},\mathbb{Z}) \neq H^{1,1}_\text{pull}(D_\text{T},\mathbb{Z}) \oplus H^{1,1}_\text{var}(D_\text{T},\mathbb{Z})$. Instead, the pullback forms and the variable forms only generate a sublattice of the full lattice of all integral group elements. The missing elements form a finite subgroup of $H^{1,1}(D_\text{T},\mathbb{Z})$ which is generated precisely by the glue vectors \cite{Denef:2007vg}.\footnote{As an illustrative example, consider the case $D_\text{T}=D_b$. Assuming that $H^{1,1}(X,\mathbb{Z})$ consists of integer combinations of $D_b$ and $D_s$, the pullback forms inherited by $H^{1,1}(D_b,\mathbb{Z})$ are $\iota^*_b D_b$ and integer multiples thereof (since $\iota^*_b D_s$ vanishes). Now, if $\iota^*_b D_b$ were the generator of an integral subgroup $H^{1,1}_\text{pull}(D_b,\mathbb{Z})$, there would have to be a basis 2-cycle over which this form integrates to 1, and the Poincar\'e dual of this 2-cycle would have to equal $n \iota^*_b D_b$ for some $n\in\mathbb{Z}$. However, $\int_{D_b} \iota^*_b D_b \w n \iota^*_b D_b = n \kappa_{bbb}$, which cannot equal 1 whenever $\kappa_{bbb}\neq 1$. Hence, it is clear in this case that $\iota^*_b D_b$ does not generate an integral subgroup and therefore $H^{1,1}(D_b,\mathbb{Z}) \neq H^{1,1}_\text{pull}(D_b,\mathbb{Z}) \oplus H^{1,1}_\text{var}(D_b,\mathbb{Z})$.}

\end{itemize}
The upshot of the discussion is that $m_b$, $m_s$ need not be integers unless $D_b$, $D_s$ are an integral basis of $H^{1,1}(X,\mathbb{Z})$, $D_\text{T}$ is ample, and variable forms are turned off.
Due to these caveats, it is in general a highly non-trivial task to identify the elements of $H^{1,1}(D_\text{T},\mathbb{Z})$ and, consequently, the correct quantization of the flux numbers $f_b$, $f_s$.

Luckily, the first caveat will not play a role in this paper since, in all models we will consider, $D_b$ and $D_s$ are an integral basis of $H^{1,1}(X,\mathbb{Z})$ (with the exception of one model briefly discussed in Section \ref{sec:largeh11} which will be treated separately).
In order to bypass the other two caveats, we will furthermore restrict to elements of $H^{1,1}(D_\text{T},\mathbb{Z})$ pulled back from $H^{1,1}(X,\mathbb{Z})$. This assumption greatly simplifies the analysis and is still general enough to allow us to study a variety of different models. Note that here we follow previous works on T-brane uplifting \cite{Cicoli:2015ylx, Cicoli:2017shd, Cicoli:2021dhg}, which made the same assumption.

To summarize, the elements of $H^{1,1}(D_\text{T},\mathbb{Z})$ we consider in \eqref{sdfdlifaj} are of the form $m_b \iota^*_\text{T} D_b + m_s \iota^*_\text{T} D_s$ with $m_b, m_s\in \mathbb{Z}$. Using this together with \eqref{dt}, we arrive at
\begin{equation}
\hat{\mathcal{F}} = \left(m_b - \frac{p_b}{2}\right) \iota^*_\text{T} D_b + \left(m_s - \frac{p_s}{2}\right) \iota^*_\text{T} D_s - \iota^*_\text{T} B_2, \qquad m_b,m_s,p_s\in \mathbb{Z}, \quad p_b\in \mathbb{N}, \label{sdgfmlsgls}
\end{equation}
where we took the upper sign in \eqref{sdfdlifaj}.\footnote{One can verify that taking the lower sign eventually leads to the same constraint \eqref{srgligshligrng}.}

The last remaining piece to understand is the $B_2$ field.
Since $B_2$ is odd under $\Omega_p (-1)^{F_L}$, it naively has to satisfy $B_2 = - \sigma(B_2)$ so that only components in $H^{1,1}_-(X,\mathbb{Z})$ survive the orientifold projection. This would mean that $B_2$ has to vanish in our setup since we have $h^{1,1}_-(X)=0$ by assumption. However, as discussed, e.g., in \cite{Bachas:2008jv, Blumenhagen:2008zz, Collinucci:2008sq}, there is a discrete shift symmetry
$B_2 \to B_2 + \omega_2$, $F_2 \to F_2 - 2\pi \iota^*_\text{T}\omega_2 \mathds{1}_N$ with $\omega_2 \in H^{1,1}(X,\mathbb{Z})$ so that satisfying $B_2 = - \sigma(B_2) + H^{1,1}(X,\mathbb{Z})$ is sufficient to survive the orientifolding. This admits in particular a half-integrally quantized $B_2 \in \frac{1}{2}H^{1,1}_+(X,\mathbb{Z}_2)$. Pulling this back onto $D_\text{T}$ yields
\begin{equation}
\iota^*_\text{T} B_2 = \frac{b_b}{2} \iota^*_\text{T} D_b + \frac{b_s}{2} \iota^*_\text{T} D_s, \qquad b_b,b_s\in\{0,1\}. \label{dfoafpkofa}
\end{equation}
Combining \eqref{dfoafpkofa} and \eqref{sdgfmlsgls}, we conclude that
\begin{equation}
\hat{\mathcal{F}} = f_b \,\iota^*_\text{T} D_b + f_s \,\iota^*_\text{T} D_s = \left(m_b - \frac{p_b}{2} - \frac{b_b}{2}\right) \iota^*_\text{T} D_b + \left(m_s - \frac{p_s}{2} - \frac{b_s}{2}\right) \iota^*_\text{T} D_s. \label{gkgjieidng}
\end{equation}
Since $m_b$, $m_s$, $p_b$, $p_s$, $b_b$, $b_s$ are all integral, it follows that we can choose any half-integral $f_b$, $f_s$ by turning on an appropriate $B_2$. We will furthermore see in Section \ref{sec:lvs-np} that the consistency of the LVS requires $f_s=0$. Hence, the allowed flux numbers are
\begin{equation}
f_b \in \frac{\mathbb{Z}}{2}, \qquad f_s=0. \label{srgligshligrng}
\end{equation}
We stress again that one of the models discussed in Section \ref{sec:largeh11} is an exception to this rule (e.g., $f_b=\frac{1}{6}$ is allowed there).

Finally, a further constraint on the flux numbers comes from requiring that the solution to \eqref{df} is globally well-defined on the compact submanifold wrapped by the T-brane. As shown in \cite{Marchesano:2017kke, Marchesano:2020idg}, this is the case if\footnote{The inequalities \eqref{dgjsidgshdligs0} can be violated in the presence of a localized defect in the D7-brane worldvolume theory \cite{Marchesano:2019azf}. Such a defect can arise, e.g., when a second brane stack intersects the stack on $D_\text{T}$ and the bifundamental degrees of freedom at the intersection acquire non-trivial vevs. The resulting solution is then a bound state of both brane stacks. In our case, one could in principle consider bound states of the branes on $D_\text{T}$ and those on $D_s$ but it is not clear to us whether such configurations admit the non-perturbative superpotentials required by the LVS. We will therefore not consider them but assume following \cite{Cicoli:2015ylx, Cicoli:2017shd, Cicoli:2021dhg} that there is a T-brane on $D_\text{T}$ and an independent second brane stack (or instanton) on $D_s$ that generates $W_\text{np}$. Note that the key problem pointed out in Section \ref{sec:problem} gets \emph{worse} when \eqref{dgjsidgshdligs0} is violated (as this implies larger flux) so that defects are not promising loopholes from that perspective.}
\begin{equation}
\int_{D_\text{T}} \iota^*_\text{T} J \w \hat{\mathcal{F}} > 0, \qquad
\int_{D_\text{T}} \iota^*_\text{T} J \w \left( \iota^*_\text{T} D_\text{T} - 2\hat{\mathcal{F}} \right) \ge 0, \label{dgjsidgshdligs0}
\end{equation}
where we again used that $c_1(D_\text{T}) = -\iota^*_\text{T} D_\text{T}$ on a Calabi-Yau manifold.
This yields
\begin{equation}
f_b> 0, \qquad p_b \ge 2f_b + \frac{|t_s|p_s^2\kappa_{sss}}{t_bp_b\kappa_{bbb}},
\end{equation}
where we used \eqref{dt}, \eqref{tzptzitjli}, \eqref{srgligshligrng} and assumed $t_s<0$ in the K\"ahler cone as usual in the LVS. We thus find that the flux numbers must satisfy the following bounds:
\begin{equation}
p_b > 2f_b> 0 \quad\text{ if } p_s\neq 0,  \qquad p_b\ge 2f_b> 0 \quad\text{ if } p_s=0. \label{march}
\end{equation}

\subsection{Uplift}
\label{sec:lvs-up2}

We now explain how the T-brane setup of Section \ref{sec:lvs-up} leads to an uplift term in the LVS potential \eqref{lvspotential}.
From the point of view of the 4d effective field theory, the D7 stack on $D_\text{T}$ generates a $D$-term potential \cite{Junghans:2026jza}
\begin{equation}
V_D = \frac{2\pi}{ \int_{D_\text{T}} \iota_\text{T}^*J^2} \text{Tr}\left(\frac{\int_{D_\text{T}} \iota_\text{T}^* J \w \mathcal{F}}{4\pi\mathcal{V}} + \int_{D_\text{T}} [\hat\Phi,\hat\Phi^\dagger]
\right)^2. \label{dtermpot}
\end{equation}
Here, $\hat \Phi$ denotes the harmonic part of $\Phi$ (with respect to the gauge-covariant derivative $\bar\partial + i[A^{0,1},\cdot]$) rescaled such that $\hat \Phi = \frac{1}{\sqrt{8\pi^2\mathcal{V}}} \Phi_\text{h}$. In this normalization, the K\"ahler potential for $\hat\Phi$ is $K_\text{D7} = \text{Tr} \int_{D_\text{T}} \hat\Phi^\dagger \w \hat\Phi$. In general, $V_D$ also depends on Wilson-line moduli which however vanish in the T-brane background and are already set to zero here. For more details, we refer to \cite{Junghans:2026jza} where the general $D$-term potential of a non-Abelian D7 stack was derived by dimensionally reducing the DBI action (see also \cite{Jockers:2004yj, Jockers:2005zy, Haack:2006cy, Martucci:2006ij} for results in the case of a single D7-brane and \cite{Camara:2004jj, Beasley:2008dc, Marchesano:2010bs, Marchesano:2016cqg, Marchesano:2017kke, Cicoli:2015ylx} for related work in the non-Abelian case).

The potential is minimized for
\begin{equation}
\int_{D_\text{T}} [\hat\Phi,\hat\Phi^\dagger]= -\frac{\int_{D_\text{T}} \iota_\text{T}^* J \w \mathcal{F}}{4\pi\mathcal{V}} = \frac{\kappa_{bbb}p_b t_b f_b}{4\pi\mathcal{V}}  \begin{pmatrix} \mathds{1}_{N/2} & 0 \\
0 & -\mathds{1}_{N/2} \\ \end{pmatrix}, \label{tjtjfghjflg}
\end{equation}
where we used \eqref{dt}, \eqref{tzptzitjli0}, \eqref{tzptzitjli} and \eqref{srgligshligrng}.
Clearly, \eqref{tjtjfghjflg} is the 4d version of the $D$-term equation in \eqref{d7eom}. As discussed in Section \ref{sec:lvs-up}, the solution is a T-brane background with
$\hat\Phi \sim \left(\begin{smallmatrix} 0 & \mathcal{M} \\ 0 & 0 \\ \end{smallmatrix}\right)$. Using this, \eqref{tjtjfghjflg} implies
\begin{equation}
\int_{D_\text{T}}\hat\Phi \w\hat\Phi^\dagger = \frac{\kappa_{bbb}p_b t_b f_b}{4\pi\mathcal{V}}  \begin{pmatrix} \mathds{1}_{N/2} & 0 \\
0 & 0 \\ \end{pmatrix}, \qquad \int_{D_\text{T}}\hat\Phi^\dagger \w \hat\Phi  = \frac{\kappa_{bbb}p_b t_b f_b}{4\pi\mathcal{V}}  \begin{pmatrix} 0 & 0 \\
0 & \mathds{1}_{N/2} \\ \end{pmatrix}. \label{fdkhodfhjlidh}
\end{equation}

As stated before, turning on bulk fluxes generates extra couplings in the D7-brane worldvolume theory and thus modifies the dynamics of the brane fields \cite{Camara:2004jj, Gomis:2005wc, Junghans:2026jza}. From the 4d point of view, this generates an $F$-term potential \cite{Junghans:2026jza} (see also App.~\ref{app:tbrane})
\begin{equation}
V_F = \frac{g_s|W_0|^2}{2\mathcal{V}^2}  \text{Tr} \int_{D_\text{T}}\hat\Phi^\dagger \w \hat\Phi. \label{ftermpot}
\end{equation}
This expression is valid if the superpotential is independent of the brane moduli, as assumed in the T-brane-uplifting scenario \cite{Cicoli:2015ylx, Cicoli:2017shd, Cicoli:2021dhg}.
We refer to \cite{Junghans:2026jza} for more general expressions for $V_F$ and for the superpotential of non-Abelian D7-branes (see also \cite{Jockers:2005zy, Martucci:2006ij, Denef:2008wq, Arends:2014qca} and \cite{Marchesano:2009rz, Marchesano:2010bs} for earlier work on the Abelian/non-Abelian brane superpotential, respectively). Note that both $V_D$ and $V_F$ receive various corrections, in particular $\alpha'$ and backreaction corrections as well as corrections from integrating out KK modes. We will come back to some of these corrections in Section \ref{sec:lvs-corr}. An extensive discussion of KK corrections to the D7-brane potential is provided in \cite{Junghans:2026jza}.

The minimum of $V_D+V_F$ is now obtained as follows. For sufficiently small $g_s$ and large $\mathcal{V}$, \eqref{ftermpot} is much smaller than \eqref{dtermpot} so that $\hat\Phi$ still satisfies \eqref{fdkhodfhjlidh} in the minimum up to a subleading term.
Using \eqref{fdkhodfhjlidh} and $t_b \approx \left(\frac{6\mathcal{V}}{\kappa_{bbb}}\right)^{1/3}$ in \eqref{ftermpot}, we find
\begin{equation}
V_F = \frac{N}{2} 6^{1/3}\kappa_{bbb}^{2/3}p_b f_b \frac{g_s|W_0|^2}{8\pi\mathcal{V}^{8/3}}. \label{dsgslijgsljg}
\end{equation}
We thus observe that the T-brane generates a \emph{positive} term in the scalar potential. Identifying this with the uplift term $\frac{C_\text{up}}{\mathcal{V}^x}$ anticipated in \eqref{lvspotential}, we conclude
\begin{equation}
x=\frac{8}{3}, \qquad C_\text{up}=\frac{N}{2} \frac{6^{1/3} \kappa_{bbb}^{2/3} p_b f_b g_s |W_0|^2}{ 8\pi}. \label{cup}
\end{equation}
This agrees with the uplift term derived in \cite[(2.33)]{Cicoli:2015ylx} and with those in \cite[(4.22)]{Cicoli:2017shd}, \cite[(5.29)]{Cicoli:2021dhg} up to a factor $N$ (upon using the values of $p_b$, $\kappa_{bbb}$ in these setups). Note that the factor $N$ is important since, as we will see, the maximal volume in a T-brane-uplifted dS minimum scales like $N^{-3}$. This can be a very small number, e.g., when the T-brane bound state is made out of $N=8$ D7-branes.
Let us emphasize again that \eqref{cup} is consistent with
the dimensional reduction performed in \cite{Junghans:2026jza}, where we derived the full non-Abelian $D$-term and $F$-term potential of the D7-branes.

As we mentioned, our approximation of minimizing $V_D$ and then substituting the result into $V_F$ is only valid when $V_D$ is much steeper than $V_F$. If this is not the case, we should instead minimize the full potential $V_D+V_F$ in a single step.
To this end, one can make the ansatz $\int_{D_\text{T}}\hat\Phi \w \hat\Phi^\dagger = |\varphi|^2 \left(\begin{smallmatrix} \mathds{1}_{N/2} & 0 \\
0 & 0 \end{smallmatrix} \right) $, $\int_{D_\text{T}}\hat\Phi^\dagger \w \hat\Phi = |\varphi|^2 \left(\begin{smallmatrix} 0 & 0 \\
0 & \mathds{1}_{N/2} \end{smallmatrix} \right) $ for some complex 4d scalar field $\varphi$. Substituting this into \eqref{dtermpot}, \eqref{ftermpot}, we get
\begin{align}
V &= \frac{\pi N}{ p_b\tau_b} \left(\frac{\kappa_{bbb}p_bf_bt_b}{4\pi\mathcal{V}} -  |\varphi|^2 \right)^2 + \frac{ g_s|W_0|^2 N|\varphi|^2}{4\mathcal{V}^2} \nll
= \frac{2\pi N}{ 6^{2/3}p_b\kappa_{bbb}^{1/3}\mathcal{V}^{2/3}} \left(\frac{6^{1/3}\kappa_{bbb}^{2/3}p_bf_b}{4\pi\mathcal{V}^{2/3}} -  |\varphi|^2 \right)^2 + \frac{ g_s|W_0|^2 N|\varphi|^2}{4\mathcal{V}^2}, \label{varphipotential}
\end{align}
where we used that $t_b \approx \left(\frac{6\mathcal{V}}{\kappa_{bbb}}\right)^{1/3}$,
$\tau_b \approx \left(\frac{3\sqrt{\kappa_{bbb}} \mathcal{V}}{\sqrt{2}}\right)^{2/3}$ in the LVS and neglected terms that would only give subleading corrections to the solution compared to the effect of the $|W_0|^2$ term. One verifies that, if
\begin{equation}
\frac{ 6^{1/3} g_s |W_0|^2}{4 \kappa_{bbb}^{1/3} f_b \mathcal{V}^{2/3}} < 1 \label{gsdjlgisjgls}
\end{equation}
holds, the equation of motion $\partial_\varphi V=0$ has two solutions: the trivial solution $\varphi = 0$ and the T-brane solution $\varphi=\varphi_0\e^{i\gamma}$, where $\gamma$ is an arbitrary phase and
\begin{equation}
\varphi_0 = \sqrt{\frac{6^{1/3}\kappa_{bbb}^{2/3}p_bf_b}{4\pi\mathcal{V}^{2/3}} - \frac{ 6^{2/3}\kappa_{bbb}^{1/3}p_b g_s |W_0|^2}{16\pi \mathcal{V}^{4/3}}}.
\end{equation}
If the second term in the square root is small, we recover the approximate solution obtained by minimizing the $D$-term potential.
On the other hand, if the susy-breaking by $W_0$ becomes too large, i.e., if \eqref{gsdjlgisjgls} is violated, the T-brane solution vanishes and the only solution is $\varphi=0$. In that case, we also expect higher-$F$-term/KK corrections to become important so that we can no longer trust the potential anyway, see Section \ref{sec:lvs-corr2}.

Let us also check the stability of the T-brane by substituting $\varphi = (\varphi_0 + \delta\varphi)\e^{i\gamma}$ into the potential, where $\delta\varphi$ is a real fluctuation. Expanding up to the second order in $\delta\varphi$, we get\footnote{We have set $p_s=0$ in this expression for simplicity. For $p_s\neq 0$, a further term with the same volume scaling as the $\frac{g_s|W_0|^2}{\mathcal{V}^{2/3}}$ correction appears in the mass term.}
\begin{align}
V &=
\frac{ 6^{1/3} N\kappa_{bbb}^{2/3}p_bf_b g_s|W_0|^2}{16\pi\mathcal{V}^{8/3}} \left( 1 - \frac{ 6^{1/3} g_s|W_0|^2 }{8 \kappa_{bbb}^{1/3}f_b \mathcal{V}^{2/3}} \right) \nl + \frac{2 N \kappa_{bbb}^{1/3}f_b}{6^{1/3} \mathcal{V}^{4/3}} \left( 1 - \frac{ 6^{1/3} g_s|W_0|^2 }{4 \kappa_{bbb}^{1/3}f_b \mathcal{V}^{2/3}} \right) (\delta\varphi)^2. \label{dfajdsfsotk}
\end{align}
If \eqref{gsdjlgisjgls} is satisfied, the $(\delta\varphi)^2$ term has a positive coefficient and therefore the T-brane is meta-stable against a decay to other brane configurations, at least along this particular direction in the field space. As stated before, the phase $\gamma$ is not stabilized.
We will make further comments on the brane-moduli space of T-brane vacua in Section \ref{sec:modstab}.

\subsection{Non-perturbative Effect}
\label{sec:lvs-np}

A crucial requirement for the LVS is that a non-perturbative term $W_\text{np} \sim \e^{-a_sT_s}$ as in \eqref{w} is present in the superpotential. Such a term can either be generated by D3-brane instantons (E3-branes) or gaugino condensation on D7-branes wrapped on the divisor $D_s$ \cite{Witten:1996bn, Kachru:2003aw}.
In order for this to happen, a number of conditions must be satisfied, which we review in this subsection.
See \cite{Ibanez:2007rs, Blumenhagen:2009qh} for more detailed reviews of instantons and \cite{Louis:2012nb, McAllister:2023vgy} for discussions of some aspects of gaugino condensation. Non-perturbative effects in the LVS/T-brane context are discussed in \cite{Cicoli:2017shd, Cicoli:2021dhg}.

Let us first recall the involved gauge groups. Since $D_s$ is orientifold-even, the (continuous) gauge group of a stack of $M$ D7-branes on $D_s$ is either SO$(M)$ (if the branes wrap a pointwise-invariant divisor and thus lie on top of an O7-plane) or USp$(M)$ (if the branes do not lie on top of an O7-plane). For the instantons, it is the other way round: a stack of $n$ E3-branes on $D_s$ has the gauge group USp$(n)$ if the divisor is pointwise-invariant and SO$(n)$ otherwise \cite{Ibanez:2007rs, Blumenhagen:2008zz, Blumenhagen:2009qh}. Since symplectic groups have $n\in 2\mathbb{N}$, a single E3-brane only exists on a not-pointwise-invariant divisor and has gauge group O$(1)$.

In order to generate a non-perturbative superpotential, E3-branes need to have exactly two fermionic zero modes. Let us ignore intersections with other branes for the moment and postpone their discussion until the end of this subsection. A sufficient (but not necessary) condition for E3-branes to have exactly two fermionic zero modes is then that the gauge group is O$(1)$ and $D_s$ is a rigid divisor, i.e., $h^{1,0}(D_s)=h^{2,0}(D_s)=0$ \cite{Witten:1996bn} (see \cite{Ibanez:2007rs, Blumenhagen:2009qh} for reviews). The case of D7-branes with gaugino condensation is very similar: A non-perturbative superpotential is then generated if the worldvolume theory is pure super-Yang-Mills theory at low energies, i.e., there are no massless matter fields valued in the adjoint or (anti-)symmetric representations of the gauge group, which is again ensured by wrapping a rigid divisor \cite{Louis:2012nb, McAllister:2023vgy}.

Rigidity of a divisor implies that the arithmetic genus satisfies $\chi_0(D_s) = 1-h^{1,0}(D_s)+h^{2,0}(D_s) = 1$. On a Calabi-Yau three-fold $X$, this can be checked using the formula
\begin{equation}
\chi_0(D_s)=\frac{1}{12}\int_X\left(c_2(X) \w D_s+2D_s^3\right),
\end{equation}
where $c_2(X)$ is the second Chern class of $X$. The models we consider in this paper all satisfy $\chi_0(D_s)=1$.
More generally, also E3-branes or D7-branes on \emph{non-rigid} divisors can contribute to $W_\text{np}$ due to the effect of fluxes. Indeed, worldvolume or bulk fluxes can ``rigidify'' a non-rigid divisor by lifting unwanted fermionic zero modes or massless matter fields that would prevent the generation of a superpotential \cite{Gorlich:2004qm, Kallosh:2005yu, Bianchi:2011qh,Bianchi:2012kt}.
However, this is often difficult to check in practice so that rigid divisors are usually the easiest way to guarantee a non-perturbative superpotential.

Let us also briefly discuss higher-rank instantons, i.e., stacks of E3-branes with $n\ge 2$. Without fluxes, such instantons have at least four fermionic zero modes and thus do not contribute to $W_\text{np}$, even when the divisor is rigid \cite{Ibanez:2007rs, Blumenhagen:2009qh}.\footnote{Similarly, instantons wrapping divisors in the class $nD_s$ for $n\ge 2$ do not contribute to $W_\text{np}$ without fluxes since $\chi_0(D_s)=1$ implies $\chi_0(nD_s)\neq 1$ for $\kappa_{sss}>0$.} However, fluxes can again lift some of these zero modes in which case instantons with $n\ge 2$ can appear in $W_\text{np}$ \cite{Berglund:2012gr}.

Depending on which of the discussed effects generates $W_\text{np}\sim \e^{-a_sT_s}$, one finds
\begin{equation}
a_s = \left\{\begin{matrix}
       2\pi n & \text{(rank-}n \text{ E3 instanton)} \\
       \frac{2\pi}{\frac{M}{2}+1} & \text{(D7 stack with USp} (M) \text{ gauge group)} \\
       \frac{2\pi}{M-2} & \text{(D7 stack with SO} (M) \text{ gauge group)} \\
      \end{matrix}\right..
\end{equation}
The last two expressions follow from the general formula $a_s=\frac{2\pi}{c(G)}$, where $c(G)$ is the dual Coxeter number of the gauge group $G$ \cite{Louis:2012nb, McAllister:2023vgy}.\footnote{There are conflicting statements in the literature about which $G$ can host a non-perturbative effect (see, e.g., \cite{Cicoli:2011qg} for models with USp gauge groups and \cite{Crino:2020qwk} for a comment that only SO$(8)$ is allowed). We will not go into the intricacies of this discussion and, to be as general as possible, study models with USp$(M)$ and SO$(M)$ gauge groups of various ranks.}

Aside from too many zero modes, another possible obstruction to generating a $W_\text{np}$ is the quantization of the worldvolume flux. As discussed in Section \ref{sec:lvs-d7}, the flux is constrained by the orientifold projection. This applies to instantons and spacetime-filling branes alike (with the mentioned caveat that the gauge groups of pointwise-invariant and not-pointwise-invariant instantons are interchanged compared to the D7 case). For an $n=1$ instanton with O$(1)$ gauge group, the projection is such that only orientifold-odd flux is allowed \cite{Grimm:2011dj}. Since we assume $h^{1,1}_-(X)=0$ in this paper, this means that the flux on the instanton has to vanish (up to variable terms, which we ignore for the moment). If $W_\text{np}$ comes from D7-branes instead of an E3-brane, the orientifold projection would in principle allow orientifold-even flux (just like it does for the D7-branes on $D_\text{T}$, cf.~\eqref{evenodd1}, \eqref{evenodd2}). However, this would break the SO$(M)$ or USp$(M)$ gauge group to a subgroup, create chiral matter fields, generate a $D$-term potential and generically recombine the branes, all of which can prevent a contribution to $W_\text{np}$.
We will therefore follow the LVS literature \cite{Cicoli:2017shd, Cicoli:2021dhg} and assume vanishing flux on the D7 stack. In summary, we are led to impose\footnote{To avoid confusion, note that $\mathcal{F}_\text{D7}$ refers to the flux of the D7 stack on $D_s$, while $\mathcal{F}$ is the flux of the D7 stack on $D_\text{T}$.}
\begin{equation}
\mathcal{F}_\text{E3} = \mathcal{F}_\text{D7} = 0. \label{fe3fd7}
\end{equation}
The flux-quantization condition under this assumption becomes \cite{Freed:1999vc}
\begin{align}
& \iota^*_{s} B_2 + \frac{\iota^*_{s} D_s}{2} \in H^2(D_s,\mathbb{Z}). \label{sgsligsiggr}
\end{align}
In the LVS context, $D_s$ is typically a non-Spin divisor so that $\frac{\iota^*_{s} D_s}{2}$ is not an element of $H^2(D_s,\mathbb{Z})$.\footnote{For complex manifolds, not having a Spin structure means that $\frac{c_1(D_s)}{2}$ is not an element of $H^2(D_s,\mathbb{Z})$. We further have $c_1(D_s)=-\iota^*_{s} D_s$ since $D_s$ is a divisor on a Calabi-Yau manifold.} Hence, \eqref{sgsligsiggr} can only be satisfied if a half-integral $B_2$ with $\iota^*_s B_2=-\frac{\iota^*_s D_s}{2}$ is turned on (modulo integral forms). However, this trick does not always work since there can be situations where one is forced to set $\iota^*_s B_2=0$. This happens, e.g., if the T-brane wraps a divisor with even $p_s\neq 0$, as follows from \eqref{dfoafpkofa}--\eqref{srgligshligrng}. In such a case, an E3-brane or a D7 stack satisfying \eqref{fe3fd7} cannot generate a $W_\text{np}$ because their existence is forbidden by the flux-quantization condition. Hence,
\begin{equation}
W_\text{np} \text{ from E3/D7s with }\mathcal{F}_\text{E3} = \mathcal{F}_\text{D7} = 0 \quad \Longrightarrow \quad p_s \in \{0, 2\mathbb{Z}+1\}.
\end{equation}
There are however loopholes to this conclusion: instantons with variable flux \cite{Junghans:2026inst} or higher-rank instantons with non-Abelian flux \cite{Berglund:2012gr} can be consistent with the orientifold projection and flux quantization in spite of $\iota^*_s B_2=0$.
Such effects can ensure that a $W_\text{np}$ is generated also in models with even $p_s$.
See, e.g., \cite{Cicoli:2021dhg} for an LVS model using a rank-2 instanton.

A final constraint we need to discuss comes from zero modes or massless matter fields localized at the intersections between different branes or brane stacks (in our case, the branes or instantons hosting the non-perturbative effect on $D_s$ and the branes wrapped on $D_\text{T}$). As mentioned in the beginning, such modes have to be absent in order that a $W_\text{np}$ is generated, and this is not guaranteed by imposing rigidity of $D_s$.
An easy-to-check necessary condition is in particular that the chiral index -- i.e., the number of chiral minus anti-chiral zero modes -- vanishes at the intersection. The formula for two branes on divisors $D_1$, $D_2$ with gauge bundles $\mathcal{L}_1$, $\mathcal{L}_2$ is \cite{Blumenhagen:2007sm, Blumenhagen:2008zz, Beasley:2008dc, Blumenhagen:2009qh}
\begin{equation}
\int_{D_1 \cap D_2} \left(c_1(\mathcal{L}_1)-c_1(\mathcal{L}_2)\right) = 0. \label{sgsligdsggg}
\end{equation}
Applying this to the intersection of a D7-brane on $D_\text{T}$ (or its image) and an E3-brane on $D_s$, we obtain
\begin{equation}
\int_{D_\text{T} \cap D_s} (\hat{\mathcal{F}} \mp \mathcal{F}_\text{E3})=\int_{D_\text{T}} f_s\, \iota^*_\text{T} D_s \w \iota^*_\text{T} D_s = 0, \label{slirgslirglirnswg}
\end{equation}
where we used \eqref{tzptzitjli} and \eqref{fe3fd7}.
Hence, in order to generate a $W_\text{np}$, we have to set the $D_s$ component of $\hat{\mathcal{F}}$ to zero in \eqref{tzptzitjli},
\begin{equation}
f_s = 0. \label{psfs}
\end{equation}
One can verify that the same condition must hold if the instanton carries a variable flux and also for instantons of higher rank.\footnote{Variable flux satisfies $\mathcal{F}_\text{E3}\w \iota^*_s D_s =0$ by definition (cf.~our discussion above \eqref{tzptzitjli}) and thus does not affect \eqref{slirgslirglirnswg} \cite{Blumenhagen:2008zz}. The generalization of \eqref{sgsligdsggg} to brane stacks with higher-rank gauge bundles $V_1$, $V_2$ is
$\int_{D_1 \cap D_2} \left(\text{rk}(V_2)c_1(V_1)- \text{rk}(V_1)c_1(V_2)\right)=0$ \cite{Blumenhagen:2009qh}.
Applied to the case of D7-branes on $D_\text{T}$ and a rank-$n$ instanton on $D_s$, this yields $\int_{D_\text{T} \cap D_s} \big(n\hat{\mathcal{F}}\mp \text{Tr} \mathcal{F}_\text{E3}\big) =0$, where the second term vanishes since the orientifold projection imposes that $\text{Tr} \mathcal{F}_\text{E3}$ is orientifold-odd and thus variable, i.e., $\text{Tr} \mathcal{F}_\text{E3} \w \iota^*_s D_s=0$. One thus gets the condition $\int_{D_\text{T} \cap D_s}\hat{\mathcal{F}}=0$ as in the rank-1 case. Note, however, that this is only sensitive to the \emph{overall} chiral intersections for the whole E3 stack. Detecting chiral intersections of individual E3-branes and D7-branes or of pairs of E3-branes would require a more refined computation.} An analogous argument can also be made in the case where the non-perturbative effect comes from D7-branes with gaugino condensation instead of instantons.

We finally note that there can be further obstructions to a non-perturbative superpotential beyond those discussed in this subsection (see, e.g., \cite[footnote 21]{Cicoli:2021dhg}). We will not study such subtleties in detail. The point of this paper is that LVS models with T-branes do not yield a controlled dS uplift, and this conclusion does not change if some of these models happen to have additional problems.

\subsection{Tadpole Cancelation}
\label{sec:lvs-tadpole}

Let us also collect a few useful formulae related to tadpole cancelation.
Our T-brane setup involves a stack of $N$ D7-branes on the divisor $D_\text{T}=p_bD_b+p_sD_s$ and optionally another stack of $M$ D7-branes on the divisor $D_s$ (if the non-perturbative effect is due to gaugino condensation).
In the models we will consider, the orientifold involution further yields an O7-plane wrapped on
\begin{equation}
D_\text{O7}=r_b D_b+r_sD_s, \qquad r_b \in \mathbb{N}, \qquad r_s \in \mathbb{Z}
\end{equation}
and, in some cases, additional O3-planes counted by the number of isolated fixed points $N_\text{O3}$ of the orientifold involution. To cancel the D7 charges of the D7-branes and the O7-plane, we have to impose the tadpole condition
\begin{equation}
N p_b = 8 r_b, \qquad Np_s + M = 8 r_s. \label{d7tadp}
\end{equation}
Here, $N$ and $M$ denote the number of branes in the upstairs convention, i.e., on the covering space of the orientifold where image branes are also counted.

The D7-branes, O7-plane and O3-planes further carry a D3 charge
\begin{equation}
Q_3 = - \frac{\chi(D_\text{O7})}{6} - N\frac{\chi(D_\text{T})}{24} - M\frac{\chi(D_s)}{24} - \frac{1}{2} \text{Tr} \int_{D_\text{T}} \mathcal{F} \w \mathcal{F} - \frac{N_\text{O3}}{2}, \label{loctadpole}
\end{equation}
again in the upstairs convention.
This charge has to cancel with the D3 charges carried by bulk fluxes and D3-branes:
\begin{equation}
Q_3 + \frac{ig_s}{2} \int_X G_3 \w \bar G_3 + N_\text{D3} = 0, \label{tadpolebound}
\end{equation}
where $N_\text{D3}$ is the number of D3-branes (counting also orientifold images). The bulk fluxes are imaginary self-dual, i.e., $iG_3 = \star G_3$, so that $ i \int_X G_3 \w \bar G_3 = \int_X \star |G_3|^2 > 0$ \cite{Giddings:2001yu}. Together with $N_\text{D3} \ge 0$, this implies that $Q_3$ is negative.

A useful formula for the Euler number $\chi(D_i)$ of a divisor $D_i$ is
\begin{equation}
\chi(D_i) = \int_{D_i} \iota^*_i \left(c_2(X) + D_i^2\right),
\end{equation}
where $c_2(X)$ is the second Chern class of $X$. Writing $c_2(X) = c_{2b}D_b^2+c_{2s}D_s^2$ \footnote{Note that terms $\sim D_b\w D_s$ vanish since the only non-vanishing triple-intersections are given by \eqref{uogjokpjtkj}.},
\eqref{loctadpole} becomes
\begin{align}
Q_3 &= - \frac{(r_bc_{2b}+r_b^3)\kappa_{bbb}+(r_sc_{2s}+r_s^3)\kappa_{sss}}{6} - N\frac{(p_bc_{2b}+p_b^3)\kappa_{bbb}+(p_sc_{2s}+p_s^3)\kappa_{sss}}{24} \nl - M\frac{(c_{2s}+1)\kappa_{sss}}{24} - \frac{N}{2} p_bf_b^2 \kappa_{bbb} - \frac{N_\text{O3}}{2}. \label{zupzuipto}
\end{align}
For later convenience, we also state a formula for the number of complex-structure moduli $h^{2,1}_-(X)$. To this end, we combine the standard formula for the Euler number $\chi(X) = 2(h^{1,1}(X)-h^{2,1}(X)) = 2(2-h^{2,1}_+(X)-h^{2,1}_-(X))$ and the Lefschetz fixed-point theorem (see, e.g., \cite{Crino:2022zjk}), which in our case states $2(h^{2,1}_+(X)-h^{2,1}_-(X))=8 - \chi(D_\text{O7}) - N_\text{O3}$. This yields
\begin{align}
h^{2,1}_-(X) &= \frac{\chi(D_\text{O7})}{4} + \frac{|\chi(X)|}{4} + \frac{N_\text{O3}}{4} - 1  \nll =
\frac{c_{2b}r_b \kappa_{bbb}+\kappa_{bbb}r_b^3}{4} + \frac{c_{2s}r_s\kappa_{sss}+\kappa_{sss}r_s^3}{4}+\frac{|\chi(X)|}{4} + \frac{N_\text{O3}}{4} - 1. \label{h21minus}
\end{align}

\section{The Volume Bound}
\label{sec:problem}

In this section, we describe the main problem of the T-brane uplift. Our claim is that the string-frame volume $\mathcal{V}_s = g_s^{3/2}\mathcal{V}$ is bounded in any dS vacuum such that the possibilities to control corrections to the LVS approximation are severely limited. Indeed, using \eqref{cup}, \eqref{zjghjgjg} and $\alpha < \frac{9}{8}$ in \eqref{alpha} yields
\begin{equation}
\boxed{\mathcal{V}_s = \frac{729\pi^3\alpha^3\hat\xi}{2 a_s^3\kappa_{bbb}^2 (\frac{N}{2}p_bf_b)^3 \kappa_{sss}} < 
 39.0 \frac{|\hat\chi|}{a_s^3\kappa_{bbb}^2  (\frac{N}{2}p_bf_b)^3 \kappa_{sss}}}. \label{dsgsg}
\end{equation}
Note in particular that this does not depend on $g_s$ or $W_0$. Furthermore, $f_b$ is bounded from below since it is quantized and $a_s$ depends on the number of D7-branes on $D_s$ which, by tadpole cancelation, is tied to the O7-planes. How strongly \eqref{dsgsg} constrains the volume is therefore completely determined by the topological data of the Calabi-Yau manifold and the choice of orientifold involution.

To illustrate the strength of the bound, let us consider manifolds where $D_b$, $D_s$ are an integral basis of $H^{1,1}(X,\mathbb{Z})$ and worldvolume fluxes of the pull-back type, as explained in Section \ref{sec:lvs-flux}.
We can then further simplify \eqref{dsgsg} by using $\frac{N}{2} p_b f_b \ge 2r_b$.
This follows from tadpole cancelation on $D_b$ (which, according to \eqref{d7tadp}, implies $Np_b=8r_b$) together with flux quantization (which, according to \eqref{srgligshligrng}, implies $f_b\ge \frac{1}{2}$). Since $D_\text{O7}$ is an integral combination of $D_b$ and $D_s$, we further have $r_b\ge 1$.
The bound thus becomes
\begin{equation}
\mathcal{V}_s < 4.87 \frac{|\hat\chi|}{a_s^3 \kappa_{bbb}^2 \kappa_{sss}}. \label{bound}
\end{equation}

An immediate consequence of this is that \emph{large}-volume dS vacua are ruled out in these models if the non-perturbative effect is due to instantons (i.e., $a_s\ge 2\pi$).
To see this, note that $|\hat\chi|$ is typically of the order $|\chi|$ or smaller and that, as will be detailed in Section \ref{sec:h112}, Swiss-cheese Calabi-Yau manifolds with $h^{1,1}=2$ satisfy $\frac{|\chi|}{\kappa_{bbb}^2 \kappa_{sss}} \le 228$.
We thus find
\begin{equation}
\mathcal{V}_s \lesssim 5
\end{equation}
and therefore dS vacua with large volumes are excluded in these models! Instead, the maximal size of the Calabi-Yau manifold is of the order of the \emph{string length}. We therefore expect an infinite number of $\alpha'$ (and warping) corrections to become relevant and the LVS potential to no longer be reliable. We stress that it is the string-frame volume, not the Einstein-frame volume $\mathcal{V}=\frac{\mathcal{V}_s}{g_s^{3/2}}$, which controls the $\alpha'$ expansion. Having a large Einstein-frame volume is in principle consistent with our bound for sufficiently small $g_s$ but this is irrelevant for the degree of control. An alternative perspective to this problem is to consider the KK scale, which is of the order $M_\text{KK}\sim \mathcal{V}_s^{-1/6} \gtrsim 0.8$ in string units and thus approximately equals the string scale. We are thus at the boundary of the regime where the description in terms of a 10d supergravity theory (and the 4d effective theory derived from compactifying it on a Calabi-Yau) is valid, suggesting that instead a fully non-perturbative string-theory calculation would be required.

Let us also note that, even ignoring string corrections and focussing just on supergravity, the derivation of the standard LVS potential \eqref{lvspotential} assumes $\mathcal{V}_s\gg 1$ in various steps, e.g., for the crucial decoupling between the axio-dilaton and the K\"ahler moduli \cite{Achucarro:2008sy, Gallego:2008qi}. The former is assumed to be frozen to a constant in the LVS, although its stabilized vev is in reality a function of $T_b$, $T_s$ due to the mixing term $\sim \hat\xi$ in the K\"ahler potential. The approximation of neglecting this dependence is only valid if $\mathcal{V}_s\gg 1$, $W_0\gg W_\text{np}$ since then the equation $D_SW=0$ is approximately independent of the K\"ahler moduli. Hence, also from this perspective, it is clear that solutions with small $\mathcal{V}_s$ are not in a trustworthy regime.

Finally, there are also models for which \eqref{bound} is less constraining than in the instanton case and in fact compatible with relatively large volumes (of the order $\mathcal{V}_s \sim 10^3$) so that no immediate problem seems to arise, neither regarding $\alpha'$ corrections nor the decoupling approximation. It is obvious from \eqref{bound} that the weakest constraints on the volume are obtained for large Euler numbers, small intersection numbers and a non-perturbative effect from gaugino condensation with a large-rank gauge group (since then $a_s$ is much smaller than in the instanton case). However, we will see that even in these naively favorable models the volume is never large enough that warping corrections
are under control. Hence, these models are put into serious doubt by our volume bound as well.
Demonstrating this in detail will be the topic of the following sections.

\section{Corrections}
\label{sec:lvs-corr}

In this section, we discuss a few potentially dangerous corrections to the potential \eqref{lvspotential}. In general, many different types of corrections can appear in the effective field theory and one has to check that all of them are sufficiently suppressed, see, e.g., \cite{Cicoli:2021rub, Junghans:2022exo, Gao:2022uop} for an overview. We will not attempt to provide an exhaustive analysis here but focus on warping/backreaction corrections. Indeed, the impossibility to control such corrections in a T-brane-uplifted dS vacuum will be a key point of this paper.\footnote{As emphasized in \cite{Junghans:2022exo, Hebecker:2026hrf}, another potential threat for the LVS is due to logarithmic redefinitions of the K\"ahler coordinates, which will however not be discussed in this paper.} We will also briefly discuss higher-$F$-term corrections and point out that they are automatically controlled once the warping corrections are controlled.

\subsection{Backreaction of D3 Charge}
\label{sec:lvs-corr1}

On an approximately isotropic space, a codimension-6 object with D3 charge $q_3$ such as an O3-plane is expected to generate backreaction corrections of the order $\frac{g_sq_3}{R^4} \sim \frac{g_sq_3}{\mathcal{V}_s^{2/3}}$ at generic\footnote{Very close to the source, at distances $r\ll R$, the backreaction becomes larger and eventually diverges like $\frac{g_sq_3}{r^4}$ since, for $r\ll R$, the Green's function is the flat-space one up to subleading corrections.} points, where $R$ is the string-frame length scale characterizing the space.
On spaces with large anisotropies, one instead expects a backreaction of the order $\frac{g_sq_3 \mathcal{D}^2}{\mathcal{V}_s}$, where $\mathcal{D}$ is the diameter in the string frame \cite{Junghans:2023lpo}.\footnote{See also \cite{Moritz:2025bsi} for the case where the anisotropy is due to a large-complex-structure regime.} Since $\mathcal{D}$ can be much larger than $\mathcal{V}_s^{1/6}$, the backreaction in the anisotropic case is parametrically larger than in the isotropic case, the intuitive reason being that the effective codimension is lowered when the manifold is stretched such that it becomes very thin in some directions and very long in others. We can therefore take the isotropic case as a conservative estimate for the expected backreaction.

Aside from D3-branes and O3-planes, also D7-branes and O7-planes carry D3 charges in the presence of fluxes or curvature on their worldvolumes. Instead of sources $\sim g_s\delta^{(6)}(y)$, we then schematically have sources $\sim g_s(\mathcal{F}^2-\mathcal{R}^2)\delta^{(2)}(y)$ in the 10d equations of motion. For branes/O-planes wrapping the big divisor, this yields $\sim  \frac{g_s q_3}{\mathcal{V}_s^{2/3}}\delta^{(2)}(y)$ with $q_3 = -\frac{1}{2}\int_{D_b} \mathcal{F}^2 - \frac{\chi(D_b)}{24}$. The Green's function grows logarithmically in the distance for $r\ll R$ and approaches some maximum of the order $\ln R$ at some $r\sim R$ inbetween the branes. Neglecting the logarithmic scaling, the backreaction is thus again of the order $\frac{g_sq_3}{R^4}$ just like for the O3-planes. We expect the same for D7/O7s wrapping the small divisor since they should look like an O3-plane at distances much larger than the size of the blow-up cycle (which at large volumes is true at generic points on the Calabi-Yau).

The localized negative D3 charges from the O3-planes, D7-branes and O7-planes generically do not cancel locally with the positive charges provided by the bulk fluxes since the latter spread out over the manifold in the regime of large volumes and weak warping assumed in the LVS.\footnote{In principle, flux densities can be exponentially localized in strongly warped regions of the Calabi-Yau (Klebanov-Strassler throats \cite{Klebanov:2000hb}). One may then try to place O-planes nearby that exactly screen the local charge density of the fluxes \cite{Gao:2020xqh}. If this can be done for most of the fluxes on the Calabi-Yau, the effectively backreacting D3 charge in the Calabi-Yau bulk would be much smaller than $|Q_3|$, leading to an essentially unwarped Calabi-Yau bulk ``spiked'' with lots of strongly warped Klebanov-Strassler throats. We will not try to estimate the 4d warping corrections in this highly ungeneric situation but assume in this paper a generic distribution of a positively charged flux density and negatively charged localized sources, without local cancelations that parametrically affect the strength of the backreaction.} We therefore generically expect that the total warping corrections are of the order
\begin{equation}
\frac{g_s|Q_3|}{\mathcal{V}_s^{2/3}},
\end{equation}
where we recall that $Q_3$ is the D3 tadpole generated by O3-planes, D7-branes and O7-planes. In simple spaces such as 6-tori, one can explicitly check that this parametric behavior is correct \cite{Junghans:2023lpo}.

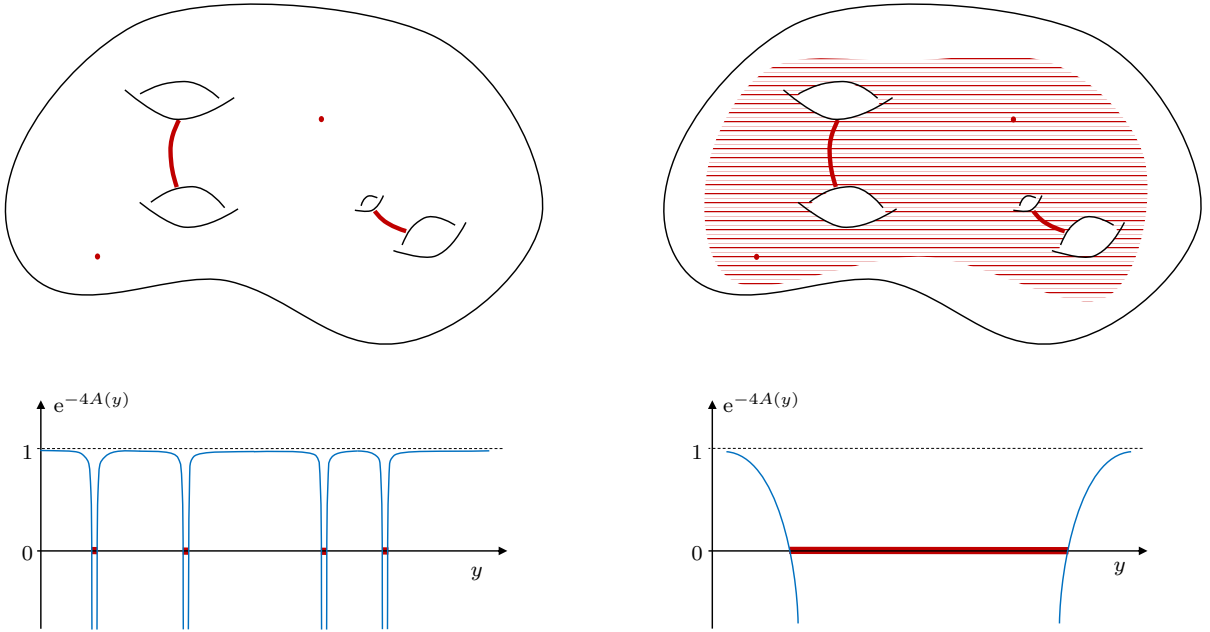
\begin{figure}[t]
\centering
\includegraphics[trim =22mm 15mm 0mm 10mm, clip, width=1.11\linewidth]{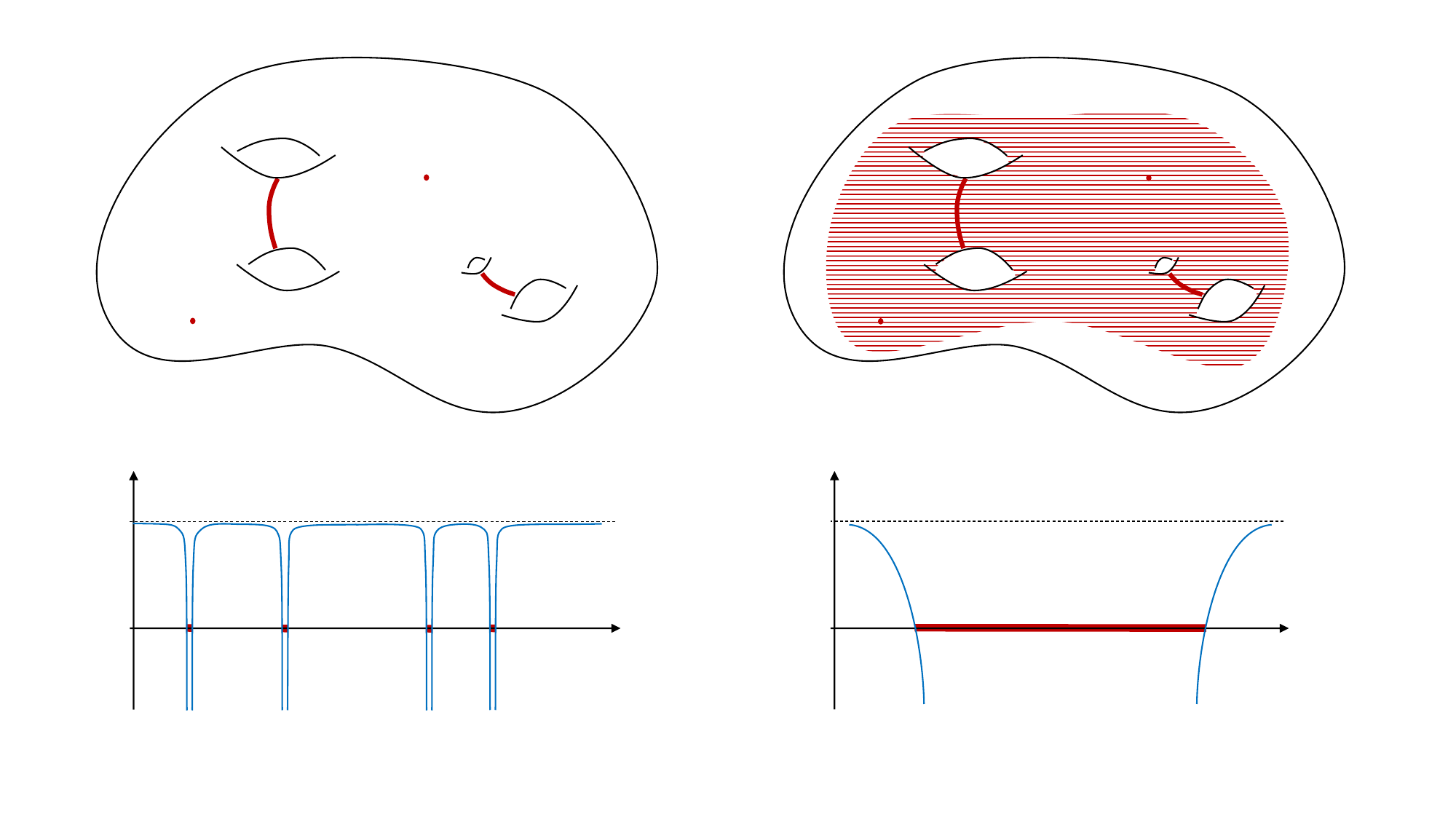}
\put(-472,98){$\scriptstyle{\e^{-4A(y)}}$}
\put(-484,80){$\scriptstyle{1}$}
\put(-484,42){$\scriptstyle{0}$}
\put(-315,36){$\scriptstyle{y}$}
\put(-220,98){$\scriptstyle{\e^{-4A(y)}}$}
\put(-232,80){$\scriptstyle{1}$}
\put(-232,42){$\scriptstyle{0}$}
\put(-73,38){$\scriptstyle{y}$}

\caption{At sufficiently large volumes and small $g_s$, backreaction corrections due to objects carrying a negative D3 charge are expected to be of the order $\frac{g_s|Q_3|}{\mathcal{V}_s^{2/3}}\ll 1$ at generic points in the bulk of the Calabi-Yau (left-hand side). Very close to the sources, the backreaction formally diverges
so that the curvature blows up and the supergravity solution breaks down, indicating that locally string corrections are important and modify the solution. This is a priori not an issue since the singular regions have a small diameter $\ll \ell_s$ and thus cannot be probed at the low energies where the supergravity EFT is valid. On the other hand, if
$\frac{g_s|Q_3|}{\mathcal{V}_s^{2/3}}\gtrsim 1$, a large part of the naive supergravity solution is eaten up by the singularities (right-hand side), indicating that the supergravity description breaks down altogether and the true solution is strongly curved and stringy.
}
\label{fig1}
\end{figure}

As pointed out in \cite{Junghans:2022exo}, a concrete example of a warping correction to the LVS potential arises for the BBHL term $\sim \int_X c_3(X) = \chi(X)$ in the K\"ahler potential (cf.~\eqref{k}, \eqref{zjghjgjg}), where
$c_3(X)$ is the Euler class of $X$. The BBHL term can be inferred from a correction to the 4d Einstein-Hilbert term that arises from dimensionally reducing the 10d $\mathcal{R}^4$ term of type IIB string theory \cite{Becker:2002nn}. In the presence of a non-trivial warp factor $\e^{-4A(y)} = 1 + \mathcal{O}(\frac{g_s|Q_3|}{\mathcal{V}_s^{2/3}})$, this is corrected as
\begin{equation}
\int_X c_3(X) = \chi(X) \to \int_X c_3(X) \e^{-4A(y)} = \chi(X) \left( 1  + \mathcal{O} \left(\frac{g_s|Q_3|}{\mathcal{V}_s^{2/3}} \right) \right).
\end{equation}
The warping correction thus amounts to replacing $\xi \to \xi(1 + C_W \frac{g_s|Q_3|}{\mathcal{V}_s^{2/3}})$ in \eqref{lvspotential}, where we included an unknown numerical coefficient $C_W$ which we assume to be $\mathcal{O}(1)$ (see below for a discussion). Hence, we expect that the LVS potential \eqref{lvspotential} receives a warping correction
\begin{equation}
\delta V = \frac{3 \hat\xi |W_0|^2}{8\sqrt{g_s}\mathcal{V}^3} C_W \frac{g_s|Q_3|}{\mathcal{V}_s^{2/3}}. \label{dsgsdligs}
\end{equation}
Note that warping corrections are also expected to appear in other places, e.g., in the $D$-term potential and the uplift term, whose derivation also assumes an absence of warping. Here we will consider the effect of the above correction as a proxy to estimate the control over warping corrections, but one should keep in mind that this is certainly not the only warping correction that will appear.

As explained in \cite{Gao:2020xqh} (see also \cite[Section 2.3]{Horer:2024hgy}), warping corrections have to be small at generic points\footnote{Note that demanding small warping at \emph{every} point would be too restrictive since the warp factor and other fields typically have singularities in the vicinity of the localized sources where, e.g., the curvature blows up. This is not an inconsistency per se as long as the singular regions are smaller than a string length and the compactification volume is much larger than that. Since supergravity is an EFT valid at energies below the string scale, a breakdown of the supergravity solution at small distances is expected and by itself not concerning.} on the Calabi-Yau since otherwise one expects a ``singular-bulk problem'', i.e. the warping is so strong that large parts of the naive geometry are eaten up by a classical singularity. Although this singularity is expected to be a supergravity artifact cured in string theory \cite{Carta:2021lqg}, this does not mean that it is harmless. Indeed, as emphasized in \cite{Gao:2020xqh}, the resulting compactification will be strongly curved and stringy and no longer amenable to a supergravity treatment. See Fig.~\ref{fig1} for an illustration. We therefore need to make sure that warping corrections of the above sort are small in candidate flux vacua.

In the LVS, the problem is exacerbated by the fact that the corrections to the vacuum energy $V_0$ and to the eigenvalue $m_2^2$ of the Hessian (defined in Section \ref{sec:lvs-km}) are by a factor $1/g_s$ stronger than the corresponding corrections to the off-shell LVS potential \cite{Junghans:2022exo}. This is due to a specific cancelation effect which was called \emph{non-perturbative no-scale structure} in \cite{Junghans:2022exo} since it is related to the exponential $g_s$ dependence of $W_\text{np}$.\footnote{The fact that $V_0$ is by a factor $g_s$ smaller than the individual off-shell terms in $V$ was already observed in \cite{Hebecker:2012aw}. The new observations in \cite{Junghans:2022exo} were that an analogous cancelation also occurs in $m_2^2$, that a no-scale structure is responsible for these cancelations and, most importantly, that they have the effect of a parametric enhancement of warping and other corrections to $V_0$ and $m_2^2$.} A consequence of this is that candidate LVS vacua are already in danger at a parametrically smaller warping than what would cause a singular-bulk problem. Explicitly, we find by adapting the results of \cite{Junghans:2022exo} to the T-brane uplift that the leading warping corrections to $V_0$ and $m_2^2$ satisfy
\begin{equation}
\frac{\delta V_0}{V_0} = - C_W \frac{3^{2/3} 10}{9(\alpha-1)} \frac{a_s \kappa_{sss}^{1/3}}{\hat\xi^{1/3}} \frac{|Q_3|}{\mathcal{V}_s^{2/3}}, \qquad
\frac{\delta (m_2^2)}{m_2^2} = C_W \frac{3^{2/3} 55}{36(\frac{9}{8}-\alpha)} \frac{a_s \kappa_{sss}^{1/3}}{\hat\xi^{1/3}} \frac{|Q_3|}{\mathcal{V}_s^{2/3}}. \label{fgfligfligd}
\end{equation}
Here $\delta V_0$, $\delta (m_2^2)$ denote the leading warping corrections to $V_0$ and $m_2^2$, respectively. Note that there is no $g_s$ suppression in these corrections, unlike in \eqref{dsgsdligs}, as predicted by the non-perturbative no-scale structure. Clearly, we have to demand $\delta V_0 \ll V_0$, $\delta (m_2^2) \ll m_2^2$ for control.

A conservative estimate of the corrections is obtained by taking $\alpha=\frac{20}{19}$ so that both terms in \eqref{fgfligfligd} are equal in size. Closer to the boundaries of the dS range $\alpha=]1,\frac{9}{8}[$, either the correction to $V_0$ or the one to $m_2^2$ would blow up.
Assuming further $C_W=\mathcal{O}(1)$ as stated above, we are led to define the following condition for control:
\begin{equation}
\lambda_W\equiv \frac{3^{2/3} 190}{9} \frac{a_s \kappa_{sss}^{1/3}}{\hat\xi^{1/3}} \frac{|Q_3|}{\mathcal{V}_s^{2/3}} \ll 1 \qquad \text{(control over warping corrections)}. \label{sljkgjgfs}
\end{equation}
In the remainder of this paper, we will take $\lambda_W$ as a proxy to measure how dangerous the warping corrections are in a given model.

Before we move on, a word of caution is in order. As usual for a parametric estimate, we do not know the precise value of $\lambda_W$ for which the unwarped approximation breaks down since we do not know the numerical prefactor $C_W$ in \eqref{fgfligfligd}. As stated before, the plausible assumption of a correction $\mathcal{O}(1) \frac{g_s|Q_3|}{\mathcal{V}^{2/3}_s}$ to the Euler number implies $C_W=\mathcal{O}(1)$ and thus motivates \eqref{sljkgjgfs}. However, we cannot exclude that $C_W$ is numerically small in certain cases so that demanding \eqref{sljkgjgfs} may be too strict.
Our philosophy regarding this loophole is that a dS construction is only under control if all ignored corrections are \emph{shown} to be negligibly small, either by explicit computation or by identifying a small parameter by which the corrections are suppressed (in this case $\lambda_W$).
The advantage of the second option is that we can bypass a full computation of the corrections as long as we know their scaling with the parameter and convince ourselves that the parameter can be tuned small.\footnote{As stated in the introduction, it is impossible in type IIB to take control parameters to zero but we can at least try to make them very small.} On the other hand, a dS model whose consistency relies purely on the hope of a small $C_W$ without computing it is in our view not under control. Indeed, it is usually not feasible to compute all warping (and other) corrections in full explicitness including numerical factors, and therefore suppressing them by small control parameters is in practice the only chance to arrive at a controlled model. This is precisely why the LVS claim of parametric control through an exponentially large volume has been studied so extensively in the last twenty years.
If we have an exponentially small control parameter as naively suggested by the LVS, we can be reasonably sure that there will be no bad surprises from unknown corrections even if we do not know all numerical factors precisely. On the other hand, if the control parameter is $\gtrsim 1$, it is clear that the existence of the dS vacua is highly sensitive to things we do not understand, making the construction speculative.
In this paper, we will therefore insist on controlling warping corrections with a small control parameter, i.e., on dS vacua in which \eqref{sljkgjgfs} is satisfied.
We will actually see that this is impossible in the models we study -- while dS vacua with relatively large $\mathcal{V}_s$ do (formally) exist, the suppression is always cancelled by the other factors in $\lambda_W$ such as $Q_3$, meaning that there is no parametric suppression of the warping corrections at all. While this does not rule out numerically small warping with $C_W\ll 1$, such hypothetical vacua cannot be established without computing the full backreaction and thus leaving the framework of the LVS.

We finally note that there is a second type of corrections which has very similar parametrics as the warping corrections but does not come from backreaction effects,
namely curvature corrections to the worldvolume theory of the D7-branes on $D_\text{T}$. For example, the gauge-kinetic function including the leading $\alpha'$ corrections is \cite{Haack:2006cy}
\begin{equation}
\text{Re} (f_\text{D7}) = \frac{1}{2} \int_{D_\text{T}} \iota^*_\text{T} J^2 - \frac{1}{2 g_s} \int_{D_\text{T}} \hat{\mathcal{F}}^2 - \frac{\chi(D_\text{T})}{24 g_s} = \frac{1}{2} \int_{D_\text{T}} \iota^*_\text{T} J^2 + \frac{Q_{3\text{T}}}{N g_s},
\end{equation}
where $Q_{3\text{T}}$ is the D3 charge of the T-brane (which is typically negative).
Another example where such $\alpha'$ corrections show up is the K\"ahler potential for the brane moduli:
\begin{equation}
K_\text{D7} =
\text{Tr} \left[1 - \frac{2}{g_s \int_{D_\text{T}} \iota^*_\text{T} J^2 } \left(\frac{1}{2}\int_{D_\text{T}} \mathcal{F}^2 + \frac{\chi(D_\text{T})}{24} \right)\right] \int_{D_\text{T}}\hat\Phi^\dagger \w \hat\Phi.
\end{equation}
For a single brane, the flux correction was derived in \cite{Jockers:2005zy} and the Euler-number one in \cite{Junghans:2014zla}, and we generalized this here to the non-Abelian case by putting a trace in front as in \cite{Junghans:2026jza}. In the regime $ \frac{2|Q_{3\text{T}}|}{g_s N \int_{D_\text{T}} \iota^*_\text{T} J^2 } \approx \frac{2^{1/3}}{N p_b 3^{2/3} \kappa_{bbb}^{1/3}}\frac{|Q_{3\text{T}}|}{\mathcal{V}_s^{2/3} } \gtrsim 1$, the gauge-kinetic function and the K\"ahler metric of the brane moduli both change their signs (assuming $Q_{3\text{T}}<0$), indicating a loss of control. From the point of view of the 8d DBI action, this corresponds to a regime where $\alpha'$ corrections become larger than the leading tension term. Although the condition for control over these corrections is numerically weaker than \eqref{sljkgjgfs}, it is interesting that the parametric scaling with the D3 charge and $\mathcal{V}_s$ is the same as for the warping corrections, which have a completely different origin.

\subsection{Higher $F$-terms}
\label{sec:lvs-corr2}

Another type of corrections that is potentially dangerous for the LVS scales like
\begin{equation}
\sim \frac{g_s|W_0|^2}{\mathcal{V}^{2/3}} = \frac{g_s^2|W_0|^2}{\mathcal{V}_s^{2/3}} \label{dfhdkf}
\end{equation}
relative to the leading terms in the LVS potential. Such corrections have been argued to arise from higher $F$-terms corresponding in 10d to $\alpha'$ corrections involving $G_3$ and from integrating out KK modes \cite{deAlwis:2012vp, Cicoli:2013swa, Ciupke:2015msa, Junghans:2026jza} (see also \cite{Conlon:2005ki, Burgess:2020qsc, Cicoli:2021rub}). As discussed in Section \ref{sec:lvs-up2}, corrections with such a scaling also arise already at the $F^2$ order from stabilizing the D7-brane moduli in the T-brane configuration (cf.~\eqref{dfajdsfsotk}).
For lack of a better name, we will collectively call all corrections which scale like \eqref{dfhdkf} ``$F$-term corrections'', regardless of their origin.

Analogously to the warping corrections, one can show \cite{Junghans:2022exo} that the $F$-term corrections to $V_0$ and $m_2^2$ are by a factor $1/g_s$ larger than in the off-shell potential, at least when they correct the BBHL term. This is again due to the previously mentioned non-perturbative no-scale structure. Concretely, one finds
\begin{equation}
\frac{\delta V_0}{V_0} = - C_F \frac{3^{2/3} 16}{27(\alpha-1)} \frac{a_s \kappa_{sss}^{1/3}}{\hat\xi^{1/3}} \frac{g_s|W_0|^2}{\mathcal{V}_s^{2/3}}, \qquad
\frac{\delta (m_2^2)}{m_2^2} = C_F \frac{3^{2/3} 22}{27(\frac{9}{8}-\alpha)} \frac{a_s \kappa_{sss}^{1/3}}{\hat\xi^{1/3}} \frac{g_s|W_0|^2}{\mathcal{V}_s^{2/3}}.
\end{equation}
Note that the corrections scale only linearly with $g_s$, while in the off-shell potential they scale quadratically with $g_s$.
To get a conservative estimate of the size of the corrections, we again take $\alpha=\frac{20}{19}$ where both corrections are equal in size. Assuming further $C_F=\mathcal{O}(1)$ then suggests the following proxy for control:
\begin{equation}
\lambda_F\equiv \frac{3^{2/3} 304}{27} \frac{a_s \kappa_{sss}^{1/3}}{\hat\xi^{1/3}} \frac{g_s|W_0|^2}{\mathcal{V}_s^{2/3}} \ll 1 \qquad \text{(control over }F\text{-term corrections)}. \label{fsgsfgsf}
\end{equation}
As discussed before, demanding small $\lambda_F$ for control would be too strict if $C_F$ were significantly smaller than $\mathcal{O}(1)$.

Interestingly, \eqref{fsgsfgsf} is automatically satisfied when the corresponding condition for the warping corrections is satisfied. This follows because, for supersymmetrically stabilized complex-structure moduli as in the LVS, $W_0$ is bounded by \cite{Denef:2004ze}
\begin{equation}
|W_0|^2 \le \frac{1}{4\pi} |Q_3|, \label{sfgjfsg}
\end{equation}
which is a straightforward consequence of the 10d inequality $|G_3^{0,3}|^2 \le |G_3|^2$ (in the conventions of \cite{Junghans:2026jza} and setting $\e^{K_\text{cs}}=1$). Using this in \eqref{sljkgjgfs} then implies \eqref{fsgsfgsf}.
In the remainder of this paper, we will therefore ignore the $F$-term corrections and focus on the constraint \eqref{sljkgjgfs} from the warping corrections.

\subsection{Backreaction of D7 Charge}
\label{sec:lvs-corr3}

Let us also briefly discuss corrections due to the backreaction of D7 charge. We will not study these corrections in detail and not demand their absence as a requirement in our analysis. Nevertheless, as pointed out in \cite{Junghans:2022exo}, they are potentially dangerous as well, which motivates searching for models with local D7/O7 tadpole cancelation, as we now explain.

Backreaction corrections due to D7 charges arise in compactifications where the D7-branes and O7-planes do not lie on top of each other. For dimensional reasons, the backreaction of codimension-2 sources cannot have a power-law volume scaling so that we generically expect corrections of the order $g_s$ (recall that the backreaction of codimension-$n$ sources scales like $\frac{g_s}{R^{n-2}}$). This estimate can be further refined as follows: at distances $r\ll R\sim \mathcal{V}_s^{1/6}$ much smaller than the size of the Calabi-Yau, the Green's function is the flat-space one and thus scales logarithmically, yielding a backreaction of the order $g_s\ln r$. At larger distances, the backreaction grows more slowly and eventually must reach a maximum since we are on a compact space. At generic points on the Calabi-Yau, we therefore expect a backreaction of the order $g_s \ln R \sim g_s \ln \mathcal{V}_s$. Hence, also the corrections to the 4d effective field theory should be of that order. This should hold unless the separation of the D7 and O7 charges is much smaller than $R$ in which case the volume scaling is milder. Note that, if the log factor occurs, this immediately implies that the D7/O7 backreaction cannot be ignored in the LVS since $g_s \ln \mathcal{V}_s \sim g_s a_s\tau_s \gtrsim 1$. However, even if one optimistically assumes that the log factor is absent, the D7/O7 backreaction is still important, as we now discuss.

As a concrete example for a backreaction correction, consider again the BBHL term \cite{Becker:2002nn}. In type IIB language, a varying dilaton $\e^{-\phi(y)}=\frac{1}{g_s}+f(y)$ corrects the term as \cite{Junghans:2022exo}
\begin{equation}
\frac{\chi(X)}{g_s} \to \int_X c_3(X) \e^{-\phi(y)} = \frac{\chi(X)}{g_s} \left( 1  + \mathcal{O} (g_s) \right).
\end{equation}
As we already noted previously, similar backreaction corrections are also expected to appear in various other places, e.g., in the $D$-term potential and the uplift term.

An F-theory argument for the $g_s$ correction was given in \cite{Minasian:2015bxa} (aside from the topological correction at the order $g_s^0$ also found there, cf.~\eqref{zjghjgjg}), and the type IIB analogue was discussed in \cite{Junghans:2022exo}. An explicit calculation of the correction was furthermore performed in \cite{Antoniadis:2019rkh} in a toroidal orbifold model where it indeed acquires a logarithmic volume scaling, in agreement with our above estimate. However, since orbifolds are special limits and our estimate was somewhat heuristic, let us be optimistic and assume that the backreaction only scales like $g_s$ without the log factor. The LVS potential \eqref{lvspotential} is then corrected such that
\begin{equation}
\hat\xi \to \hat\xi + C_\phi g_s,
\end{equation}
where $\hat\xi$ is defined as in \eqref{zjghjgjg} and $C_\phi$ is some $\mathcal{O}(1)$ coefficient.
Naively, this is under control for $g_s\ll 1$. However, the shift in $\hat\xi$ rescales the stabilized volume as \cite{Junghans:2022exo}
\begin{equation}
\mathcal{V} \to \e^{\frac{C_\phi}{(2\kappa_s)^{2/3}}} \mathcal{V}, \label{sdgsligdslgs}
\end{equation}
which is an $\mathcal{O}(1)$ factor regardless of how small $g_s$ is.

One may object that such a rescaling is not problematic since we can absorb the unknown factor into the parameters $W_0$, $g_s$, $A_s$ in \eqref{sglisgljs}, which in the context of the LVS are usually assumed to be free parameters. However, $A_s$ in reality is a fixed number which has been argued to not depend on the complex-structure moduli on a blow-up divisor \cite{Hebecker:2018fln}. In situations where the available flux tadpole is not too large (as it will be the case below in many models we consider), the possibilities to tune $W_0$, $g_s$ may be quite limited as well. Hence, $A_s$, $W_0$ and $g_s$ are not really free parameters.
A change of $\mathcal{V}$ by a factor of, say, 2 or 3 due to \eqref{sdgsligdslgs} may therefore decide whether or not a putative dS vacuum is consistent.
Unless one computes the D7/O7 backreaction fully explicitly, we therefore consider models with local D7/O7 tadpole cancelation more desirable from the point of view of ensuring a controlled approximation. However, as stated before, we will not impose this as a requirement and also study models with non-local D7/O7 tadpole cancelation below.

\section{Models with gaugino condensation}
\label{sec:h112}

We now study in more detail the consequences of the bound \eqref{dsgsg} in explicit Calabi-Yau orientifolds with $h^{1,1}=2$. As stated before, we are particularly interested in how well warping corrections can be controlled (as estimated by the minimally attainable value of our proxy $\lambda_W$ defined in \eqref{sljkgjgfs}) and how large the tadpole $|Q_3|$ can be made. We will assume that the non-perturbative superpotential is generated by gaugino condensation since, as we saw in Section \ref{sec:problem}, models with an instanton-generated superpotential tend to have very small volumes.
It will furthermore again be convenient to consider manifolds where $D_b$, $D_s$ are an integral basis and restrict to elements pulled back from $H^2(X,\mathbb{Z})$ in the flux-quantization condition.
As explained in Section \ref{sec:lvs-flux}, these assumptions simplify the analysis but are still general enough to allow us to study a large set of different models. We expect similar results to hold in models violating these assumptions but leave a thorough analysis of this question for future work. 

The Swiss-cheese Calabi-Yau manifolds with $h^{1,1}=2$ that arise from the Kreuzer-Skarke list \cite{Kreuzer:2000xy} were classified in \cite{Gray:2012jy} (see also \cite{Altman:2014bfa, Altman:2017vzk}). There are 22 such manifolds\footnote{Some of them have the same Hodge, Chern and intersection numbers and may therefore actually be topologically equivalent \cite{Chandra:2023afu, Gendler:2023ujl}. We will not be bothered with determining this and count each manifold as distinct. \label{sgligsfg}}, which we will label as $M_{2,i}$ following the notation in Table 11 of \cite{AbdusSalam:2020ywo}.
Most of the relevant topological data of these manifolds can be looked up in that table or in the database of \cite{Altman:2014bfa, Altman:2017vzk} (available at \url{http://www.rossealtman.com/toriccy}), and all further data we need (in particular $c_2(X)$ in the $D_b$, $D_s$ basis) can be straightforwardly computed from the former.

In order to identify the manifolds for which \eqref{bound} yields the weakest constraints on the volume, we consider the parameter
\begin{equation}
\gamma \equiv \frac{|\chi|}{\kappa_{bbb}^2 \kappa_{sss}}.
\end{equation}
Using that $|\hat\chi|$ is typically of the same order or smaller than $|\chi|$, \eqref{bound} yields $\mathcal{V}_s \lesssim \frac{5 \gamma}{a_s^3}$.
We will therefore focus on manifolds with $\gamma > 1$ from now on since they have the best chances of admitting large volumes and perturbative control.
Computing $\gamma$ for all 22 manifolds, it turns out that only 11 of them satisfy $\gamma > 1$.
These manifolds have the following properties:
\\

\noindent\begin{tabular}{ l | l l l l l }
CY &  $|\chi|$ & $\kappa_{sss}$ & $\kappa_{bbb}$ & $\gamma$ & $c_2(X)$ \\
\hline
$M_{2,3}$ & $144$ & $1$ & $3$ & $16$ & $14D_b^2+10D_s^2 $ \\
$M_{2,4}$ & $144$ & $1$ & $3$ & $16$ & $14D_b^2+10D_s^2$ \\
$M_{2,5}$ & $144$ & $1$ & $3$ & $16$ & $14D_b^2+10D_s^2 $ \\
$M_{2,7}$ & $164$ & $2$ & $5$ & $\frac{82}{25}$ & $10D_b^2+4D_s^2 $ \\
$M_{2,13}$ & $168$ & $2$ & $3$ & $\frac{28}{3}$ & $14D_b^2+4D_s^2$  \\
$M_{2,18}$ & $176$ & $3$ & $5$ & $\frac{176}{75}$ & $10D_b^2+2D_s^2$  \\
$M_{2,19}$ & $180$ & $3$ & $3$ & $\frac{20}{3}$ & $14D_b^2+2D_s^2$  \\
$M_{2,25}$ & $228$ & $1$ & $1$ & $228$ & $34D_b^2+10D_s^2$  \\
$M_{2,26}$ & $236$ & $1$ & $2$ & $59$ & $22D_b^2+10D_s^2$  \\
$M_{2,35}$ & $252$ & $2$ & $1$ & $126$ & $34D_b^2+4D_s^2$  \\
$M_{2,36}$ & $260$ & $2$ & $2$ & $\frac{65}{2}$ & $22D_b^2+4D_s^2$  \\
\end{tabular}
\\[1em]

\noindent One can verify that $D_b$, $D_s$ are an integral basis in all of these manifolds (cf.~App.~\ref{app:toric}). One also easily verifies that $\chi_0(D_s)=\frac{1}{12}\int_X(c_2(X)\w D_s+2D_s^3)=1$ for all manifolds. Hence, a necessary condition for the rigidity of $D_s$ is satisfied in all models (cf.~Section \ref{sec:lvs-np}).

Using the toric data of the database of \cite{Altman:2014bfa, Altman:2017vzk}, we computed in App.~\ref{app:toric} all possible orientifoldings of these 11 manifolds that are obtained by reflections on one of the irreducible basis divisors of the ambient varieties. We discarded orientifolds which do not have an O7-plane on $D_b$ since they are not compatible with a T-brane uplift due to the requirement of tadpole cancelation.
We also discarded orientifolds for which imposing orientifold invariance makes the Calabi-Yau hypersurface singular. As detailed in the appendix, such singularities are often signaled by fractional O3-plane or O7-plane numbers. Discarding these cases, we are left with 28 different orientifolds for which we perform our analysis. As a further consistency check, we verified that $h^{2,1}_-$ is integral for each of these orientifolds. Some of the orientifolds we kept may still have more subtle singularities which cannot be detected by our simple checks and would require a more careful analysis. We refrain from performing such an analysis since our main point is that all models have large warping corrections, which would of course still be true in case some of them have to be discarded.

The next step is then to compute, for each orientifold, the possible T-brane configurations, i.e., the values of $p_b$, $p_s$ and $f_b$ allowed by the various constraints discussed in Section \ref{sec:lvs}. Since there are many ways to place the D7-branes and choose a flux consistently with these requirements, this leads to 605 different T-brane models with candidate dS vacua. In each of these models, one can then compute
the tadpole $|Q_3|$ (using \eqref{zupzuipto}) as well as the bounds on $\mathcal{V}_s$ and $\lambda_W$ (using \eqref{dsgsg} and \eqref{sljkgjgfs}).
The results of our scan are discussed in the following subsections.

\subsection{General Results}
\label{sec:h112gen}

\begin{figure}[t]
\centering
\includegraphics[trim = 0mm 0mm 0mm 0mm, clip, width=0.48\linewidth]{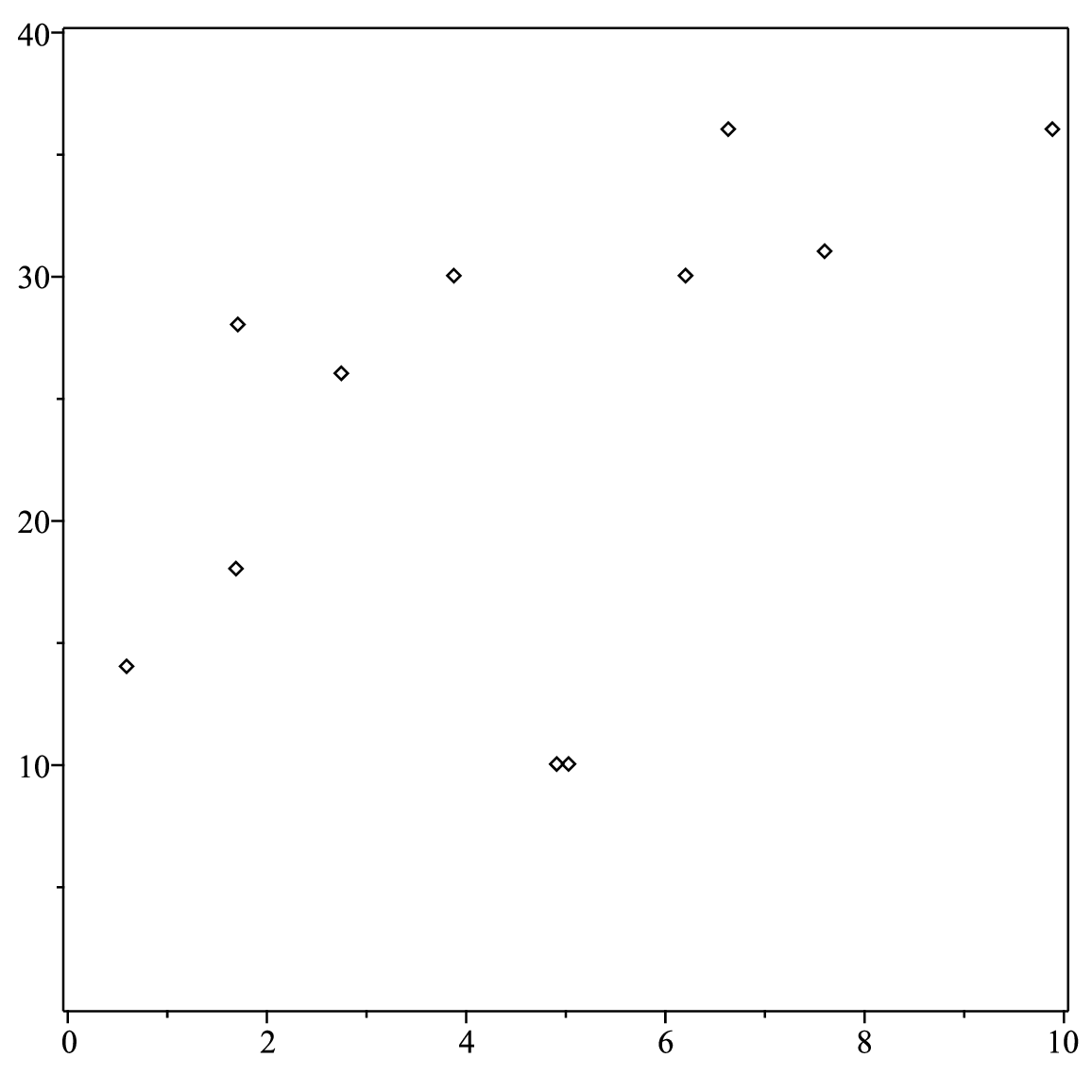} \,\,\,\,\, \includegraphics[trim = 0mm 0mm 0mm 0mm, clip, width=0.48\linewidth]{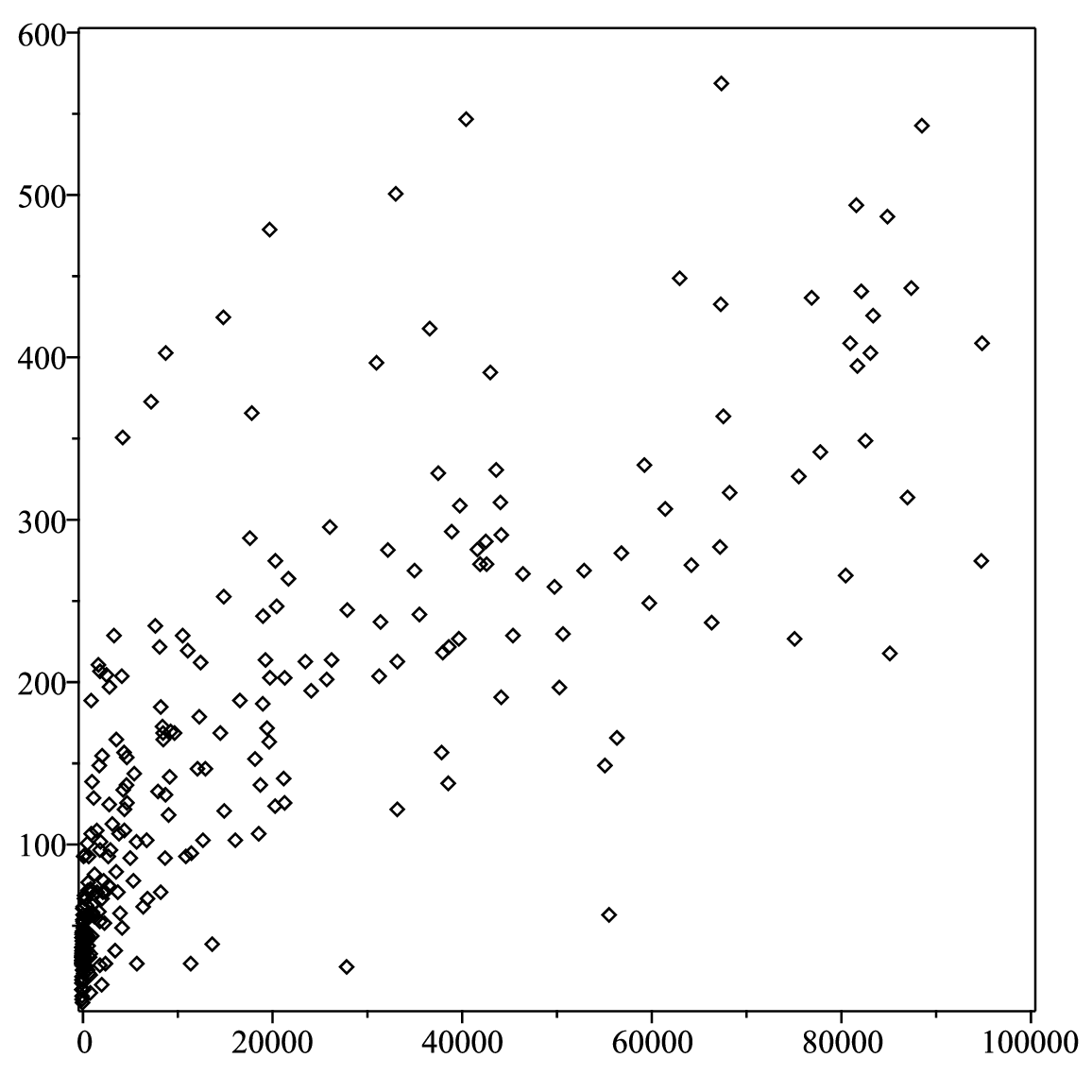}
\put(-348,-2){$\scriptstyle{\text{min}(\lambda_W)}$}
\put(-455,110){$\scriptstyle{|Q_3|}$}
\put(-118,-2){$\scriptstyle{\text{min}(\lambda_W)}$}
\put(-227,110){$\scriptstyle{|Q_3|}$}

\caption{Plots of the lower bound on the control parameter $\lambda_W$ versus the tadpole $|Q_3|$ for models with $\text{min}(\lambda_W)\le 10$ (left) and for models with $\text{min}(\lambda_W)\le 10^5$ (right), showing that small warping corrections correlate with small tadpoles. Each point corresponds to a different T-brane model.
}
\label{fig2}
\end{figure}

As shown in Fig.~\ref{fig2}, the vast majority of the models have a very large $\lambda_W$ parameter several orders of magnitude above 1, suggesting that warping corrections are far from being under control.
There are however also a few models with $\lambda_W=\mathcal{O}(1)$. In particular, out of the 605 models we studied, 11 admit a $\lambda_W$ smaller than 10. The five models with the smallest $\lambda_W$ are
\\

\noindent\begin{tabular}{l | l | l l l l l l l | l l l l l l }
ID & CY & $p_b$ & $p_s$ & $r_b$ & $r_s$ & $f_b$ & $N$ & $M$ & $N_\text{O3}$ & $|Q_3|$ & $h^{2,1}_-$ & $\text{max}(\mathcal{V}_s)$ & $\text{min}(\lambda_W)$ \\
\hline

NL1 & $M_{2,5}$ & $2$ & $-5$ & $1$& $1$ & $\frac{1}{2}$ & $4$ & $28$ & $0$ & $14$ & $49$ & $5.22\cdot 10^3$ & $0.602$ \\
NL2 & $M_{2,5}$ & $4$ & $-7$ & $1$ & $1$ & $\frac{1}{2}$ & $2$ & $22$ & $0$ & $18$ & $49$ & $2.38\cdot 10^3$ & $1.70$ \\
NL3 & $M_{2,25}$ & $4$ & $-3$ & $1$ & $1$ & $\frac{1}{2}$ & $2$ & $14$ & $2$ & $28$ & $68$ & $7.61\cdot 10^3$ & $1.72$ \\
NL4 & $M_{2,25}$ & $2$ & $-1$ & $1$& $1$ & $\frac{1}{2}$ & $4$ & $12$ & $2$ & $26$ & $68$ & $4.40\cdot 10^3$ & $2.76$ \\
NL5 & $M_{2,5}$ & $2$ & $-3$ & $1$& $1$ & $\frac{1}{2}$ & $4$ & $20$ & $0$ & $30$ & $49$ & $1.73\cdot 10^3$ & $3.89$ \\
\end{tabular}
\\[1em]

\noindent Here ``NL'' means that the D7/O7 tadpoles are cancelled non-locally (see Section \ref{sec:h112loc} for models with local tadpole cancelation).

One might wonder whether these models with $\lambda_W=\mathcal{O}(1)$ can be marginally ok regarding the warping corrections.
An interesting observation in this context is that smaller warping corrections correlate with smaller tadpoles. In Fig.~\ref{fig2}, this is shown for the subset of models with $\lambda_W<10$ and for a larger subset with $\lambda_W<10^5$. As we will discuss in Section \ref{sec:h112tune}, we therefore expect that there is a trade-off between control and consistent moduli stabilization. In particular, the models in the above table have tadpoles $|Q_3| \le 30$ and should therefore be taken with a grain of salt. Recall that these are the upstairs tadpoles, i.e., they are even smaller by a further factor 2 on the orientifold.

Let us also remark that the T-branes in the NL1 and NL2 models have a negative Euler number $\chi(D_\text{T})<0$ and, as a consequence, contribute a \emph{positive} D3 charge $Q_{3\text{T}}=-N\frac{\chi(D_\text{T})}{24}-\frac{1}{2}\text{Tr}\int_{D_\text{T}}\mathcal{F}^2>0$ to the tadpole. This is quite unusual since D7-branes and O7-planes typically contribute negative D3 charges and thus act like O3-planes in the tadpole condition. Although we do not see an inconsistency with a negative $\chi(D_\text{T})$ per se\footnote{Nef divisors satisfy $c_2(X) \w D \ge 0$ and $D^3 \ge 0$ \cite{Katz:2020ewz} and therefore $\chi(D)\ge 0$ but this is not necessarily true for effective divisors.}, it suggests that $\lambda_W$ is not a good proxy for the warping corrections in these models since it estimates the backreaction to be proportional to $|Q_3|$. Since the positive D3 charge of the T-brane and the negative D3 charges of the other localized sources do not locally cancel in these models, it is unclear why the warping should only be sensitive to the difference between these charges. As explained in Section \ref{sec:lvs-corr1}, a more plausible estimate is that the backreaction is proportional the total negative D3 charge (or, equivalently, the total positive one). This suggests that one should use a modified proxy $\tilde\lambda_W$ in these models in which $Q_3$ is replaced by $Q_{3\text{neg}}=Q_3-Q_{3\text{T}} = - \frac{\chi(D_\text{O7})}{6}- \frac{M \chi(D_s)}{24}-\frac{N_\text{O3}}{2}$ which does not contain the T-brane contribution. If one measures the warping corrections using this modified proxy in the NL1 and NL2 models, one finds $\tilde\lambda_W > 0.953$, which is somewhat larger than the bound obtained using $\lambda_W$.

Regardless of which proxy one prefers to estimate the corrections, our analysis in any case suggests that LVS dS vacua with T-brane uplifting can at best lie \emph{at the boundary of control} where the LVS potential becomes highly sensitive to the precise numerical coefficients of the warping corrections, in sharp contrast with the common belief of an exponentially small control parameter (cf.~the discussion in Section \ref{sec:lvs-corr1}).

\subsection{Tuning Problem for Small Tadpoles}
\label{sec:h112tune}

As discussed above, the models with the smallest $\lambda_W$ (and thus the best control over warping corrections) have quite small tadpoles $|Q_3|\le 30$.
Let us now elaborate a bit more on why this may be problematic.
The crucial point is that the LVS is based on the assumption that the axio-dilaton, the complex-structure moduli and the brane moduli are stabilized by fluxes and that this happens in such a way that $g_s$ and $W_0$ take precisely the values required for a dS minimum.

One may wonder whether a possible obstacle to this assumption is the tadpole conjecture \cite{Bena:2020xrh}, which postulates that $|Q_3|$ has to satisfy a certain lower bound that depends linearly on the number of stabilized moduli. However, there is no strong motivation to consider the tadpole conjecture in our context since it is known that the bound can be violated in compactifications with non-Abelian gauge groups and/or $W_0\neq 0$ \cite{Aspinwall:2005ad, Braun:2008pz, Braun:2014ola, Braun:2023pzd, Lust:2022mhk}.\footnote{The proposed bound is also violated in compactifications with a small number of moduli \cite{Plauschinn:2021hkp} and in non-geometric examples \cite{Becker:2022hse, Becker:2024ayh, Rajaguru:2024emw, Becker:2024ijy}.} The T-brane models studied in this paper have both of these features which suggests that small tadpoles can in principle be fine.

However, it is not sufficient to stabilize the axio-dilaton and the complex-structure moduli \emph{anywhere} in the moduli space. Instead, a trustworthy type IIB description requires $g_s\ll 1$ since otherwise the genus expansion breaks down, leading to large loop and instanton corrections and also large backreaction corrections from the D7-branes and O7-planes. Recall further from Sections \ref{sec:lvs-km} and \ref{sec:lvs-corr1} that the uplift parameter $\alpha$ lies in the interval $]1,\frac{9}{8}[$ in a dS vacuum and that we should not come too close to the boundaries of this interval since otherwise either corrections to the vacuum energy or the Hessian blow up.
Hence, $\alpha$ is fixed up to a few percent, which corresponds to a very thin band in the $(g_s,W_0)$-plane. Stabilizing the moduli such that we land within this band is expected to require a certain amount of fine-tuning. This is only possible if there is enough freedom in choosing the fluxes or, in other words, if $|Q_3|$ is not too small.

To be more concrete, note that \eqref{sglisgljs}, \eqref{sglisgljs2} and \eqref{dsgsg} imply
\begin{equation}
|W_0| = \frac{1458\pi^3\alpha^3\hat\xi^{2/3} \hat A_s}{3^{1/3} a_s^2\kappa_{bbb}^2 (4r_bf_b)^3 \kappa_{sss}^{2/3} g_s} \e^{- a_s \frac{3^{2/3} \hat\xi^{2/3} \kappa_{sss}^{1/3}}{ 2 g_s}+\frac{a_s\chi(D_s)}{24g_s}}
\end{equation}
up to subleading terms in $g_s$, where we also used the tadpole condition $Np_b=8r_b$. We further defined $A_s = \hat A_s \e^{\frac{a_s\chi(D_s)}{24g_s}}$ in order to make explicit the $g_s$ dependence of $A_s$ due to a curvature correction to the instanton/D7-brane action which is proportional to the Euler number $\chi(D_s)$ of the small divisor.

In a given Calabi-Yau orientifold model, we thus have $|W_0| = \frac{c_1}{g_s} \e^{-c_2/g_s} \alpha^3$ for some fixed numbers $c_1$, $c_2$. Hence, for $\alpha$ in the dS range and a given choice of $g_s$, $|W_0|$ is fixed up to a few percent. According to \eqref{sfgjfsg}, the allowed parameter region is in addition bounded by
\begin{equation}
|W_0|^2 \le \frac{1}{4\pi}|Q_3|.
\end{equation}
Note that this inequality is automatically satisfied in every flux vacuum in which the complex-structure moduli and the axio-dilaton are stabilized supersymmetrically. The only tuning requirements on $g_s$, $W_0$ are therefore hitting the small-coupling regime and the dS region.

\begin{figure}[t]
\centering
\includegraphics[trim = 0mm 0mm 0mm 0mm, clip, width=0.5\linewidth]{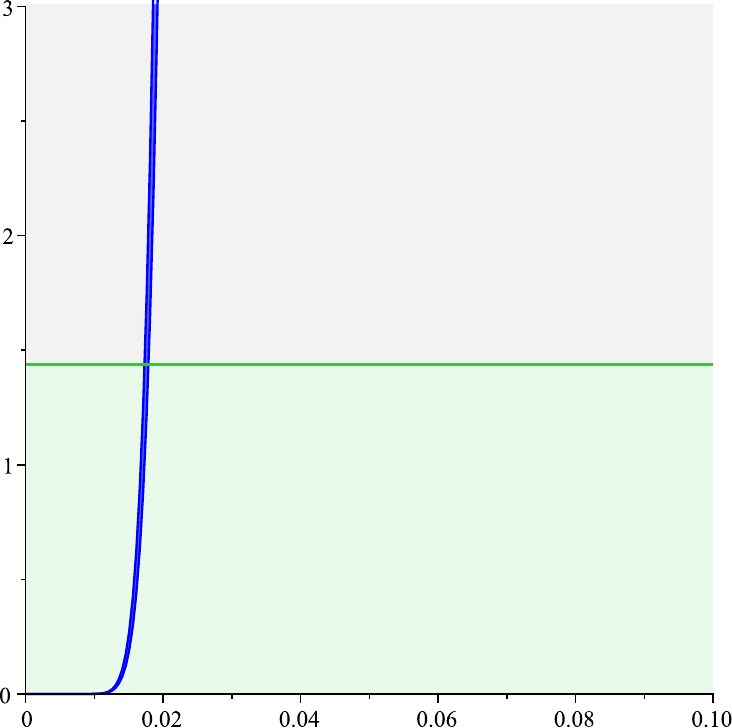}
\put(-110,-10){$\scriptstyle{g_s}$}
\put(-240,110){$\scriptstyle{|W_0|}$}

\caption{Plot of the $(g_s,W_0)$-plane for the model NL4 of Section \ref{sec:h112gen}. General flux vacua lie in the green region defined by $|W_0|^2 \le \frac{1}{4\pi} |Q_3|$, and dS vacua lie in the thin blue band.}
\label{fig3}
\end{figure}

As an example, consider the model NL4 of Section \ref{sec:h112gen}, which corresponds to the parameter choices $\hat\xi=0.543$, $a_s= \frac{2\pi}{7}$, $f_b=\frac{1}{2}$, $r_b=\kappa_{bbb}=\kappa_{sss}=1$, $\chi(D_s)=11$, $Q_3=-26$. As is standard in the literature, we also assume $\hat A_s=1$ for concreteness. Using these values in the above expressions, we find
\begin{equation}
|W_0| = \frac{3.24\cdot 10^3}{g_s} \e^{-\frac{0.210}{g_s}} \alpha^3, \qquad |W_0|^2 \le \frac{26}{4\pi}.
\end{equation}
The resulting dS region in the $(g_s,W_0)$-plane is plotted in Fig.~\ref{fig3}. As anticipated, this region is indeed a very thin band, suggesting that landing in it requires some freedom in the fluxes and, accordingly, a not too small tadpole. Note that the dS region satisfies $g_s \lesssim 0.02$ so that the requirement of a small string coupling does not lead to an additional constraint in this case.

To summarize, we expect that consistent dS vacua require an at least moderately large tadpole (even assuming that the tadpole conjecture does not apply in our context). This suggests that the naively best-controlled solutions in Section \ref{sec:h112gen} may have to be discarded and that the control problems observed there would become even worse.

In principle, constructing Calabi-Yau orientifolds with very large tadpoles $\gtrsim \mathcal{O}(10^3)$ is not a problem, as emphasized, e.g., in \cite{Crino:2022zjk}.
However, we have seen that, in LVS models with a T-brane uplift, there is a trade-off between achieving control over warping/$\alpha'$ corrections on the one side and large $|Q_3|$ on the other side.
This did not have to be the case: although it naively seems obvious that making $|Q_3|$ large will also make $\lambda_W\sim\frac{|Q_3|}{\mathcal{V}_s^{2/3}}$ large, this is not necessarily true since the vev of $\mathcal{V}_s$ may a priori depend non-trivially on the fluxes. For the anti-brane uplift, one indeed finds that $\frac{|Q_3|}{\mathcal{V}_s^{2/3}}$ tends to get \emph{smaller} in dS solutions with larger tadpoles because then there is more freedom to find solutions with large volumes \cite{Gao:2022fdi, Junghans:2022kxg}. For the T-brane uplift, it is the other way round, i.e., we cannot obtain large $|Q_3|$ and small $\frac{|Q_3|}{\mathcal{V}_s^{2/3}}$ at the same time.

\subsection{Local Tadpole Cancelation}
\label{sec:h112loc}

As explained in Section \ref{sec:lvs-corr3}, models with non-local D7/O7 tadpole cancelation have the disadvantage that, regardless of how small $\lambda_W$ is, there are in any case unsuppressed dilaton-backreaction corrections which modify the vev of the volume modulus by $\mathcal{O}(1)$ factors. Whether or not this is a serious problem is debatable but let us in any case discuss what happens if we demand local tadpole cancelation as an additional constraint.

Since, by assumption of the scenario, we have a stack of $M$ D7-branes on $D_s$, canceling their tadpole locally means that we require O7-planes on $D_s$ as well and therefore we have to set $r_s=1$. In order to cancel the tadpole of the T-brane, we furthermore require an O7-plane wrapping $D_b$ or a multiple thereof, i.e., $r_b\neq 0$. The only possibility consistent with $r_s=1$ is $r_b=1$ in our models, cf.~App.~\ref{app:toric}. With this orientifold choice, the T-brane stack must then wrap $D_\text{T}=D_b$ for local tadpole cancelation, i.e., we have $p_b=1$, $p_s=0$. The constraint \eqref{march} further implies $f_b=\frac{1}{2}$, and the tadpole conditions \eqref{d7tadp} yield $N=M=8$. In summary, demanding that O7/D7 charges cancel locally amounts to the parameter choices
\begin{equation}
p_b=r_b=r_s=1, \qquad p_s=0, \qquad f_b=\frac{1}{2}, \qquad N=M=8. \label{zioutzot}
\end{equation}
As shown in App.~\ref{app:toric}, only five of the 11 Calabi-Yau manifolds in our list admit an orientifolding with $r_b=r_s=1$. Since the brane configuration (i.e., the choice for $p_b$, $p_s$ and $f_b$) is unique, this means that only five of the 605 T-brane models we scanned have local tadpole cancelation. Using \eqref{zioutzot}, we find that these models have the following properties:
\\

\noindent\begin{tabular}{l | l | l l l l l l }
ID & CY & $N_\text{O3}$ & $|Q_3|$ & $h^{2,1}_-$ & $\text{max}(\mathcal{V}_s)$ & $\text{min}(\lambda_W)$ \\
\hline
L1 & $M_{2,25}$ & $2$ & $25$ & $68$  & $951$ & $12.3$ \\
L2 & $M_{2,36}$ & $0$ & $30$ & $78$  & $134$ & $66.0$ \\
L3 & $M_{2,5}$ & $0$ & $31$ & $49$  & $64.1$ & $108$  \\
L4 & $M_{2,13}$ & $1$ & $31$ & $55$  & $37.3$ & $187$ \\
L5 & $M_{2,18}$ & $0$ & $37$ & $59$  & $9.06$ & $652$ \\
\end{tabular}
\\[1em]

\noindent We thus see that $\lambda_W \gg 1$ in all models with local D7/O7 tadpole cancelation, suggesting that warping corrections are far from being under control. Furthermore, the D3 tadpoles are relatively small in all models. This is perhaps not surprising, as the D3 tadpole in the local case is known to be much more constrained than in the non-local case \cite{Crino:2022zjk}.

\section{Models with $h^{1,1}>2$ }
\label{sec:largeh11}

So far, our focus was on models with $h^{1,1}=2$. Let us now move on to the case $h^{1,1}>2$. We will not perform an exhaustive analysis in this section but only make a few observations.
We consider a Swiss-cheese Calabi-Yau with $\mathcal{N}=h^{1,1}-1\ge 2$ small divisors such that
\begin{equation}
\mathcal{V} = \kappa_b \tau_b^{3/2} - \sum_{i=1}^\mathcal{N} \kappa_{si} \tau_{si}^{3/2},
\end{equation}
where we denote the volumes of the small divisors by $\tau_{si}$ and their triple-intersection numbers by $\kappa_{iii}$ with $\kappa_{si} = \frac{\sqrt{2}}{3\sqrt{\kappa_{iii}}}$. We further assume that each of the small divisors hosts a non-perturbative effect leading to a superpotential
\begin{equation}
W = W_0 + \sum_{i=1}^\mathcal{N} A_{si} \e^{-a_{si} \tau_{si}}.
\end{equation}
For simplicity, we will consider the situation where $\kappa_{si}$, $A_{si}$ and $a_{si}$ take the same values for all $i$ and where all $\tau_{si}$ are stabilized at the same value. Writing
\begin{equation}
\kappa_{si} = \frac{\kappa_s}{\mathcal{N}}, \qquad \kappa_{iii} = \mathcal{N}^{2} \kappa_{sss}, \qquad A_{si} = \frac{A_s}{\mathcal{N}}, \qquad a_{si}=a_s, \qquad \tau_{si}=\tau_s \qquad \forall i, \label{dfdfsdforrdg}
\end{equation}
one then finds that the volume and the superpotential take the same form as in the $h^{1,1}=2$ case:
\begin{equation}
\mathcal{V} = \kappa_b \tau_b^{3/2} - \kappa_{s} \tau_{s}^{3/2}, \qquad W = W_0 + A_{s} \e^{-a_{s} \tau_{s}}.
\end{equation}
One furthermore verifies that the scalar potential is given by \eqref{lvspotential} and so also takes the same form as for $h^{1,1}=2$.

This implies in particular that the bound \eqref{dsgsg} on $\mathcal{V}_s$ holds for arbitrary $h^{1,1}$, with the caveat that $\kappa_{sss}$ in the bound now refers to the ``effective'' triple-intersection number defined in \eqref{dfdfsdforrdg}. Since $\kappa_{sss} \sim \mathcal{N}^{-2}$ and the bound on $\mathcal{V}_s$ scales inversely with $\kappa_{sss}$, one may naively think that the bound is relaxed at large $\mathcal{N}$. However, the bound also depends on $|\chi(X)|$, which gets smaller for larger $\mathcal{N}$ (since $|\chi(X)|=2(h^{2,1}-\mathcal{N}-1)$). Furthermore, the warping corrections we want to minimize do not only depend on the volume but also on the tadpole which is expected to get larger when many cycles are present on which D7-branes/O7-planes can contribute. It is therefore a priori unclear whether large $\mathcal{N}$ makes the problem with the warping corrections better or worse. We will refrain from a general analysis of this question here since there is a more immediate problem with large $\mathcal{N}$. Indeed, the volumes of the small divisors satisfy
\begin{equation}
\tau_{si} = \tau_s = \frac{\hat\xi^{2/3}}{(2\kappa_s)^{2/3}g_s} = \frac{\hat\xi^{2/3}}{\mathcal{N}^{2/3} (2\kappa_{si})^{2/3}g_s} \label{hdjhdlj}
\end{equation}
up to subleading $g_s$ corrections. This scaling behavior with respect to $\mathcal{N}$ was already pointed out in \cite{Cicoli:2021dhg}. However, unlike that reference, we view this as problematic since $\tau_s\to 0$ at large $\mathcal{N}$.
Because of \eqref{sglisgljs}, it is unclear how to ensure a large volume in this regime.
We also expect to lose control over the instanton expansion since $S_\text{DBI}\sim \tau_{si}-\frac{\chi(D_{si})}{24g_s}\approx -\frac{\chi(D_{si})}{24g_s} <0$  so that higher-rank instantons contributing to the superpotential are in danger of dominating over the lowest-rank ones. Similarly, the (curvature-corrected) gauge kinetic function of a D7-brane wrapping one of the small divisors has the wrong sign at large $\mathcal{N}$, Re$(f_\text{D7})<0$.
Finally, another possible problem are loop corrections \cite{Gao:2022uop}, which may blow up for $\tau_{si}\to 0$. For these various reasons, we do not think that large $\mathcal{N}$ can lead to viable models.

Let us therefore now consider the case where $h^{1,1}$ is a bit larger than 2 but not too large. We will not perform a systematic scan of such manifolds here but only discuss two explicit examples with $h^{1,1}=4$. The first example was worked out and found to have dS vacua in \cite{Cicoli:2021dhg}. We refer to \cite{Cicoli:2021dhg} for the details of the toric construction and only state the topological data relevant for us:
\begin{equation}
\hat\chi=-56, \qquad r_b=2, \qquad a_s\ge 2\pi, \qquad \kappa_{bbb}=\kappa_{iii} = 4 \quad (i=1,2,3),
\end{equation}
where the bound on $a_s$ arises because the model of \cite{Cicoli:2021dhg} does not have a D7 stack wrapped on $D_s$ and therefore the non-perturbative effect is due to instantons.

Furthermore, the volume $\tau_{s3}$ of one of the small divisors is taken to zero in \cite{Cicoli:2021dhg} in order to be able to place D3-branes at a singularity. In this limit, $\tau_{s3}$ does not appear in the volume anymore so that there are effectively only two small divisors to be stabilized, i.e., $\mathcal{N}=2$. Each of these divisors is del Pezzo and hosts a non-perturbative effect.
Hence, we can use the bound derived in Section \ref{sec:problem} if we take for $\kappa_{sss}$ the effective intersection number
\begin{equation}
\kappa_{sss}=\frac{\kappa_{iii}}{\mathcal{N}^2} = 1.
\end{equation}
Plugging this into \eqref{bound}, we find
\begin{equation}
\mathcal{V}_s < 0.00860.
\end{equation}
Hence, the string-frame volume is very small in any dS vacuum and we do not expect that $\alpha'$ and warping corrections are under control.\footnote{As mentioned in Section \ref{sec:lvs-up2}, our uplift term differs from the one used in \cite{Cicoli:2021dhg} by a factor $N$. Using the uplift term of \cite{Cicoli:2021dhg}, we would obtain a somewhat larger (but still quite small) bound, namely $\mathcal{V}_s<N^3 0.00860 = 4.40$. This latter bound is consistent with the Einstein-frame volumes listed in \cite[Table 4]{Cicoli:2021dhg} after converting them to the string frame.}

We also note that the curvature-corrected instanton actions on the small divisors are negative in this example,
\begin{equation}
S_\text{DBI}= 2\pi \left(\tau_{si} - \frac{\chi(D_{si})}{24g_s}\right) = - \frac{0.366}{g_s},
\end{equation}
where we used \eqref{hdjhdlj} and that $\chi(D_{si})=12-\kappa_{iii}$ for a del Pezzo divisor. As explained above, this suggests that the instanton expansion may not be under control.

A very similar model, which also has $h^{1,1}=4$, was studied in \cite{Cicoli:2017shd}. The relevant topological data are
\begin{gather}
\hat\chi=-24, \qquad p_b=r_b=3, \qquad N=8, \qquad a_s= 2\pi, \notag \\ \kappa_{bbb}= \kappa_{333} = 9, \qquad \kappa_{iii} = 1 \quad (i=1,2).
\end{gather}
As in the previous model, the divisor volume $\tau_{s3}$ is taken to zero so that, from the point of view of moduli stabilization, we effectively have one big and two small divisors, i.e., $\mathcal{N}=2$.
The effective triple-intersection number is thus
\begin{equation}
\kappa_{sss}=\frac{\kappa_{iii}}{\mathcal{N}^2} = \frac{1}{4}.
\end{equation}
A subtlety in this model is that, contrary to all other models discussed in this paper, $D_b$ and $D_s$ are not an integral basis of $H^{1,1}(X,\mathbb{Z})$, which has the consequence that the flux on the T-brane is not bounded by $f_b\ge\frac{1}{2}$ but $f_b\ge \frac{1}{6}$, as explained in \cite{Cicoli:2017shd} (see also our discussion in Section \ref{sec:lvs-flux}).
Using these data in \eqref{dsgsg}, we obtain
\begin{equation}
\mathcal{V}_s < 0.0233. \label{sligsglslgi}
\end{equation}
Hence, we again find a very small string-frame volume.\footnote{As mentioned in Section \ref{sec:lvs-up2}, our uplift term differs from the one used in \cite{Cicoli:2017shd} by a factor $N$. Using the uplift term of \cite{Cicoli:2017shd}, we would obtain a somewhat larger (but still quite small) bound, namely $\mathcal{V}_s < N^3 0.0233 = 11.9$. However, this latter bound is still at odds with \cite[Table 1]{Cicoli:2017shd}, which yields much larger values $\mathcal{V}_s=\mathcal{O}(300)$ when converting the Einstein-frame volumes displayed there to the string frame. The remaining discrepancy seems to originate from a disagreement between \cite[Table 1]{Cicoli:2017shd} and \cite[(4.27)]{Cicoli:2017shd} when plugging in the corresponding values of $\tau_2=\tau_3\equiv \tau_s$ and $\mathcal{C}_\text{up}$ (in the notation of \cite{Cicoli:2017shd}).}
We also again observe in this model that the instanton action is negative,
\begin{equation}
S_\text{DBI}= 2\pi \left(\tau_{si} - \frac{\chi(D_{si})}{24g_s}\right) = - \frac{2.26}{g_s},
\end{equation}
where we used \eqref{hdjhdlj} and that $\chi(D_{si}) = 12-\kappa_{iii} = 11$.

In summary, we argued that the control problems of the T-brane uplift are likely to be present on manifolds with $h^{1,1}>2$ as well, and we analyzed two explicit examples in which this indeed turns out to be the case.

\section{Comments on Brane-moduli Stabilization}
\label{sec:modstab}

In this section, we discuss a potential problem with -- or rather a challenge for -- the stabilization of the D7-brane moduli in vacua with T-branes.
This problem is independent of the control problems discussed in the previous sections, i.e., it would occur even in (hypothetical) solutions without large $\alpha'$ or warping corrections. It also does not depend on the vacuum energy and so occurs for dS, Minkowski and AdS.

Our main point is that the combined $D$-term and $F$-term potential for $\hat\Phi$ derived in Section \ref{sec:lvs-up2} leaves many of the brane moduli unstabilized. This is a source of concern since any unknown correction, regardless of how small it is, can destabilize these moduli and possibly lead to decays to other configurations or runaways towards uncontrolled regimes.
To see the problem,
consider for simplicity the case $N=2$, i.e., a stack of two D7-branes on $D_\text{T}$ with gauge group USp$(2)\simeq $ SU$(2)$. As explained in Section \ref{sec:lvs-d7}, we can then decompose the field $\hat\Phi$ as
\begin{equation}
\hat\Phi = \begin{pmatrix} \hat\Phi^0_- + \hat\Phi^1_+ &  \hat\Phi^2_+ + i \hat\Phi^3_+ \\ \hat\Phi^2_+ - i \hat\Phi^3_+ & \hat\Phi^0_- -  \hat\Phi^1_+ \end{pmatrix},
\end{equation}
where the $\hat\Phi^a$'s are harmonic $(2,0)$-forms (with respect to an appropriate gauge-covariant derivative operator) and the subscripts $\pm$ indicate that they are orientifold-even or odd. Expanding this into cohomology classes yields
\begin{equation}
\hat\Phi = \begin{pmatrix} \zeta^a \omega_a + \rho^\alpha \omega_\alpha & \varphi^m \omega_m \\ \xi^q \omega_q & \zeta^a \omega_a - \rho^\alpha \omega_\alpha \\ \end{pmatrix}. \label{ansatzphisu2}
\end{equation}
Here, $\zeta^a$, $\rho^\alpha$, $\varphi^m$ and $\xi^q$ are complex moduli with indices $a,b=1,\ldots, h^{2,0}_-(D_\text{T})$, $\alpha,\beta=1,\ldots,$ $h^{2,0}_+(D_\text{T})$, $m,n=1,\ldots, \text{dim} H^2_+(D_\text{T},\mathcal{L}^{-2})$ and $q,r=1,\ldots, \text{dim} H^2_+(D_\text{T},\mathcal{L}^{2})$. Note that $H^2(D_\text{T},\mathcal{L}^{\pm 2})$ denotes the cohomology group of $(0,2)$-forms valued in the line bundle $\mathcal{L}^{\pm 2}$ and the subscripts $\pm$ stand for orientifold-even/odd as before. The $\omega_\bullet$'s are harmonic $(2,0)$-forms which are a basis of $H^{2,0}_\pm(D_\text{T})$ or complex conjugate to a basis of $H^2 (D_\text{T},\mathcal{L}^{\pm 2})$, respectively. We will ignore the Wilson-line moduli for simplicity (see \cite{Junghans:2026jza} for more details), which are counted by $h^{0,1}_\pm(D_\text{T})$ and $\text{dim} H^1_+(D_\text{T},\mathcal{L}^{\pm 2})$ and should in principle also be stabilized.

Substituting \eqref{ansatzphisu2} into \eqref{dtermpot}, \eqref{ftermpot}, we obtain the potential
\begin{align}
V &= 
C_1\left(C_2 - G_{m\bar n} \varphi^m \bar\varphi^{\bar n} + G_{q\bar r} \xi^q \bar\xi^{\bar r} \right)^2 \nl 
+ C_3 \left(2G_{a \bar b} \zeta^a \bar\zeta^{\bar b} + 2G_{\alpha\bar \beta} \rho^\alpha \bar\rho^{\bar \beta} + G_{m\bar n} \varphi^m \bar\varphi^{\bar n} + G_{q\bar r} \xi^q \bar\xi^{\bar r}\right), \label{potentialbranemoduli}
\end{align}
where $G_{\bullet \bar \bullet}\equiv \int_{D_\text{T}} \omega_\bullet \w \bar \omega_{\bar \bullet}$ and
\begin{equation}
C_1 = \frac{4\pi}{ 6^{2/3}\kappa_{bbb}^{1/3}p_b\mathcal{V}^{2/3}}, \qquad C_2 = \frac{6^{1/3}\kappa_{bbb}^{2/3}p_bf_b}{4\pi\mathcal{V}^{2/3}}, \qquad C_3 = \frac{ g_s|W_0|^2 }{2\mathcal{V}^2}.
\end{equation}
The first line in \eqref{potentialbranemoduli} comes from the $D$-term potential \eqref{dtermpot} and the second line is the contribution from the $F$-term potential \eqref{ftermpot}.
Note that \eqref{potentialbranemoduli} reduces to our earlier potential \eqref{varphipotential} if we set all fields except $\varphi^m$ to zero and consider a collective mode with $|\varphi|^2 = G_{m\bar n} \varphi^m \bar\varphi^{\bar n}$.

Assuming the $D$-term part of the potential \eqref{potentialbranemoduli} is steep enough ($C_1C_2-\frac{C_3}{2}>0$), there are extrema with $\varphi^m\neq 0$ satisfying
\begin{equation}
\zeta^a=\rho^\alpha=\xi^q=0, \qquad G_{m\bar n} \varphi^m \bar\varphi^{\bar n}=C_2-\frac{C_3}{2C_1}.
\end{equation}
These are precisely the T-brane vacua we have been discussing throughout this paper. We thus see that the $\zeta^a$, $\rho^\alpha$ and $\xi^q$ moduli are all stabilized at the origin but only a single real combination of the $\varphi^m$ moduli is stabilized. The remaining $2\, \text{dim} H^2_+(D_\text{T},\mathcal{L}^{-2})-1$ real combinations are unstabilized. Although we focussed on the SU$(2)$ case here, it is clear that similar conclusions also apply for brane stacks with other gauge groups.

One may try to resolve this problem by turning on terms in the superpotential that depend on the $\varphi^m$ moduli. Recall that we assumed that the superpotential does not depend on the brane moduli (following \cite{Cicoli:2015ylx, Cicoli:2017shd, Cicoli:2021dhg}) since this was sufficient for the purpose of generating the uplift term. A superpotential can in principle be generated in several different ways: by turning on $(2,0)$-form worldvolume fluxes \cite{Jockers:2005zy, Gomis:2005wc, Lust:2005bd, Martucci:2006ij}, by D7/O7 backreaction effects \cite{Lust:2005bd, Denef:2008wq, Arends:2014qca} or by instanton corrections whose Pfaffian prefactors depend on the brane moduli \cite{Blumenhagen:2007sm}.

Consistency of the $D$-term requires that the superpotential is gauge-invariant \cite{Villadoro:2005yq} (see also \cite{Cicoli:2015ylx}), i.e.,
\begin{equation}
q_b \partial_{T_b} W + 2 \varphi^m \partial_{\varphi^m} W - 2 \xi^q \partial_{\xi^q} W = 0
\end{equation}
must hold for all moduli values. Here, $q_b$ is the U$(1)$ charge of $T_b$ (with respect to the U$(1)$ along which we turned on the worldvolume flux) and we used that the $\varphi^m$ and $\xi^q$ moduli have U$(1)$ charges $\pm 2$ (since they are the coefficients of bundle-valued $(2,0)$-forms). The above constraint is satisfied, e.g., for polynomial terms
\begin{equation}
W \supset a_{mq}\varphi^m\xi^q + b_{mnqr}\varphi^m\varphi^n\xi^q\xi^r + \ldots
\end{equation}
or non-perturbative terms
\begin{equation}
W \supset c_{m\ldots n q\ldots r} \underbrace{\varphi^m \cdots \varphi^n}_{x \text{ times}} \underbrace{\xi^q \cdots \xi^r}_{y \text{ times}} \e^{-a_b T_b}
\end{equation}
with $2x-2y= a_bq_b$. It would be important to understand to what extent such terms can be generated in the superpotential without significantly affecting the uplift mechanism explained in Section \ref{sec:lvs-up2}.
Alternatively, a potential for the $\varphi^m$'s might also be generated by $\alpha'$ corrections in the K\"ahler potential. As stated before, all of these possibilities may in principle go either way: they could stabilize or destabilize the moduli. We leave a detailed analysis of such questions for future work. In any case, even if the moduli-stabilization issue turns out to be merely a technical challenge rather than a serious obstacle, we believe that the control problem pointed out in the earlier sections of this paper is by itself already severe enough to invalidate the consistency of T-brane-uplifted dS vacua.

\section{Conclusion}
\label{sec:concl}

In this paper, we studied dS vacua obtained through a T-brane uplift in the LARGE-volume scenario (LVS).
Since the LVS with anti-brane uplift is known to have control issues \cite{Junghans:2022exo, Gao:2022fdi, Junghans:2022kxg, Hebecker:2022zme, Schreyer:2022len, Schreyer:2024pml, ValeixoBento:2023nbv}, a natural suspicion is that similar problems occur for the T-brane uplift as well, and we indeed found this to be the case.

We emphasized that a controlled dS construction requires that corrections which are not explicitly computed must be shown to be suppressed by a small control parameter. Naively, the LVS provides an exponentially small control parameter for both $\alpha'$ and warping corrections, namely the inverse of the volume. A common lore is that the volume is exponentially large in the LVS for moderately small $g_s$ and that therefore the scenario is excellently controlled. However, we showed that the string-frame volume in T-brane-uplifted dS vacua is in fact bounded by a number which only depends on topological data of the Calabi-Yau orientifold and does therefore not have much tuning freedom. This bound immediately rules out large volumes in many models where the non-perturbative effect comes from D3-brane instantons. In such models, the string-frame volume is often of the order 1 in string units or even much smaller than that, implying that there is \emph{no control at all}. We furthermore argued that the control parameter that potentially suppresses the warping corrections is not simply $1/\mathcal{V}_s$ but contains other factors, in particular the tadpole $Q_3$, which can be large and thus cancel the naive volume suppression even in models with gaugino condensation where our bound admits relatively large volumes and would thus naively not be that worrisome. Explicitly, we argued that a plausible proxy for controlling warping corrections is the condition $\lambda_W = \frac{3^{2/3} 190}{9} \frac{a_s \kappa_{sss}^{1/3}}{\hat\xi^{1/3}} \frac{|Q_3|}{\mathcal{V}_s^{2/3}} \ll 1$.

In order to assess whether this condition can be satisfied in T-brane-uplifted dS vacua, we performed a detailed study of eleven Swiss-cheese Calabi-Yau manifolds with $h^{1,1}=2$ that can be constructed from the Kreuzer-Skarke list.
We constructed all orientifolds of these manifolds which arise from reflection involutions with respect to the irreducible divisors of the ambient toric variety, resulting in 28 different Calabi-Yau orientifolds suitable for the LVS. We then computed all possible D-brane and flux configurations yielding a T-brane uplift in these orientifolds, leading in total to $\mathcal{O}(600)$ different models with a T-brane uplift and candidate dS vacua.

Our analysis showed that, while large volumes are indeed possible in these models, the volumes are never large enough to allow a suppression of the warping corrections, assuming they are reliably estimated by our proxy $\lambda_W$. Indeed, in all models we studied, $\lambda_W \gtrsim \mathcal{O}(1)$
so that we expect large warping corrections to the vacuum energy and the moduli masses. In fact, we found that the majority of the models have a huge $\lambda_W$ several orders of magnitude above 1 and only a handful of models has a $\lambda_W$ near the boundary of control $\lambda_W=\mathcal{O}(1)$. Another interesting observation was that the latter models have very small tadpoles with $|Q_3| \le 30$ (upstairs). This raises the question whether the axio-dilaton and all complex-structure moduli can consistently be stabilized in these models and at the same time $g_s$, $W_0$ can be fine-tuned such that a dS vacuum is obtained. If this is not the case, the true lower bound on $\lambda_W$ is much larger than $1$, increasing the control problems even more.

The problems we found have been overlooked previously due to a combination of reasons: First, our analysis revealed that previous derivations of the uplift term missed a factor $N$ (which counts the number of D7-branes combining into the T-brane). The stabilized volume in a dS vacuum scales with this factor like $1/N^3$ and is thus highly sensitive to it. Second, many papers studying the LVS only compute the Einstein-frame volumes of the claimed dS solutions, but the relevant parameter for controlling corrections is the string-frame one, which is by a factor $g_s^{3/2}$ smaller. This is in particular true for $\alpha'$ corrections. Also the warping corrections we studied are not suppressed by $1/\mathcal{V}^{2/3}=g_s/\mathcal{V}_s^{2/3}$ as one might have expected but actually only by $1/\mathcal{V}_s^{2/3}$ due to the non-perturbative no-scale structure \cite{Junghans:2022exo} of the LVS potential. Third, we argued that the warping corrections do not only scale with the inverse volume but also with the tadpole $|Q_3|$ which tends to cancel any naive volume suppression that might arise. Hence, compared to what one might have guessed, the warping corrections are larger by a factor $|Q_3| N^2 /g_s$ and thus much more dangerous than previously anticipated.

Another interesting observation, which we only discussed very briefly, is that there is a second type of corrections which mimicks the parametric behavior of the warping corrections but has a completely different origin, namely curvature corrections in the worldvolume theory of the D7-branes. This dual-threat situation is reminiscent of the anti-brane uplift where similarly not only warping corrections in the bulk but also local curvature corrections on the anti-brane can lead to control problems, as pointed out in \cite{Junghans:2022exo} and further studied in \cite{Junghans:2022kxg, Hebecker:2022zme, Schreyer:2022len, Schreyer:2024pml}.

In the last section of this paper, we also briefly discussed an additional challenge for vacua with T-branes, namely that some of the brane moduli are left unstabilized if one follows the assumptions of the original scenario. This challenge is independent of the control problems for dS vacua and needs to be addressed in any putative T-brane vacuum, regardless of the sign of the vacuum energy.

Regarding future work, it would be interesting to elaborate on several aspects of the issues raised in this paper. For one thing, it would be interesting to perform a more detailed study of Calabi-Yau manifolds with $h^{1,1}>2$. We already gave an argument that large $h^{1,1}$ does not lead to controlled vacua and moreover analyzed two explicit models with $h^{1,1}=4$ constructed previously in \cite{Cicoli:2017shd, Cicoli:2021dhg}, which we found to have very small volumes. Nevertheless, it is possible that there are other models with $h^{1,1}>2$ in which the control problems observed in this paper are somewhat relaxed. It would be nice to prove a general lower bound on $\lambda_W$ which either rules out the existence of controlled models entirely for all $h^{1,1}$ or instead gives a concrete hint which properties are needed to achieve an at least smallish $\lambda_W$. It would also be interesting to study setups with more general worldvolume fluxes, in particular by allowing elements of $H^{2}(D_\text{T},\mathbb{Z})$ which are not pulled back from $H^{2}(X,\mathbb{Z})$. Identifying all such fluxes is a non-trivial task but may lead to interesting effects, e.g.,
a somewhat different quantization of the flux numbers $f_b$, $f_s$ (which may affect the size of the uplift term) and the generation of a superpotential for the brane moduli (which may resolve the stabilization challenge discussed in Section \ref{sec:modstab}).

More generally, given that control over warping and curvature corrections seems to be a prevalent issue in many different dS scenarios, it would also be important to make progress on computing such corrections explicitly. Instead of having to suppress them, one could then incorporate them into the potential and check whether the corrected potential still admits dS vacua. For the warping corrections,
this requires a computation of the Green's function of the Laplacian with respect to the Calabi-Yau metric. In toroidal orbifold limits, this can be done explicitly \cite{Andriot:2019hay, Shandera:2003gx, CourantHilbert}. Recent progress for smooth metrics was made in \cite{Lust:2026mys}.

Finally, we find it quite striking that very similar control problems seem to exist in the different variants of the LVS with T-brane and anti-brane uplift and also in other scenarios like KKLT, classical dS models and even supercritical dS models which are technically quite different from the LVS. In our view, this strongly suggests an underlying principle that string theory does not like dS vacua, at least in the computationally accessible weakly curved regime, and it would clearly be important to understand on general grounds if and why such a principle could be true.

\section*{Acknowledgments}

I would like to thank Christoph Mayrhofer for a useful discussion.
This research was funded by the German Research Foundation (Deutsche Forschungsgemeinschaft) under the project number 516370439.

\appendix

\section{Normalization of the T-brane Uplift}
\label{app:tbrane}

In this appendix, we verify the normalization of the uplift term generated by the T-brane.
To this end, the following kinetic and potential terms in the 4d effective field theory are relevant:
\begin{align}
\mathcal{L}_\text{kin}-V_D-V_F &\supset - C_1 \,\text{Tr} \int_{D_\text{T}} (D_\mu\hat\Phi^\dagger)\w ( D^\mu \hat\Phi) - C_2 \,\text{Tr} \left( C_3 \int_{D_\text{T}} \iota^*_\text{T} J \w \mathcal{F} + \int_{D_\text{T}} [\hat\Phi,\hat\Phi^\dagger]
\right)^2 \nl - C_4 \frac{g_s|W_0|^2}{\mathcal{V}^2} \,\text{Tr} \int_{D_\text{T}} \hat\Phi^\dagger\w \hat\Phi. \label{gsjigsjigsjg}
\end{align}
Recall that $\hat\Phi$ and $\mathcal{F}$ are $N\times N$ matrices whose entries are 2-forms on the divisor $D_\text{T}$. We furthermore denote by $\iota^*_\text{T}$ the pullback onto that divisor.

Without loss of generality, we normalize $\hat\Phi$ such that, in 4d Planck units,
\begin{equation}
C_1=1. \label{kefafkfaflaf}
\end{equation}
In \cite{Junghans:2026jza}, the $D$-term potential of a non-Abelian D7 stack was derived by dimensionally reducing the DBI action. The result is
\begin{equation}
C_2 = \frac{2\pi}{\int_{D_\text{T}} \iota_\text{T}^* J^2}, \qquad C_3 = \frac{1}{4\pi\mathcal{V}}.
\end{equation}
The $F$-term potential generated by the D7-branes was also derived in \cite{Junghans:2026jza} by dimensional reduction, yielding
\begin{equation}
C_4 = \frac{1}{2}.
\end{equation}
An alternative is to determine $C_4$ using the K\"ahler potential. For a single brane and assuming $h^{1,1}_-(X)=0$, one has \cite{Jockers:2004yj}
\begin{equation}
K \supset - \ln \left( S+\bar S - f(Z_i,\bar Z_i) \int_{D_\text{T}} \hat\Phi^\dagger \w \hat\Phi \right) \approx - \ln \left( S+\bar S\right) + \frac{f(Z_i,\bar Z_i)}{S+\bar S} \int_{D_\text{T}} \hat\Phi^\dagger \w \hat\Phi, \label{dsgksfgjsgslijfgs}
\end{equation}
where the axio-dilaton is defined as $S=S_0+\frac{f}{2}\int_{D_\text{T}} \hat\Phi^\dagger\w \hat\Phi$, $S_0= \frac{1}{g_s}+i C_0$ and the $Z_i$'s are the complex-structure moduli.
This expression is valid up to various corrections (e.g., terms of higher orders in $\hat\Phi$ and $1/\mathcal{V}$) which we neglect here. Also note that $\int_{D_\text{T}} \hat\Phi^\dagger\w \hat\Phi$ should be viewed as a function of the 4d moduli. For a single brane, we can write $\hat\Phi = \zeta^a \omega_a$ in terms of complex moduli $\zeta^a$ and some basis of harmonic $(2,0)$-forms. For brane stacks with $N\ge 2$, one has to expand $\hat\Phi$ in certain bundle cohomologies, see \cite{Junghans:2026jza} for an extensive discussion. Spelling out the moduli dependence explicitly is then rather cumbersome (cf.~Section \ref{sec:modstab}) so that it is more convenient to keep it implicit in the following.

A key assumption of the LVS is that integrating out $S$ and the $Z_i$'s is equivalent to setting them to constants in the superpotential and K\"ahler potential. As explained in \cite{Achucarro:2008sy, Gallego:2008qi}, this is justified if these moduli approximately decouple from the K\"ahler and brane moduli. This is indeed the case in the regime of small $\hat\Phi$ we are interested in: Although solving $D_S W=D_{Z_i} W=0$ yields $\hat\Phi$-dependent vevs for $S$, $Z_i$ due to the mixing in \eqref{dsgksfgjsgslijfgs}, taking this into account only gives subleading corrections of the order $\mathcal{O}(\hat\Phi^2 (D_\mu\hat\Phi)^2)$ to the kinetic terms. The same is true in the $F$-term potential. Indeed, one can check that taking into account the $\hat\Phi$ dependence of the $S$, $Z_i$ vevs yields terms of the order $V_F \supset \frac{\hat\Phi^2}{\mathcal{V}^3}, \frac{\hat\Phi^4}{\mathcal{V}^2}$, whereas we will see momentarily that the leading terms depending on $\hat\Phi$ are of the order $V_F \supset \frac{\hat\Phi^2}{\mathcal{V}^2}$. Hence, also here the mixing is negligible. Note that we assume that the mixing comes purely from \eqref{dsgksfgjsgslijfgs}. In particular, we adopt the assumption of \cite{Cicoli:2015ylx} and subsequent works on T-brane uplifting that the superpotential is independent of $\hat\Phi$.\footnote{As discussed in Section \ref{sec:modstab}, turning on a superpotential for the brane moduli may help to stabilize them but can at the same time affect the uplift term. One would furthermore have to redo the check of the decoupling approximation in that case.}

We can therefore neglect the mixing between $\hat\Phi$ and the bulk moduli and set them to constants. Hence, after integrating out $S$ and the $Z_i$'s, the K\"ahler potential for $\hat\Phi$ is
\begin{equation}
K_\text{D7} = \frac{g_s f}{2} \int_{D_\text{T}} \hat\Phi^\dagger \w \hat\Phi.
\end{equation}
Here, $f$ is an arbitrary constant which can be set to any value by a field rescaling. In the following, we set $f=\frac{2}{g_s}$ since this normalizes $\hat\Phi$ such that its kinetic term agrees with \eqref{gsjigsjigsjg}, \eqref{kefafkfaflaf}.
Furthermore, the natural generalization from the single-brane case to a brane stack is to put a trace in front of the K\"ahler potential. We thus arrive at
\begin{equation}
K_\text{D7} = \text{Tr} \int_{D_\text{T}} \hat\Phi^\dagger \w \hat\Phi.
\end{equation}
This is indeed confirmed in \cite{Junghans:2026jza}.

We now consider the $F$-term potential, which in 4d Planck units is $V_F=\e^K(|D_{I}W|^2-3|W|^2)$. Since, by assumption of the LVS, $S$ and the $Z_i$'s are stabilized supersymmetrically, the index $I$ only runs over the K\"ahler and brane moduli. One can check that all terms involving derivatives with respect to the K\"ahler moduli cancel with the $-3|W|^2$ term due to the well-known no-scale structure, up to terms whose $\hat\Phi$ dependence is negligibly small.
The leading $F$-term potential for $\hat\Phi$ is given by $\e^K |D_I W|^2$ where $I$ now only runs over the brane moduli.
Since the superpotential is independent of $\hat\Phi$ by assumption of the scenario, we have
\begin{equation}
V_F = \e^K |D_I W|^2 = \e^K K^{I \bar J}K_I K_{\bar J} |W|^2  = \frac{g_s |W_0|^2}{2\mathcal{V}^2} \text{Tr}\int_{D_\text{T}} \hat\Phi^\dagger\w \hat\Phi,
\end{equation}
where we used $\e^K \approx \frac{g_s}{2\mathcal{V}^2}$, $|W_0|\gg |W_\text{np}|$ and $K^{I \bar J}K_I K_{\bar J} = K_\text{D7}$ at the quadratic order in $\hat\Phi$.
Comparing to \eqref{gsjigsjigsjg}, we conclude
\begin{equation}
C_4= \frac{1}{2},
\end{equation}
in agreement with the (more general) result of \cite{Junghans:2026jza}. We have thus fixed the coefficient of the $F$-term potential in \eqref{gsjigsjigsjg} without having to derive it directly from the D7-brane action. This is possible due to the fact that the coefficients of the kinetic term and the $F$-term potential for $\hat\Phi$ are both determined by the K\"ahler potential, i.e., $C_4$ is fixed by the structure of 4d $\mathcal{N}=1$ supergravity once we choose $C_1$.

\section{Swiss-Cheese Calabi-Yau Orientifolds with $h^{1,1}=2$}
\label{app:toric}

In this appendix, we provide some details on the Swiss-Cheese Calabi-Yau manifolds used in our scan in Section \ref{sec:h112}. Aside from stating their relevant topological data, we determine all orientifolds of these manifolds which are obtained by reflection involutions of the form
\begin{equation}
\text{I}i: x_i\to -x_i
\end{equation}
on one of the projective coordinates of the ambient variety $A$. Such involutions have codimen-sion-1 and codimension-3 fixed loci on $X$, i.e., O7-planes and O3-planes. We compute, for each involution, the invariant locus $D_\text{O7}$ of the O7-plane and the O3-plane number $N_\text{O3}$. We discard an orientifold if $D_\text{O7}=D_s$ since then tadpole cancelation implies that we cannot put a T-brane on $D_b$.
We also perform a few consistency checks to identify possible singularities of the orientifold-invariant Calabi-Yau hypersurfaces. In particular, we verify that the number of complex-structure moduli $h^{2,1}_-$ is integral in all orientifolds we keep. We further observe that a singularity is signaled when codimension-3 fixed loci of the orientifold involution appear to have fractional O3-plane numbers or when codimension-1 fixed loci appear to have fractional O7-plane numbers.
We show that both phenomena are indeed related to singularities in several examples and discard orientifolds exhibiting such a behavior. An exhaustive analysis of all possible singularities is beyond the scope of this work so that some of the orientifolds we keep may still have more subtle singularities missed by our checks.
See also \cite{Crino:2022zjk} for a related discussion.

\subsection{ $M_{2,3}$}

The relevant topological data of the manifold are
\begin{align}
& \chi = -144, && \kappa_{sss}=1, && \kappa_{bbb}=3, && c_{2b} = 14, && c_{2s} = 10.
\end{align}
The weight matrix is
\\

\noindent\begin{tabular}{l l l l l l | l }
$x_1$ & $x_2$ & $x_3$ & $x_4$ & $x_5$ & $x_6$ & CY \\
\hline
$1$ & $1$ & $0$ & $1$ & $2$ & $1$ & $6$ \\
$1$ & $1$ & $1$ & $0$ & $1$ & $2$ & $6$
\end{tabular}
\\[1em]

\noindent with Stanley-Reisner (SR) ideal $\{x_3x_6,x_1x_2x_4x_5\}$. The irreducible toric divisors $D_i$ satisfy
\begin{equation}
D_1 = D_2 = D_b-D_s, \quad D_3=D_s, \quad D_4 = D_b-2D_s, \quad D_5 = 2D_b-3D_s, \quad D_6 = D_b
\end{equation}
up to linear equivalence.
The divisor specifying the Calabi-Yau hypersurface is $D_\text{CY}=6D_b-6D_s$.
Here and in the following, we denote by $D_i$ both the divisors on the ambient variety $A$ and their pullbacks onto $X$.\footnote{Calabi-Yau manifolds can in principle have additional divisors that do not descend from the ambient space but this does not happen for the manifolds in this appendix. The condition ensuring this is called favorability, see, e.g., \cite{Altman:2014bfa, McAllister:2024lnt} for discussions. According to the database of \cite{Altman:2014bfa, Altman:2017vzk}, all Calabi-Yau manifolds with $h^{1,1}=2$ obtained from the Kreuzer-Skarke list are indeed favorable.} Note that all divisors are linear combinations of the two linear equivalence classes $D_b$, $D_s$ (as expected for $h^{1,1}=2$). Since these combinations are integral, $D_b$, $D_s$ span an integral basis of $H_{2,2}(X,\mathbb{Z})$, which is the case for all manifolds discussed in this appendix.
Effective divisors are obtained by considering non-negative integer linear combinations of the $D_i$'s.

The possible orientifold choices are
\\

\noindent\begin{tabular}{l | l l l | l}
involution & $D_\text{O7}$ & $N_\text{O3}$ & $h^{2,1}_-$ & viable? \\
\hline
I1/I2 & $D_b-D_s$ & $\int_X (D_2D_4D_6+D_2D_3D_5)=6$ & 45 & \checkmark \\
I3 & $D_s$ &  &  & \texttimes \\
I4 & $D_b-2D_s$ &  &  & \texttimes \\
I5 & $2D_b-3D_s$ &  &  & \texttimes \\
I6 & $D_b$ &  &  & \texttimes
\end{tabular}
\\[1em]

\noindent Let us explain how the various entries in the table are obtained. We first consider the case I1. Due to the projective identifications following from the weight matrix, the involution $x_1 \to -x_1$ can equivalently be viewed as $(x_2,x_4,x_6)\to -(x_2,x_4,x_6)$ or $(x_2,x_3,x_5)\to -(x_2,x_3,x_5)$. The fixed points are thus $x_1=0$, $x_2=x_4=x_6=0$ and $x_2=x_3=x_5=0$. One verifies that none of them lie in the SR ideal and all of them intersect with the Calabi-Yau hypersurface. They thus correspond to an O7-plane wrapped on $D_1=D_b-D_s$ and six O3-planes since $N_\text{O3}=\int_X (D_2D_4D_6+D_2D_3D_5) =
\kappa_{bbb}+3\kappa_{sss}=6$. The I2 involution works analogously.

For the other involutions,
it turns out that the requirement of orientifold invariance enforces singularities on the Calabi-Yau hypersurface.
To see this, we first consider the I3 and I5 involutions. Using the projective identifications, one can show that both involutions have fixed points at $x_1=x_2=x_4=x_6=0$. The associated number of O3-planes can naively be computed by intersecting the corresponding divisors, but this does not yield an integer, $\int_A D_1D_2D_4D_6=\frac{1}{2}$, suggesting a singularity. To see that this is indeed the correct interpretation, consider first a generic Calabi-Yau hypersurface in $A$. The defining polynomial reads $P_\text{CY}= x_3^3 x_5^3 + \mathcal{O}(\epsilon^2)$ at $x_{1,2,4,6}\sim \epsilon$. Since $x_3,x_5\neq 0$ at $\epsilon=0$ due to the SR ideal, this implies $P_\text{CY}|_{\epsilon=0} \neq 0$. Hence, the fixed points do not intersect with a generic Calabi-Yau. However, for the I3 and I5 orientifolds, the $x_3^3 x_5^3$ term in $P_\text{CY}$ is projected out so that now $P_\text{CY}|_{\epsilon=0}=0$, i.e., the fixed points are located on the Calabi-Yau as soon as we impose orientifold invariance. At the same time, one also finds $\partial_{x_i}P_\text{CY}|_{\epsilon=0} = 0$ for all $x_i$.
Hence, both $P_\text{CY}$ and its differential $\d P_\text{CY}$ vanish at $\epsilon=0$, which indicates that the hypersurface is singular there \cite{Crino:2022zjk}.
We thus see that the fractional O3-plane number is related to a singularity as anticipated.
We therefore discard the I3 and I5 orientifolds.

We can repeat the analysis for the I4 and I6 involutions, which have fixed points at $x_1=x_2=x_3=x_5=0$. The intersection number of the corresponding divisors is again fractional, $\int_A D_1D_2D_3D_5=\frac{1}{2}$, so that we again expect a singularity. For a generic Calabi-Yau, we have $P_\text{CY}= x_4^3 x_6^3 + \mathcal{O}(\epsilon^2)$ at $x_{1,2,3,5}\sim \epsilon$. Since $x_4,x_6\neq 0$ at $\epsilon=0$ due to the SR ideal, the fixed points lie outside of a generic Calabi-Yau. However, for the I4 and I6 orientifolds, the $x_4^3x_6^3$ term in $P_\text{CY}$ is projected out so that $P_\text{CY}|_{\epsilon=0} = 0$, $\partial_{x_i}P_\text{CY}|_{\epsilon=0} = 0$. We thus conclude that the I4 and I6 orientifolds are singular as well and discard them.

\subsection{ $M_{2,4}$}

The relevant topological data of the manifold are
\begin{align}
& \chi = -144, && \kappa_{sss}=1, && \kappa_{bbb}=3, && c_{2b} = 14, && c_{2s} = 10.
\end{align}
The weight matrix is
\\

\noindent \begin{tabular}{l l l l l l | l }
$x_1$ & $x_2$ & $x_3$ & $x_4$ & $x_5$ & $x_6$ & CY \\
\hline
$1$ & $1$ & $0$ & $1$ & $2$ & $1$ & $6$ \\
$1$ & $1$ & $1$ & $0$ & $1$ & $2$ & $6$
\end{tabular}
\\[1em]

\noindent with SR ideal $\{x_4x_5,x_1x_2x_3x_6\}$. The irreducible toric divisors $D_i$ satisfy
\begin{equation}
D_1 = D_2 = D_b-D_s, \quad D_3=D_b-2D_s, \quad D_4 = D_s, \quad D_5 = D_b, \quad D_6 = 2D_b-3D_s.
\end{equation}
The divisor specifying the Calabi-Yau hypersurface is $D_\text{CY}=6D_b-6D_s$. Note that the above data are equivalent to $M_{2,3}$ upon exchanging $x_3\leftrightarrow x_4$, $x_5\leftrightarrow x_6$.

The possible orientifold choices are
\\

\noindent\begin{tabular}{l | l l l | l}
 involution & $D_\text{O7}$ & $N_\text{O3}$ & $h^{2,1}_-$ & viable? \\
\hline
I1/I2 & $D_b-D_s$ & $\int_X (D_2D_4D_6+D_2D_3D_5)=6$ & 45 & \checkmark \\
I3 & $D_b-2D_s$ &  &  & \texttimes \\
I4 & $D_s$ &  &  & \texttimes \\
I5 & $D_b$ &  &  & \texttimes \\
I6 & $2D_b-3D_s$ & & & \texttimes 
\end{tabular}
\\[1em]

\noindent The singularity checks are analogous to $M_{2,3}$, leading to the conclusion that the I3, I4, I5 and I6 orientifolds are singular.

\subsection{ $M_{2,5}$}

The relevant topological data of the manifold are
\begin{align}
& \chi = -144, && \kappa_{sss}=1, && \kappa_{bbb}=3, && c_{2b} = 14, && c_{2s} = 10.
\end{align}
The weight matrix is
\\

\noindent\begin{tabular}{l l l l l l | l }
$x_1$ & $x_2$ & $x_3$ & $x_4$ & $x_5$ & $x_6$ & CY \\
\hline
$0$ & $1$ & $1$ & $1$ & $2$ & $1$ & $6$ \\
$1$ & $2$ & $2$ & $3$ & $4$ & $0$ & $12$
\end{tabular}
\\[1em]

\noindent with SR ideal $\{x_1x_4,x_2x_3x_5x_6\}$. The irreducible toric divisors $D_i$ satisfy
\begin{equation}
D_1 = D_s, \quad D_2=D_3=D_b-D_s, \quad D_4 = D_b, \quad D_5 = 2D_b-2D_s, \quad D_6 = D_b-3D_s.
\end{equation}
The divisor specifying the Calabi-Yau hypersurface is $D_\text{CY}=6D_b-6D_s$.

The possible orientifold choices are
\\

\noindent\begin{tabular}{l | l l l | l}
 involution & $D_\text{O7}$ & $N_\text{O3}$ & $h^{2,1}_-$ & viable? \\
\hline
I1 & $D_b+D_s$ & $0$ & 49 & \checkmark \\
I2/I3 & $D_b-D_s$ & $\int_X (D_3D_4D_6+D_1D_3D_6)=6$ & 45 & \checkmark \\
I4 & $D_b+D_s$ & $0$ & 49 & \checkmark \\
I5 & $2D_b-2D_s$ & & & \texttimes \\
I6 & $D_b-3D_s$ & $\int_X (D_2D_3D_4+D_1D_2D_3)=4$ & 33 & \checkmark
\end{tabular}
\\[1em]

\noindent
Note that the I5 involution has fractional fixed points at $x_2=x_3=x_4=x_6=0$ and $x_1=x_2=x_3=x_6=0$ which are not in the SR ideal. For a generic Calabi-Yau, we would have $P_\text{CY}= x_5^3 + \mathcal{O}(\epsilon^2)$ at $x_{2,3,4,6}\sim \epsilon$ or $x_{1,2,3,6}\sim \epsilon$. Since $x_5\neq 0$ at $\epsilon=0$ due to the SR ideal, the fixed points would lie outside of a generic Calabi-Yau. However, for the I5 orientifold, the $x_5^3$ term in $P_\text{CY}$ is projected out so that $P_\text{CY}|_{\epsilon=0} = 0$, $\partial_{x_i}P_\text{CY}|_{\epsilon=0} = 0$. We thus conclude that the I5 orientifold is singular.

\subsection{ $M_{2,7}$}

The relevant topological data of the manifold are
\begin{align}
& \chi = -164, && \kappa_{sss}=2, && \kappa_{bbb}=5, && c_{2b} = 10, && c_{2s} = 4.
\end{align}
The weight matrix is
\\

\noindent\begin{tabular}{l l l l l l | l }
$x_1$ & $x_2$ & $x_3$ & $x_4$ & $x_5$ & $x_6$ & CY \\
\hline
$0$ & $1$ & $1$ & $1$ & $1$ & $1$ & $5$ \\
$1$ & $1$ & $1$ & $1$ & $0$ & $2$ & $6$
\end{tabular}
\\[1em]

\noindent with SR ideal $\{x_1x_6,x_2x_3x_4x_5\}$. The irreducible toric divisors $D_i$ satisfy
\begin{equation}
D_1 = D_s, \quad D_2=D_3=D_4=D_b-D_s, \quad D_5 = D_b-2D_s, \quad D_6 = D_b.
\end{equation}
The divisor specifying the Calabi-Yau hypersurface is $D_\text{CY}=5D_b-4D_s$.

The possible orientifold choices are
\\

\noindent\begin{tabular}{l | l l l | l}
 involution & $D_\text{O7}$ & $N_\text{O3}$ & $h^{2,1}_-$ & viable? \\
\hline
I1 & $D_s+?$ & &  & \texttimes \\
I2/I3/I4 & $D_b-D_s$ & $\int_X D_1D_3D_4+\int_A D_3D_4D_5D_6=3$ & 52 & \checkmark \\
I5 & $D_b-2D_s$ & $\int_A D_2D_3D_4D_6=1$ & 46 & \checkmark \\
I6 & $D_b+?$ & & & \texttimes
\end{tabular}
\\[1em]

\noindent Note that the I6 involution has fractional fixed points at $x_1=x_2=x_3=x_4=0$ which are not in the SR ideal. For a generic Calabi-Yau, we would have $P_\text{CY}= x_5^2 x_6^3 + \mathcal{O}(\epsilon^2)$ at $x_{1,2,3,4}\sim \epsilon$. Since $x_5,x_6\neq 0$ at $\epsilon=0$ due to the SR ideal, the fixed points would lie outside of a generic Calabi-Yau. However, for the I6 orientifold, the $x_5^2x_6^3$ term in $P_\text{CY}$ is projected out so that $P_\text{CY}|_{\epsilon=0} = 0$, $\partial_{x_i}P_\text{CY}|_{\epsilon=0} = 0$. We thus conclude that the I6 orientifold is singular.

In addition, also the I1 orientifold is singular. Unlike in the I6 case, this is not related to a fractional O3-plane number but to a fractional O7-plane number.
The fixed points of I1 are located at $x_1=0$, $x_2=x_3=x_4=0$ and $x_5=x_6=0$. The first two sets correspond to an O7-plane wrapped on $D_s$ and three O3-planes, but the interpretation of the last one is less clear. Naively, one might think that it is an O5-plane since it arises at the intersection of $D_5$ and $D_6$, which would have codimension 2 on a generic Calabi-Yau. However, on the orientifold-invariant Calabi-Yau, this degenerates to a codimension-1 surface and singularities appear.

To see this, consider again a generic Calabi-Yau first. We then have $P_\text{CY} = P_5 x_1 + P_4 x_6 + \hat P_4x_1^2x_5 + \mathcal{O}(\epsilon^2)$ at $x_{5,6}\sim \epsilon$, where $P_5$, $P_4$, $\hat P_4$ are degree-5 and degree-4 polynomials of $x_2$, $x_3$, $x_4$. It follows that
\begin{gather}
P_\text{CY}|_{\epsilon=0} = P_5x_1, \qquad  \partial_{x_5} P_\text{CY}|_{\epsilon=0} = \hat P_4 x_1^2, \qquad \partial_{x_6} P_\text{CY}|_{\epsilon=0} = P_4, \notag \\ \partial_{x_{1}} P_\text{CY}|_{\epsilon=0} = P_5, \qquad \partial_{x_{2,3,4}} P_\text{CY}|_{\epsilon=0} = \partial_{x_{2,3,4}} P_5x_1. \label{dgslijdgsdijgs}
\end{gather}
Since $x_1\neq 0$ and $x_{2,3,4}$ cannot all vanish at $\epsilon=0$ due to the SR ideal, at least one of the above expressions is non-zero on a generic Calabi-Yau. On the other hand, for the I1 orientifold, the $P_5x_1$ term is projected out in $P_\text{CY}$ so that all expressions in \eqref{dgslijdgsdijgs} vanish at $x_5=x_6=P_4=\hat P_4=0$. Note that this in general has solutions with $x_{2,3,4}\neq 0$ and is therefore not excluded by the SR ideal. We can therefore conclude that the Calabi-Yau has singularities in the I1 case. One can further check that $D_5$ and $D_6$ do not intersect transversally but in a codimension-1 surface on the invariant Calabi-Yau. This surface can however not be identified with a $D_i$ divisor on $X$, as already suggested by applying the naive formula $D_5\cap D_6 = \frac{1}{5}D_b \cap D_\text{CY}$, which looks like a fractional O7-plane. An analogous problem can be shown to arise for the I6 orientifold (which we already excluded above) at $x_1=x_5=0$. Below, we will furthermore observe the same behavior for some orientifolds of $M_{2,19}$.

\subsection{ $M_{2,13}$}

The relevant topological data of the manifold are
\begin{align}
& \chi = -168, && \kappa_{sss}=2, && \kappa_{bbb}=3, && c_{2b} = 14, && c_{2s} = 4.
\end{align}
The weight matrix is
\\

\noindent\begin{tabular}{l l l l l l | l }
$x_1$ & $x_2$ & $x_3$ & $x_4$ & $x_5$ & $x_6$ & CY \\
\hline
$0$ & $0$ & $0$ & $1$ & $1$ & $0$ & $2$ \\
$1$ & $1$ & $1$ & $0$ & $1$ & $2$ & $6$
\end{tabular}
\\[1em]

\noindent with SR ideal $\{x_4x_5,x_1x_2x_3x_6\}$. The irreducible toric divisors $D_i$ satisfy
\begin{equation}
D_1 = D_2 = D_3 = D_b - D_s, \quad D_4 = D_s, \quad D_5 = D_b, \quad D_6 = 2D_b-2D_s.
\end{equation}
The divisor specifying the Calabi-Yau hypersurface is $D_\text{CY}=6D_b-4D_s$.

The possible orientifold choices are
\\

\noindent\begin{tabular}{l | l l l | l}
 involution & $D_\text{O7}$ & $N_\text{O3}$ & $h^{2,1}_-$ & viable? \\
\hline
I1/I2/I3 & $D_b-D_s$ & $\int_X (D_2D_3D_5+D_2D_3D_4)=5$ & 51 & \checkmark \\
I4 & $D_b+D_s$ & $\int_X D_1D_2D_3=1$ & 55 & \checkmark \\
I5 & $D_b+D_s$ & $\int_X D_1D_2D_3=1$ &  55 & \checkmark \\
I6 & $2D_b-2D_s$ & & & \texttimes
\end{tabular}
\\[1em]

\noindent Note that the I6 involution has fractional fixed points at $x_1=x_2=x_3=x_4=0$ and $x_1=x_2=x_3=x_5=0$ which are not in the SR ideal. For a generic Calabi-Yau, we would have $P_\text{CY}= x_5^2x_6^2 + P_1 x_4x_5 x_6^2 + P_0 x_4^2x_6^3 + \mathcal{O}(\epsilon^2)$ at $x_{1,2,3}\sim \epsilon$. Since $x_6\neq 0$ at $\epsilon=0$ and $x_4$, $x_5$ cannot both vanish due to the SR ideal, the fixed points would lie outside of a generic Calabi-Yau. However, for the I6 orientifold, the $x_4^2x_6^3$ term in $P_\text{CY}$ is projected out so that $P_\text{CY}|_{\epsilon=x_5=0} = 0$, $\partial_{x_i}P_\text{CY}|_{\epsilon=x_5=0} = 0$. We thus conclude that the I6 orientifold is singular.

\subsection{ $M_{2,18}$}

The relevant topological data of the manifold are
\begin{align}
& \chi = -176, && \kappa_{sss}=3, && \kappa_{bbb}=5, && c_{2b} = 10, && c_{2s} = 2.
\end{align}
The weight matrix is
\\

\noindent\begin{tabular}{l l l l l l | l }
$x_1$ & $x_2$ & $x_3$ & $x_4$ & $x_5$ & $x_6$ & CY \\
\hline
$0$ & $1$ & $1$ & $1$ & $1$ & $1$ & $5$ \\
$1$ & $0$ & $0$ & $0$ & $1$ & $0$ & $2$
\end{tabular}
\\[1em]

\noindent with SR ideal $\{x_1x_5,x_2x_3x_4x_6\}$. The irreducible toric divisors $D_i$ satisfy
\begin{equation}
D_1 = D_s, \quad D_2=D_3=D_4=D_6=D_b-D_s, \quad D_5 = D_b.
\end{equation}
The divisor specifying the Calabi-Yau hypersurface is $D_\text{CY}=5D_b-3D_s$.

The possible orientifold choices are
\\

\noindent\begin{tabular}{l | l l l | l}
 involution & $D_\text{O7}$ & $N_\text{O3}$ & $h^{2,1}_-$ & viable? \\
\hline
I1 & $D_b+D_s$ & $0$ & 59 & \checkmark \\
I2/I3/I4/I6 & $D_b-D_s$ & $\int_A (D_3D_4D_5D_6+D_1D_3D_4D_6)=2$ & 55 & \checkmark \\
I5 & $D_b+D_s$ & $0$ &  59 & \checkmark \\
\end{tabular}
\\[1em]

\noindent We did not observe singularities for any of the involutions on this manifold.

\subsection{ $M_{2,19}$}

The relevant topological data of the manifold are
\begin{align}
& \chi = -180, && \kappa_{sss}=3, && \kappa_{bbb}=3, && c_{2b} = 14, && c_{2s} = 2.
\end{align}
The weight matrix is
\\

\noindent\begin{tabular}{l l l l l l | l }
$x_1$ & $x_2$ & $x_3$ & $x_4$ & $x_5$ & $x_6$ & CY \\
\hline
$0$ & $1$ & $1$ & $1$ & $1$ & $2$ & $6$ \\
$1$ & $0$ & $0$ & $0$ & $1$ & $1$ & $3$
\end{tabular}
\\[1em]

\noindent with SR ideal $\{x_1x_5,x_2x_3x_4x_6\}$. The irreducible toric divisors $D_i$ satisfy
\begin{equation}
D_1 = D_s, \quad D_2=D_3=D_4=D_b-D_s, \quad D_5 = D_b, \quad D_6 = 2D_b-D_s.
\end{equation}
The divisor specifying the Calabi-Yau hypersurface is $D_\text{CY}=6D_b-3D_s$.

The possible orientifold choices are
\\

\noindent\begin{tabular}{l | l l l | l}
 involution & $D_\text{O7}$ & $N_\text{O3}$ & $h^{2,1}_-$ & viable? \\
\hline
I1 & $D_s+?$ & & & \texttimes \\
I2/I3/I4 & $D_b-D_s$ & $\int_X D_3D_4D_5+\int_A D_1D_3D_4D_6=4$ & 54 & \checkmark \\
I5 & $D_b+?$ & &  & \texttimes \\
I6 & $2D_b-D_s$ & & & \texttimes
\end{tabular}
\\[1em]

\noindent Note that the I6 involution has fractional fixed points at $x_2=x_3=x_4=x_5=0$ which are not in the SR ideal. For a generic Calabi-Yau, we would have $P_\text{CY}= x_6^3 + \mathcal{O}(\epsilon^2)$ at $x_{2,3,4,5}\sim \epsilon$. Since $x_6\neq 0$ at $\epsilon=0$ due to the SR ideal, the fixed points would lie outside of a generic Calabi-Yau. However, for the I6 orientifold, the $x_6^3$ term in $P_\text{CY}$ is projected out so that $P_\text{CY}|_{\epsilon=0} = 0$, $\partial_{x_i}P_\text{CY}|_{\epsilon=0} = 0$. We thus conclude that the I6 orientifold is singular.

We also find that the I1 and I5 orientifolds are singular. This is related to fractional O7-planes, analogously to what we discussed for $M_{2,7}$. For I1, the orientifold involution has fixed points at $x_1=0$ and $x_5=x_6=0$. The latter would be a codimension-2 surface on a generic Calabi-Yau and thus naively corresponds to an O5-plane. However, on the orientifold-invariant Calabi-Yau, the surface degenerates and becomes singular.
For a generic Calabi-Yau, we have $P_\text{CY} = P_4 x_1^2 x_6 + P_5 x_1^2 x_5 + P_6 x_1^3 + \mathcal{O}(\epsilon^2)$ at $x_{5,6}\sim \epsilon$, where the $P_n$'s are degree-$n$ polynomials in $x_2$, $x_3$, $x_4$. Hence,
\begin{gather}
P_\text{CY}|_{\epsilon=0} = P_6x_1^3, \qquad  \partial_{x_5} P_\text{CY}|_{\epsilon=0} = P_5 x_1^2, \qquad \partial_{x_6} P_\text{CY}|_{\epsilon=0} = P_4x_1^2, \notag \\ \partial_{x_{1}} P_\text{CY}|_{\epsilon=0} = 3P_6x_1^2, \qquad \partial_{x_{2,3,4}} P_\text{CY}|_{\epsilon=0} = \partial_{x_{2,3,4}} P_6 x_1^3.
\end{gather}
These expressions cannot all vanish generically due to the SR ideal. However, for the I1 orientifold, the $P_6x_1^3$ term in $P_\text{CY}$ is projected out so that all expressions vanish at $x_5=x_6=P_4=P_5=0$. This can be solved for $x_{2,3,4}\neq 0$ and thus does not lie in the SR ideal. We thus conclude that the I1 orientifold is singular. As in the $M_{2,7}$ examples, the singularity is signaled by the fact that the fixed locus is not given by an O7-plane wrapping an integral divisor on $X$ since $D_5\cap D_6 = \frac{1}{3}D_b\cap D_\text{CY}$. The discussion of the I5 orientifold is analogous, where the problematic fixed locus is now at $x_1=x_6=0$.

\subsection{ $M_{2,25}$}

The relevant topological data of the manifold are
\begin{align}
& \chi = -228, && \kappa_{sss}=1, && \kappa_{bbb}=1, && c_{2b} = 34, && c_{2s} = 10.
\end{align}
The weight matrix is
\\

\noindent\begin{tabular}{l l l l l l | l }
$x_1$ & $x_2$ & $x_3$ & $x_4$ & $x_5$ & $x_6$ & CY \\
\hline
$0$ & $0$ & $1$ & $0$ & $1$ & $2$ & $4$ \\
$1$ & $1$ & $0$ & $2$ & $1$ & $5$ & $10$
\end{tabular}
\\[1em]

\noindent with SR ideal $\{x_3x_5,x_1x_2x_4x_6\}$. The irreducible toric divisors $D_i$ satisfy
\begin{equation}
D_1 = D_2 = D_b - D_s, \quad D_3=D_s, \quad D_4 = 2D_b - 2D_s, \quad D_5 = D_b, \quad D_6 = 5D_b-3D_s.
\end{equation}
The divisor specifying the Calabi-Yau hypersurface is $D_\text{CY}=10D_b-6D_s$.

The possible orientifold choices are
\\

\noindent\begin{tabular}{l | l l l | l}
 involution & $D_\text{O7}$ & $N_\text{O3}$ & $h^{2,1}_-$ & viable? \\
\hline
I1/I2 & $D_b-D_s$ & $\int_X (D_2D_5D_6+D_2D_3D_6)=8$ & 64 & \checkmark \\
I3 & $D_b+D_s$ & $\int_X D_1D_2D_6=2$ & 68 & \checkmark \\
I4 & $2D_b-2D_s$ & &  & \texttimes \\
I5 & $D_b+D_s$ & $\int_X D_1D_2D_6=2$ & 68 & \checkmark \\
I6 & $5D_b-3D_s$ & $\int_X (D_1D_2D_5+D_1D_2D_3)=2$ & 116 & \checkmark
\end{tabular}
\\[1em]

\noindent Note that the I4 involution has fractional fixed points at $x_1=x_2=x_3=x_6=0$ and $x_1=x_2=x_5=x_6=0$ which are not in the SR ideal. For a generic Calabi-Yau, we would have
$P_\text{CY}= x_4^3 x_5^4 + \mathcal{O}(\epsilon^2)$ at $x_{1,2,3,6}\sim \epsilon$ and $P_\text{CY}= x_3^4 x_4^5 +\mathcal{O}(\epsilon^2)$ at $x_{1,2,5,6}\sim \epsilon$.
Since $x_4\neq 0$ at $\epsilon=0$ and $x_3$, $x_5$ cannot both vanish due to the SR ideal, it follows $P_\text{CY}|_{\epsilon=0} \neq 0$ so that the fixed points would lie outside of a generic Calabi-Yau. However, for the I4 orientifold, the $x_3^4 x_4^5$ and $x_4^3 x_5^4$ terms in $P_\text{CY}$ are projected out so that $P_\text{CY}|_{\epsilon=0} = 0$, $\partial_{x_i}P_\text{CY}|_{\epsilon=0} = 0$. We thus conclude that the I4 orientifold is singular.

\subsection{ $M_{2,26}$}

The relevant topological data of the manifold are
\begin{align}
& \chi = -236, && \kappa_{sss}=1, && \kappa_{bbb}=2, && c_{2b} = 22, && c_{2s} = 10.
\end{align}
The weight matrix is
\\

\noindent\begin{tabular}{l l l l l l | l }
$x_1$ & $x_2$ & $x_3$ & $x_4$ & $x_5$ & $x_6$ & CY \\
\hline
$0$ & $1$ & $1$ & $1$ & $1$ & $4$ & $8$ \\
$1$ & $1$ & $1$ & $0$ & $2$ & $5$ & $10$
\end{tabular}
\\[1em]

\noindent with SR ideal $\{x_1x_5,x_2x_3x_4x_6\}$. The irreducible toric divisors $D_i$ satisfy
\begin{equation}
D_1 = D_s, \quad D_2=D_3=D_b-D_s, \quad D_4 = D_b-2D_s, \quad D_5 = D_b, \quad D_6 = 4D_b-3D_s.
\end{equation}
The divisor specifying the Calabi-Yau hypersurface is $D_\text{CY}=8D_b-6D_s$.

The possible orientifold choices are
\\

\noindent\begin{tabular}{l | l l l | l}
 involution & $D_\text{O7}$ & $N_\text{O3}$ & $h^{2,1}_-$ & viable? \\
\hline
I1 & $D_s$ & $\int_X (D_2D_3D_6+D_4D_5D_6)=13$ & 64 & \texttimes \\
I2/I3 & $D_b-D_s$ & $\int_X (D_1D_3D_6+D_3D_4D_5)=5$ & 68 & \checkmark \\
I4 & $D_b-2D_s$ & &  & \texttimes \\
I5 & $D_b$ &  &  & \texttimes \\
I6 & $4D_b-3D_s$ & $\int_X D_1D_2D_3=1$ & 120 & \checkmark
\end{tabular}
\\[1em]

\noindent Note that the I1 and I6 involutions have fractional fixed points at $x_2=x_3=x_4=x_5=0$ which are not in the SR ideal. For a generic Calabi-Yau, we have $P_\text{CY}= x_6^2 +\mathcal{O}(\epsilon^4)$ at $x_{2,3,4,5}\sim \epsilon$. Since $x_6\neq 0$ at $\epsilon=0$ due to the SR ideal, the fixed points lie outside of a generic Calabi-Yau. The same is true for the orientifolds since the $x_6^2$ term is not projected out. Hence, we cannot conclude on a singularity. However, we still exclude the I1 orientifold since it does not have an O7-plane on $D_b$.

The I4 and I5 involutions have fractional fixed points at $x_1=x_2=x_3=x_6=0$ which are not in the SR ideal.
For a generic Calabi-Yau, we would have $P_\text{CY}= x_4^3 x_5^5 +\mathcal{O}(\epsilon^2)$ at $x_{1,2,3,6}\sim \epsilon$. Since $x_4,x_5\neq 0$ at $\epsilon=0$ due to the SR ideal, the fixed points would lie outside of a generic Calabi-Yau. However, for the I4 and I5 orientifolds, the $x_4^3 x_5^5$ term in $P_\text{CY}$ is projected out so that $P_\text{CY}|_{\epsilon=0} = 0$, $\partial_{x_i}P_\text{CY}|_{\epsilon=0} = 0$. We thus conclude that the I4 and I5 orientifolds are singular.

\subsection{ $M_{2,35}$}

The relevant topological data of the manifold are
\begin{align}
& \chi = -252, && \kappa_{sss}=2, && \kappa_{bbb}=1, && c_{2b} = 34, && c_{2s} = 4.
\end{align}
The weight matrix is
\\

\noindent\begin{tabular}{l l l l l l | l }
$x_1$ & $x_2$ & $x_3$ & $x_4$ & $x_5$ & $x_6$ & CY \\
\hline
$0$ & $0$ & $1$ & $1$ & $1$ & $3$ & $6$ \\
$1$ & $1$ & $0$ & $1$ & $2$ & $5$ & $10$
\end{tabular}
\\[1em]

\noindent with SR ideal $\{x_3x_4,x_1x_2x_5x_6\}$. The irreducible toric divisors $D_i$ satisfy
\begin{equation}
D_1 = D_2 = D_b-D_s, \quad D_3=D_s, \quad D_4 = D_b, \quad D_5 = 2D_b-D_s, \quad D_6 = 5D_b-2D_s.
\end{equation}
The divisor specifying the Calabi-Yau hypersurface is $D_\text{CY}=10D_b-4D_s$.

The possible orientifold choices are
\\

\noindent\begin{tabular}{l | l l l | l}
 involution & $D_\text{O7}$ & $N_\text{O3}$ & $h^{2,1}_-$ & viable? \\
\hline
I1/I2 & $D_b-D_s$ & $\int_X (D_2D_4D_6+D_2D_3D_5)=7$ & 70 & \checkmark \\
I3 & $D_s$ &  & & \texttimes \\
I4 & $D_b$ & $\int_X (D_3D_5D_6+D_1D_2D_6)=5$ & 72 & \checkmark \\
I5 & $2D_b-D_s$ & &  & \texttimes \\
I6 & $5D_b-2D_s$ & $\int_X D_1D_2D_4=1$ & 128 & \checkmark
\end{tabular}
\\[1em]

\noindent Note that the I3 and I5 involutions have fractional fixed points at $x_1=x_2=x_4=x_6=0$ which are not in the SR ideal. For a generic Calabi-Yau, we would have $P_\text{CY}= x_3x_5^5 +\mathcal{O}(\epsilon^2)$ at $x_{1,2,4,6}\sim \epsilon$. Since $x_3,x_5\neq 0$ at $\epsilon=0$ due to the SR ideal, the fixed points would lie outside of a generic Calabi-Yau. 
However, for the I3 and I5 orientifolds, the $x_3x_5^5$ term in $P_\text{CY}$ is projected out so that $P_\text{CY}|_{\epsilon=0} = 0$, $\partial_{x_i}P_\text{CY}|_{\epsilon=0} = 0$. We thus conclude that the I3 and I5 orientifolds are singular.

The I4 and I6 involutions have fractional fixed points at $x_1=x_2=x_3=x_5=0$ which are not in the SR ideal.
For a generic Calabi-Yau, we have $P_\text{CY}= x_6^2 +\mathcal{O}(\epsilon^2)$ at $x_{1,2,3,5}\sim \epsilon$. Since $x_6 \neq 0$ at $\epsilon=0$ due to the SR ideal, the fixed points lie outside of a generic Calabi-Yau. The same is true for the orientifolds since the $x_6^2$ term is not projected out. Hence, we cannot conclude on a singularity.

\subsection{ $M_{2,36}$}

The relevant topological data of the manifold are
\begin{align}
& \chi = -260, && \kappa_{sss}=2, && \kappa_{bbb}=2, && c_{2b} = 22, && c_{2s} = 4.
\end{align}
The weight matrix is
\\

\noindent\begin{tabular}{l l l l l l | l }
$x_1$ & $x_2$ & $x_3$ & $x_4$ & $x_5$ & $x_6$ & CY \\
\hline
$0$ & $1$ & $1$ & $1$ & $1$ & $4$ & $8$ \\
$1$ & $0$ & $0$ & $0$ & $1$ & $2$ & $4$
\end{tabular}
\\[1em]

\noindent with SR ideal $\{x_1x_5,x_2x_3x_4x_6\}$. The irreducible toric divisors $D_i$ satisfy
\begin{equation}
D_1 = D_s, \quad D_2=D_3=D_4=D_b-D_s, \quad D_5 = D_b, \quad D_6 = 4D_b-2D_s.
\end{equation}
The divisor specifying the Calabi-Yau hypersurface is $D_\text{CY}=8D_b-4D_s$.

The possible orientifold choices are
\\

\noindent\begin{tabular}{l | l l l | l}
 involution & $D_\text{O7}$ & $N_\text{O3}$ & $h^{2,1}_-$ & viable? \\
\hline
I1 & $D_b+D_s$ & $\int_X D_2D_3D_4=0$ & 78 & \checkmark \\
I2/I3/I4 & $D_b-D_s$ & $\int_X (D_3D_4D_5+D_1D_3D_4)=4$ & 74 & \checkmark \\
I5 & $D_b+D_s$ & $\int_X D_2D_3D_4=0$ & 78 & \checkmark \\
I6 & $4D_b-2D_s$ & $0$ & 132 & \checkmark
\end{tabular}
\\[1em]

\noindent There are no singularities related to fractional fixed points on this manifold. Note that the I6 involution has fractional fixed points at $x_2=x_3=x_4=x_5=0$ and $x_1=x_2=x_3=x_4=0$ which are not in the SR ideal. For a generic Calabi-Yau, we have $P_\text{CY}= x_6^2 +\mathcal{O}(\epsilon^2)$ at $x_{2,3,4}\sim \epsilon$. Since $x_6\neq 0$ at $\epsilon=0$ due to the SR ideal, the fixed points lie outside of a generic Calabi-Yau. The same is true for the orientifold since the $x_6^2$ term is not projected out. Hence, we cannot conclude on a singularity.

\bibliographystyle{utphys}
\bibliography{groups}

\end{document}